\documentclass[fleqn,10pt,showpacs]{revtex4}

\usepackage{amsmath}
\usepackage{amssymb}
\usepackage{graphicx,psfrag,epsfig}
\usepackage{afterpage}
\usepackage{mathrsfs,simplewick}

\newcommand{\be}{\begin{equation}} \newcommand{\ee}{\end{equation}}
\newcommand{\ba}{\begin{eqnarray}} \newcommand{\ea}{\end{eqnarray}}
\newcommand{\bea}{\begin{eqnarray}} \newcommand{\eea}{\end{eqnarray}}
\newcommand{\bean}{\begin{eqnarray*}} \newcommand{\eean}{\end{eqnarray*}}

\newcommand{\st}{{\scriptscriptstyle T}}

\def\slash#1{\setbox0=\hbox{$#1$}               % set a box for #1
        \dimen0=\wd0                            % and get its size
        \setbox1=\hbox{/} \dimen1=\wd1          % get size of /
        \ifdim\dimen0>\dimen1                   % #1 is bigger
        \rlap{\hbox to \dimen0{\hfil/\hfil}}    % so center / in box
        #1                                      % and print #1
        \else                                   % / is bigger
        \rlap{\hbox to \dimen1{\hfil$#1$\hfil}} % so center #1
        /                                       % and print /
        \fi}                                    %

\begin{document}

\title{Gauge links for Transverse Momentum Dependent Correlators at tree-level}

\author{M.G.A.~Buffing}
\email{m.g.a.buffing@vu.nl}
\affiliation{
Nikhef and Department of Physics and Astronomy, VU University Amsterdam,\\
De Boelelaan 1081, NL-1081 HV Amsterdam, the Netherlands}

\author{P.J.~Mulders}
\email{mulders@few.vu.nl}
\affiliation{
Nikhef and
Department of Physics and Astronomy, VU University Amsterdam,\\
De Boelelaan 1081, NL-1081 HV Amsterdam, the Netherlands}

\begin{abstract}
In this paper we discuss the incorporation of gauge links in hadronic 
matrix elements that describe the soft hadronic physics in high energy 
scattering processes. In this description the matrix elements appear in 
soft correlators and they contain non-local combinations of quark and 
gluon fields.  In our description we go beyond the collinear approach 
in which case also the dependence on transverse momenta of partons is 
taken into consideration. The non-locality in the transverse direction 
leads to a complex gauge link structure for the full process, in which 
color is entangled, even at tree-level. We show that at tree-level in 
a 1-parton unintegrated (1PU) situation, in which only the transverse 
momentum of one of the initial state hadrons is relevant, one can get 
a factorized expression involving transverse momentum dependent (TMD) 
distribution functions. We point out problems at the level of two 
initial state hadrons, even for relatively simple processes such as 
Drell-Yan scattering.
\end{abstract}
%\date{\today}
\pacs{12.38.-t; 13.85.Ni; 13.85.Qk}

\maketitle

\section{Introduction}

In the description of high energy processes involving hadrons one aims
at isolating the underlying hard process in terms of partons, 
quarks and gluons. The purpose of this can be twofold. The
aim might be to study the quark and gluon structure of hadrons, 
or it might be to account for the soft hadronic physics
to study unknown details in the hard process, for instance involving
physics beyond the Standard Model. In both cases one identifies a number
of soft functions among them distribution functions $f^{H\rightarrow i}(x)$ 
and fragmentation functions $D^{h\rightarrow i}(z)$ for 
quarks ($i = q$, where $q$ is an (anti-)quark flavor) and gluons ($i = g$), 
which have a natural interpretation as the probability
of finding a parton $i$ with momentum fractions $x$ in a hadron $H$
or as measure of the number of hadrons $h$ with momentum fractions $z$ in the 
`decay' of a parton $i$. We will refine the definitions of these fractions 
below. To some level of accuracy, one can express observables such as
cross sections and asymmetries in terms of these distribution and
fragmentation functions.
Going beyond the collinear treatment, one includes the 
dependence on transverse momenta. These momenta can serve as
degrees of freedom in the connection between hadrons and partons very
much like spin degrees of freedom. In the case of fragmentation, 
transverse momentum is in essence just the mismatch between 
parton momentum $k$ and hadron momentum $K_h$, or better between the 
fraction of the parton momentum $zk$ and $K_h$.
This, at least intuitively, corresponds for jet fragmentation 
to identifying the parton momentum with an appropriately
defined jet direction. Also for initial state hadrons one can include
dependence on transverse momentum, which is the 
mismatch between an appropriate fraction of the hadron momentum $xP$ 
and the parton momentum $p$.
In a high energy scattering process, one is able to use the presence
of a hard scale to identify parton momenta (integration variables)
with accessible combinations of external momenta. This is best known 
for the momentum fractions, but it is also possible for the 
transverse momenta. 

In the present paper, we remain at tree-level (to be made more explicit),
which implies that the intuitive language can in most cases also be used 
at the level of the matrix elements of quark and gluon fields that constitute 
correlators~\cite{Collins:1981uw}, which in turn are parametrized in terms 
of the beforementioned soft functions connecting partons and 
hadrons~\cite{Jaffe:1991kp}.
A complication that arises already at tree-level, is that the appropriate 
combinations of
quark and gluon fields in the correlators need to be gauge-invariant 
combinations. This would trivially be the case for local products of fields, 
but already in the collinear treatment which studies the dependence
on the momentum fractions $x$ and $z$, the partonic field combinations are
non-local along a light-like direction conjugate to the parton momentum.
Then one finds that gluon fields with polarizations along the momentum 
direction, which naturally appear in a twist analysis of leading
operators, need to be resummed to make up the required path ordered 
exponentials or Wilson lines connecting the non-local parton 
fields~\cite{Efremov:1978xm}. Although this involves
an infinite number of additional gluons, we still refer to this
resummation as tree-level, since one resums leading combinations of
coupling constant and field ($gA$). Although their momentum is integrated
over, these gluons don't appear in loops but as additional gluons connecting
the soft and hard parts and as such are at the same level as other partons. 
The procedure is in fact ensuring that the non-locality is color gauge
invariant, i.e.\ the replacement $i\partial_\mu \rightarrow iD_\mu =
i\partial_\mu + gA_\mu$.
The importance of the directions of gauge links was realized in
Ref.~\cite{Collins:2002kn}. Even in the case of a light-like
non-locality, the links are either past- or future-pointing, but in
the absence of transverse separation this feature becomes irrelevant
in the squared amplitude. 
In the case of transverse momentum dependent 
(TMD) correlators~\cite{Ralston:1979ys} and correspondingly TMD 
soft functions~\cite{Mulders:1995dh,Boer:1997nt}, one also must 
account for a transverse non-locality, requiring more complicated
Wilson lines. These Wilson lines have been extensively 
studied~\cite{Belitsky:2002sm,Boer:2003cm,Bomhof:2004aw,Bomhof:2006dp,Bomhof:2007xt,Gao:2010sd}.
The TMDs lead to a rich phenomenology of azimuthal 
asymmetries~\cite{Mulders:1995dh,Boer:1997nt,Bacchetta:2006tn,Barone:2010zz}. 
The distinction of past- and future-pointing gauge links provides
a natural explanation of single spin asymmetries at the partonic level.
Through the gauge links, time-reversal odd (T-odd) parts are incorporated
in the TMD correlators and the soft functions in their parametrization
within a field theoretical framework of Quantum Chromodynamics (QCD).
Beyond tree-level, many complications 
arise~\cite{Cherednikov:2010uy,Collins:2011,Aybat:2011zv}, 
of which at present certainly not all implications have been investigated.
Although higher orders in QCD may invalidate any tree-level results, we
will follow here the {\em diagrammatic approach} outlined in the above
that provides us with the basic field theoretical picture which needs 
verification in an all-order QCD treatment.

In section~\ref{section2} we introduce some of the basics of TMDs needed
in the rest of the paper.
As said, a particularly interesting feature for the TMD soft functions 
entering the description of hard processes already at tree-level 
is the non-trivial nature of the Wilson lines connecting the non-local
field operators in the matrix element. This structure of Wilson lines, 
arising from both collinearly and some of the transversely polarized 
gluon fields, now becomes dependent on the color flow in the hard process.
In particular, in those situations that multiple color flow possibilities
exist, this gives rise to an entanglement that can spoil factorization
already at tree-level. 
How one gets the basic tree-level entangled result in a diagrammatic
approach is outlined in section~\ref{section3}.

The aim of this paper is to show that this entanglement simplifies
for the 1-parton unintegrated (1PU) case, by which we refer to a situation 
in which only the transverse momentum in one of the hadrons
is manifest. What remains is at tree-level a factorized expression
with a correlator that still does have process dependence, which is 
contained in a non-trivial process-dependent gauge link. This is
sometimes referred to as {\em generalized factorization}.
The proof is given in section~\ref{section4}. We will also show how one 
can proceed if transverse
parton momenta in several hadrons are involved. In such cases one 
can consider weighted asymmetries expressed in terms of transverse 
moments of the TMD functions. These weighted asymmetries actually
again involve only collinear functions, but these functions are given by matrix 
elements of higher twist operators, among them gluonic pole matrix elements
or Efremov-Teryaev-Qiu-Sterman (ETQS) functions.
In section~\ref{section5}, we will use transverse moments to analyse 
single and double weighted asymmetries and illustrate
this for Drell-Yan (DY) scattering and for a process with
quark-quark scattering as underlying hard
partonic process. The complications and non-universality always 
involves the gluonic pole matrix elements, which have been extensively
studied~\cite{Efremov:1981sh,Efremov:1984ip,Qiu:1991pp,Qiu:1991wg,Qiu:1998ia,Kanazawa:2000hz,Eguchi:2006mc,Koike:2006qv}. Since such matrix elements
vanish for fragmentation~\cite{Gamberg:2008yt,Meissner:2008yf,Gamberg:2010uw}, 
we will not have to worry
about the transverse momentum in the final state. We will comment on
this further in our conclusions.

\section{Collinear and Transverse Momentum Dependent (TMD) 
Correlators\label{section2}}

We have split up this paper in a number of sections, in which we discuss 
in a diagrammatic expansion the inclusion of all gluon fields that contribute
at leading order and tree-level. The starting point is a hard subprocess, 
for which we will consider as the most generic example
a two to two process with a truncated amplitude 
$\mathscr M (p_1,p_2;k_1,k_2)$, from which the wave functions of
the partons (Dirac spinors $u(p_1)$ for quarks, or polarizations 
$\epsilon(p_1)$ for gluons), are omitted. Rather than through 
wave functions, the external partons are accounted for through
quark or gluon correlation or spectral functions, which are built 
from matrix elements of the form $\langle X\vert \psi(\xi)\vert P\rangle$
involving hadron states $\vert P\rangle$ rather than a free parton wave
function $\langle 0\vert\psi(\xi)\vert p\rangle$. This immediately brings
in the need to also consider multi-parton matrix elements with the same 
states, such as $\langle X\vert A^\mu(\eta)\,\psi(\xi)\vert P\rangle$.

\begin{figure}[b]
\epsfig{file=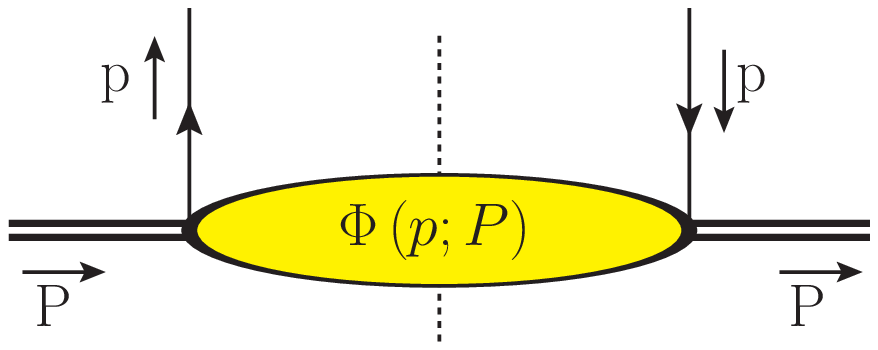,width=0.27\textwidth}
\hspace{1.5cm}
\epsfig{file=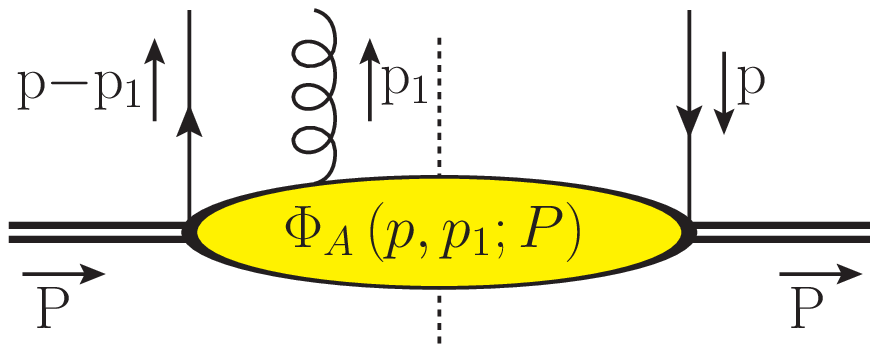,width=0.27\textwidth}
\hspace{1.5cm}
\epsfig{file=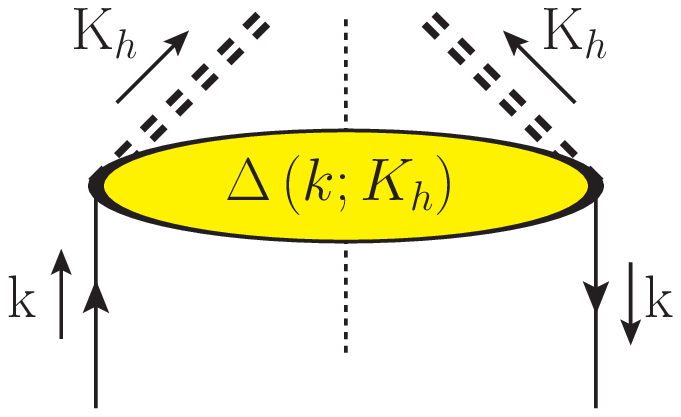,width=0.21\textwidth}
\\[0.2cm]
(a)\hspace{6.0cm} (b)\hspace{5.5cm} (c)
\caption{\label{fig1}
The pictorial momentum space representation of quark-quark correlators for 
distribution functions (a), a
quark-quark-gluon correlator (b) and a quark-quark correlator for
fragmentation functions (c).} 
\end{figure}

These matrix elements appear as squared contributions in the 
correlators (including Dirac space indices $i$ and $j$),
\bea
\Phi_{ij}(p;P) & = &
\sum_X\int \frac{d^3P_X}{(2\pi)^3\,2E_X}
\ \langle P\vert \overline\psi_j(0)\vert X\rangle
\,\langle X\vert \psi_i(0)\vert P\rangle\,\delta^4(p+P_X-P)
\nonumber \\ & = &
\frac{1}{(2\pi)^4}\int d^4\xi\ e^{i\,p\cdot \xi}
\ \langle P\vert \overline\psi_j(0)\,\psi_i(\xi)\vert P\rangle ,
\eea
pictorially represented in Fig.~\ref{fig1}(a). 
Usually, a summation over color indices is understood. This means that
we will have $\Phi(p) = {\rm Tr}_c\bigl[\Phi(p)\bigr]$, where
$\Phi_{ij}(p)$ is considered also a matrix in color space,
made explicit 
$\Phi_{ij;rs} \propto \psi_{ir}(\xi)\,\overline\psi_{js}(0)$.
Including gluon fields one has quark-quark-gluon correlators like
\be
\Phi^\mu_{A\,ij}(p,p_1;P) =
\frac{1}{(2\pi)^8}\int d^4\xi\,d^4\eta
\ e^{i\,(p-p_1)\cdot \xi}\ e^{i\,p_1\cdot \eta}
\ \langle P\vert \overline\psi_j(0)\,A^\mu(\eta)\,\psi_i(\xi)\vert P\rangle,
\label{quarkgluonquark}
\ee
illustrated in Fig.~\ref{fig1}(b), 
and similarly matrix elements with more partons.
The color structure of the field combination $\psi_r(\xi)\overline\psi_s(0)$ 
in the quark-quark-gluon correlator now actually has a color octet structure,
denoted (when appropriate) as $\Phi_8 = {\rm Tr}_c[\Phi\,T^a]T^a$.
Using for $A^\mu = A^{\mu a}T^a$ a matrix-valued field we have
$\Phi_A^\mu = {\rm Tr}_c[\Phi_8 A^\mu]$.
In some cases, it will be convenient to explicitly use the momentum 
space fields defined as 
$\psi(p) \equiv \int d^4\xi\ e^{i\,p\cdot \xi}\,\psi(\xi)$ and
$A_\mu(p) \equiv \int d^4\xi\ e^{i\,p\cdot \xi}\,A_\mu(\xi)$,
which in the case of free fields would have parton and anti-parton
contributions multiplying on-shell factors 
$(2\pi)\,\delta(p^2-m^2)\,\theta(\pm p^0)$.
For the correlators, we get
\bea
&&
(2\pi)^4\delta^4(p-p^\prime)\,\Phi_{ij}(p;P)
= \frac{1}{(2\pi)^4}
\,\langle P\vert \overline\psi_j(p^\prime)\,\psi_i(p)\vert P\rangle,
\\&&
(2\pi)^4\delta^4(p-p^\prime)\,\Phi^\mu_{A\,ij}(p,p_1;P)
= \frac{1}{(2\pi)^8}\,\langle P\vert \overline\psi_j(p^\prime)\,A^\mu(p_1)
\,\psi_i(p-p_1)\vert P\rangle.
\eea
The corresponding correlator describing fragmentation into hadrons
is for quarks given by
\begin{eqnarray}
\Delta_{ij}(k;K_h) & = & \sum_X \frac{1}{(2\pi)^4}
\int d^4\xi\ e^{-ik\cdot \xi}\,
\langle 0 \vert \psi_i(0) \vert K_h, X \rangle
\langle K_h,X \vert \overline \psi_j(\xi)
\vert 0 \rangle \nonumber \\
& = & \frac{1}{(2\pi)^4}\int d^4\xi\ e^{-ik\cdot \xi}\,
\langle 0 \vert \psi_i (0) a_h^\dagger
a_h \overline \psi_j(\xi) \vert 0 \rangle,
\label{frag}
\end{eqnarray}
pictorially represented by the blob in Fig.~\ref{fig1}(c).
An averaging over color indices is implicit,
thus $\Delta(k) = \frac{1}{N_c}{\rm Tr}_c\bigl[\Delta(k)\bigr]$ with
again $\Delta(k)$ a diagonal matrix in color space. 
The second expression in the above involves hadronic creation and annihilation 
operators $a_h^\dagger\vert 0\rangle = \vert K_h\rangle$.
In a momentum space representation for the operators, we have
\be
(2\pi)^4\delta^4(k-k^\prime)\,\Delta_{ij}(k;K_h)
= \frac{1}{(2\pi)^4}\sum_X \langle 0 \vert \psi_i(k) \vert K_h, X \rangle
\langle K_h,X \vert \overline \psi_j(k^\prime) \vert 0 \rangle.
\ee
In fragmentation correlators, one no longer
deals with plane wave hadronic states, but with out-states
$\vert K_h,X\rangle$.

\begin{figure}[b]
\epsfig{file=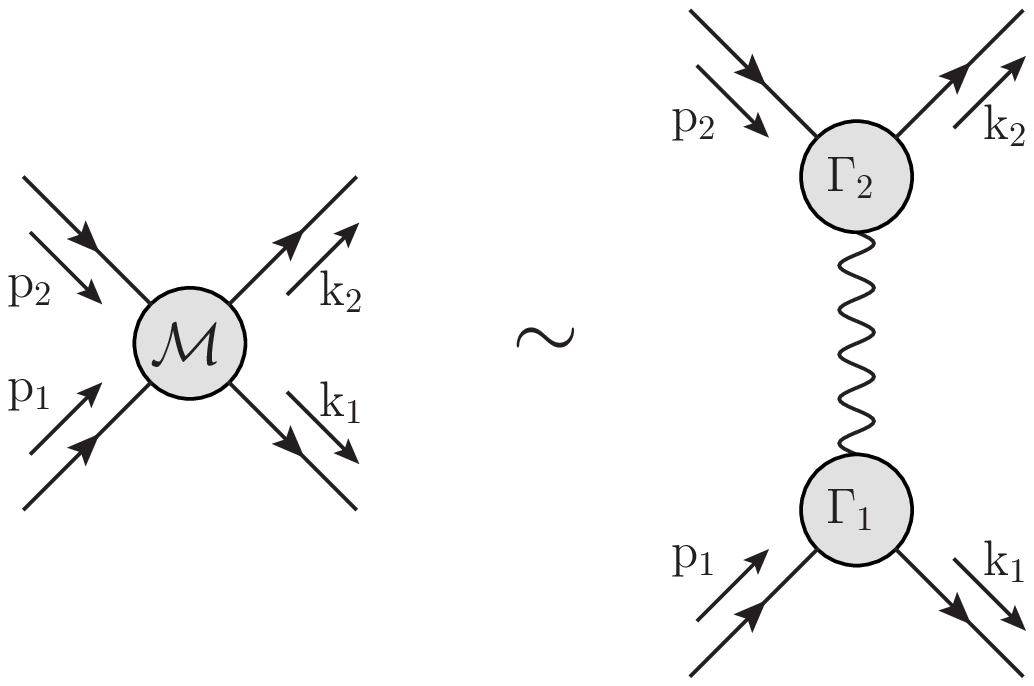,width=0.34\textwidth}
\hspace{2cm}
\epsfig{file=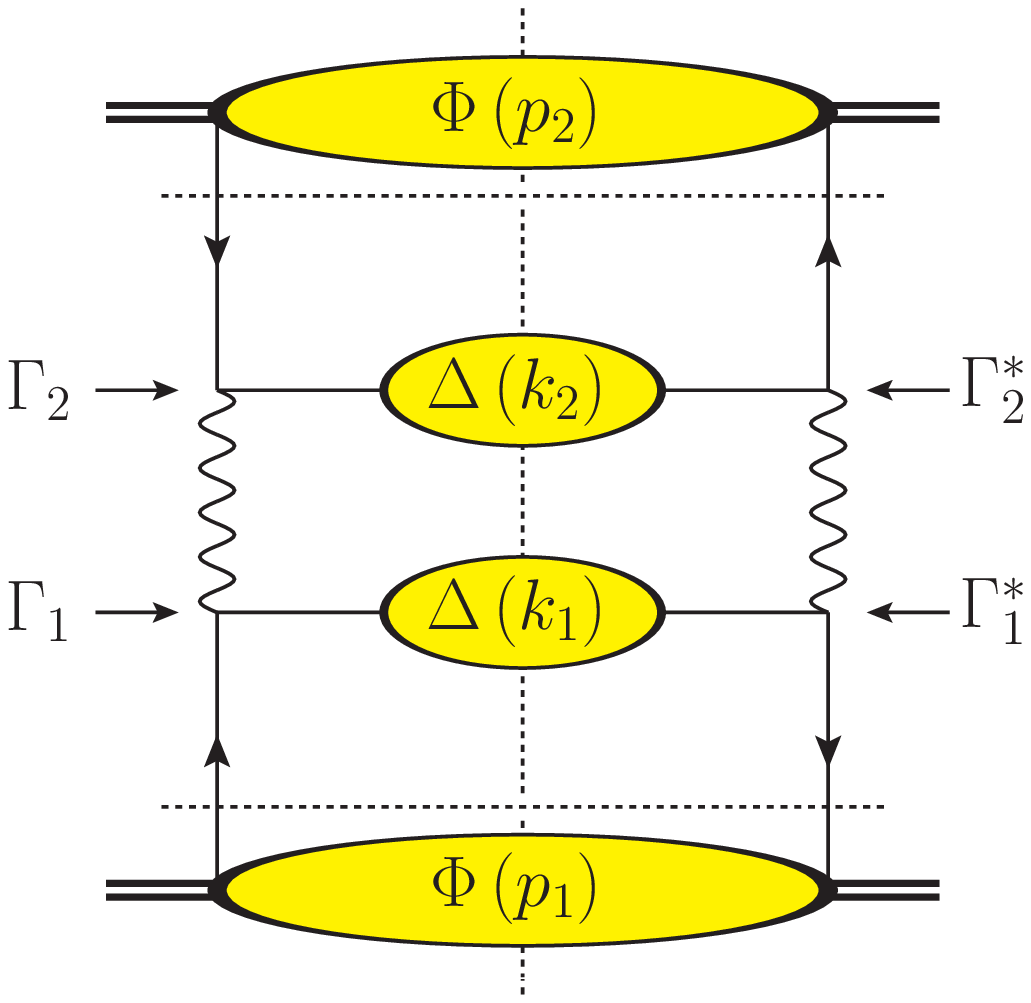,width=0.3\textwidth}
\\[0.2cm]
\mbox{}\hspace{1cm}(a)\hspace{6.5cm} (b)
\caption{\label{fig2}
For the purpose of illustrating the structure of Wilson lines, we use the
hard amplitude with one particular color flow for the quark lines as
shown in (a). The squared amplitude needed for the cross section 
of the scattering process initiated by two hadrons with momenta 
$P_1$ and $P_2$ is shown in (b). }
\end{figure}
We want to give an expression for the cross section of the (semi)-inclusive
process $H_1(P_1) + H_2(P_2) \rightarrow h_1(K_1) + h_2(K_2) + \dots$
in a kinematic regime where $P_1{\cdot}p_1 \sim P_2{\cdot}p_2 
\sim K_1{\cdot}k_1
\sim K_2{\cdot}k_2$ are small (we will refer to this scale as
the squared hadronic mass scale, $M^2 \sim 1$ GeV$^2$) as compared to
the usual hard invariants in the full or the partonic process such as
$s \approx 2\,P_1{\cdot}P_2$, $t_1 \approx -2\,K_1{\cdot}P_1$,
$\hat s \approx 2\,p_1{\cdot}p_2$, $\hat t \approx -2\,k_1{\cdot}p_1$
(we will refer to this scale as the squared hard scale, $Q^2 \gg M^2$).
We note in passing that in cases where heavy quarks are involved, 
those quark masses of course have to be included. 
In a hard process as described here, we aim for a description in which 
the squared partonic amplitude $\vert \mathscr M\vert^2$ is convoluted 
with the correlators $\Phi(p,P)$, $\Delta(k,K_h)$, etc.
In order to illustrate the use of the correlators, assume
an amplitude $\mathscr M \propto \Gamma_1\,\Gamma_2$ 
as illustrated in Fig.~\ref{fig2}(a). The simplest tree-level diagrammatic
contributions to the cross section that can be written down is shown 
in Fig.~\ref{fig2}(b) and is of the form
\be
d\sigma \sim 
{\rm Tr}_c\bigl[\Phi(p_1)\,\Gamma_1^\ast\,\Delta(k_1)\,\Gamma_1\bigr]
\,{\rm Tr}_c\bigl[\Phi(p_2)\,\Gamma_2^\ast\,\Delta(k_2)\,\Gamma_2\bigr],
\label{facto1}
\ee
where ${\rm Tr}_c\bigl[\ldots\bigr]$ parts are traced over color.
This expression still needs to be integrated over the parton momenta,
which will be discussed below.
In the case that the vertices $\Gamma$ don't have any color structure,
one can, because of the simple color singlet structure of $\Phi$ and $\Delta$ 
in the quark-quark correlators, perform the color trace separately 
for $\Phi$ and $\Delta$,
${\rm Tr}_c[\Phi(p)\,\Gamma^\ast\,\Delta(k)\,\Gamma] 
= {\rm Tr}_c[\Phi(p)]\tfrac{1}{N_c}{\rm Tr_c}[\Delta(k)]
\,\Gamma\,\Gamma^\ast$
(one summed and one averaged) and the cross section can be written in 
terms of the color-traced entities 
\be
d\sigma \sim 
\Phi(p_1)\,\Phi(p_2)
\,\underbrace{\Gamma_1\,\Gamma_1^\ast\,\Gamma_2\,\Gamma_2^\ast}_{\hat\Sigma}
\,\Delta(k_1)\,\Delta(k_2),
\label{facto1a}
\ee
where the remaining contractions are Dirac space and Lorentz indices, which
have been suppressed in both Eqs~\ref{facto1} and \ref{facto1a}.
These expressions of course should be extended with all possible
correlators containing quark and gluon fields, in which cases color
traces become more complicated because the quark-quark part of the
correlator can have a color octet structure ($\Phi_8$).
The restriction to hard kinematics limits the number 
of diagrammatic contributions, although even at leading order, there 
still are many gluon contributions as will be discussed in 
Section~\ref{section3}. These will complicate the color tracing, 
an effect which is particularly important when dependence on transverse 
momenta is considered, which precisely is our goal in this paper. 
We will only work out leading contributions in an $M/Q$ expansion,
although the separation of various orders requires care, as we will
argue in Section~\ref{section3}.
The $M^2/Q^2$ effects certainly cannot be calculated in our diagrammatic
approach. At that level, there are many contributions that spoil 
already at tree-level
the possibility to write down in a consistent way a cross section in the 
form of Eq.~\ref{facto1} and certainly to perform the color traces as in
Eq.~\ref{facto1a}.

For parton momenta relevant in a hadron correlator (hadron momentum $P$)
we make the Sudakov decomposition,
\be
p = x\,P + p_\st + \underbrace{(p{\cdot}P - x\,M^2)}_{\sigma}\,n,
\ee
where the role of the (approximately) light-like vector $n$, satisfying
$P{\cdot}n = 1$ can come from any of the hard (external) momenta, 
e.g.\ $n = K_h/K_h{\cdot}P$
or $n = k/k{\cdot}P$ (provided $k{\cdot}P \sim K_h{\cdot}P \sim Q^2$). 
The momentum fraction $x = p{\cdot}n = p^n$ is $\mathscr O(1)$. For
any contractions with vectors outside the correlator $\Phi(p,P)$ one
has $P \sim Q$, $p_\st \sim M$ and $n \sim 1/Q$.
Note that if $n$ is an exact light-like vector, 
one can construct {\em two} exact conjugate null-vectors,
\be
n_+ = P -\tfrac{1}{2}\,M^2\,n \quad \mbox{and} \quad n_- = n ,
\label{lcdef}
\ee
satifying
$n_+{\cdot}n_- = 1$ and $n_+^2 = n_-^2 = 0$, that can be used to define
light-cone components
$a^\pm = a{\cdot}n_\mp$ (thus $x = p{\cdot}n_- = p^+$).
The symmetric and antisymmetric `transverse' projectors are defined as
\bea
&& g_\st^{\mu\nu} = g^{\mu\nu} - n_+^{\{\mu}n_-^{\nu\}}
%= g^{\mu\nu} - \frac{P^{\{\mu}n^{\nu\}}}{P{\cdot}n}
%+ \frac{M^2}{(P{\cdot}n)^2}\,n^\mu n^\nu
= g^{\mu\nu} - P^{\{\mu}n^{\nu\}}+M^2\,n^\mu n^\nu
\approx g^{\mu\nu} - P^{\{\mu}n^{\nu\}},
\\
&& \epsilon_\st^{\mu\nu} = \epsilon^{n_+n_-\mu\nu}
= \epsilon^{-+\mu\nu}
= \epsilon^{Pn\mu\nu}.
\eea
Since transverse momentum dependence is a central issue in this paper,
we have to worry about different $n$ vectors. With 
$\Delta n = n^\prime - n \sim 1/Q$ one has at $\mathscr O(Q^0)$
that $\Delta x \approx \Delta p_\st^2 \approx 0$ although the 
transverse momentum itself does change at order $\mathscr O(1)$, 
$\Delta p_\st = -\Delta x\,P$.

The integration over parton momenta,
\be
\int d^4p = \int dx\,d^2p_\st\,d\sigma
=\int d(p{\cdot}n)\,d^2p_\st\,d(p{\cdot}P),
\ee
is insensitive to the particular $n$ vector.
In view of the relative importance of the components in this integration,
one can, upon neglecting any $M^2/Q^2$ contributions in the cross section,
integrate within a soft correlator over $p{\cdot}P$ (i.e.\ $p^-$) to 
obtain the TMD correlator
\be
\Phi(x,p_\st;n) = \int d\,p{\cdot}P\ \Phi(p;P)
= \left. \int \frac{d\,\xi{\cdot}P\,d^2\xi_\st}{(2\pi)^3}\ e^{i\,p\cdot \xi}
\ \langle P\vert \overline\psi(0)\,\psi(\xi)\vert P\rangle \right|_{LF}\ ,
\ee
which we will still consider as the unintegrated correlator.
On the left-hand side the dependence on hadron momentum $P$ has been
suppressed.
In the TMD correlator the non-locality is restricted to the 
light-front (LF: $\xi{\cdot}n = \xi^+ = 0$) and the correlator depends 
on $x = p{\cdot}n$ and $p_\st$. 
The light-cone correlators are the collinear correlators containing the
parton distribution functions depending only on the light-cone momentum
fraction $x$, obtained upon integration over both $p{\cdot}P$ and $p_\st$,
\be
\Phi(x;n) = \int d\,p{\cdot}P\,d^2p_\st\ \Phi(p;P)
= \left. \int \frac{d\,\xi{\cdot}P}{(2\pi)} \ e^{i\,p\cdot \xi}
\ \langle P\vert \overline\psi(0)\,\psi(\xi)\vert P\rangle \right|_{LC}\ ,
\ee
where the subscript LC refers to light-cone, implying $\xi{\cdot}n$ = $\xi_\st$
= 0. This integration is generally allowed in hard 
processes up to $M^2/Q^2$ contributions and also up to contributions coming 
from the tails, e.g.\ logarithmically divergent contributions proportional to
$\alpha_s(p_\st^2)/p_\st^2$ tails~\cite{Bacchetta:2008xw} and relevant
when looking at evolution~\cite{Henneman:2001ev}. 
Such contributions, however, require next-to-leading order (NLO) QCD, 
which goes beyond the tree-level resummations that we discuss in this paper. 
In diagrammatic language they for instance involve ladder graphs describing
emission of gluons into the final state, relevant for the evolution of 
the correlators.
The collinear correlators are relevant in hard processes in which 
only hard scales (large invariants $\sim Q^2$ or ratios thereof, angles, 
rapidities) are measured.
If one considers hadronic scale observables (correlations or transverse
momenta in jets, slightly off-collinear configurations) one will need
the TMD correlators for a full treatment. 

The correlators encompass the information on the soft parts. They
depend on the hadron and quark momenta $P$ and $p$ (and in general also
spin vectors). Depending on the Lorentz and Dirac structure of the matrix 
elements involved one can look for the pieces in the correlator
that show up as the most dominant matrix elements among the 
contributions in the hard process.
Including also gluon fields, the Fourier transform of matrix elements 
with a maximal number of contractions with $n$,
\be
\langle\ \overline\psi(0)\slash n\psi(\xi)\ \rangle
\qquad \mbox{and} \qquad
\langle\ G^{n\alpha}(0)G^{n\beta}(\xi)\ \rangle,
\label{nonlocalfields}
\ee
(the latter with transverse indices $\alpha$ and $\beta$) are the
dominant combinations that appear in the correlators. They are the
dominant ones because the contractions with
$n$ lower the canonical dimension of the operator combination, 
minimizing the power of $M$ that after contractions of open indices
inevitably is the scale of the hadronic matrix elements. 
The two matrix elements above have canonical dimension two. 
The corresponding local matrix elements, 
$\overline\psi(0)\,\slash n\,\psi(0)$ and
$G^{n\alpha}(0)\,G^{n\beta}(0)$ for quarks and gluons, 
respectively, are color gauge-invariant (twist 2) operators,
the non-local combinations in Eq.~\ref{nonlocalfields} are not gauge
invariant. Expanded into local operators, the expansion would involve 
operator combinations with derivatives such as 
$\overline\psi(0)\,\slash n\,\partial^n\ldots \partial^n\,\psi(0)$.
Color gauge invariance in the correlators requires in the
local matrix elements covariant derivatives or in the non-local
matrix elements the presence of a gauge link connecting the two fields. 
For the light-cone correlators the gauge link corresponds to
the inclusion of an arbitrary number of `leading' gluon fields 
$A^n(\eta)$ in the field combinations in Eq.~\ref{nonlocalfields} 
which are resummed into a gauge link
$\overline\psi(0)\,W^{[n]}_{[0,\xi]}\,\psi(\xi)$
= ${\rm Tr}_c\bigl[W^{[n]}_{[0,\xi]}\,\psi(\xi)\overline\psi(0)\bigr]$, 
given by
\be
W^{[n]}_{[0,\xi]}
= {\mathscr P}\,\exp\left(-i\int_0^\xi d\,\eta{\cdot}P\ n{\cdot}A(\eta)\right).
\label{glcol}
\ee
Including this gauge link, the non-local operator combinations 
\be
\left.\langle\ \overline\psi(0)\slash n\,W^{[n]}_{[0,\xi]}\psi(\xi)
\ \rangle\right|_{LC}
\qquad \mbox{and} \qquad
\left.\langle\ G^{n\alpha}(0)\,W^{[n]}_{[0,\xi]}
G^{n\beta}(\xi)\,W^{[n]}_{[\xi,0]}\ \rangle\right|_{LC},
\label{nonlocalfields-2}
\ee
can be expanded into twist two operators 
$\overline\psi(0)\,\slash n\,D^n\ldots D^n\,\psi(0)$ and
$G^{n\alpha}(0)\,D^n\ldots D^n\,G^{n\beta}(0)$ for quarks and gluons, 
respectively (number of $D^n$'s is the spin of these operators).
Also TMD correlators require a gauge link, but the separation of the
two fields is no longer a simple light-like one and they involve
derivatives with transverse indices. It is important to realize that
in principle any gauge link with an arbitrary path gives a 
gauge-invariant combination. What is the appropriate link contributing
at leading order (in $M/Q$) in a given
hard scattering process, however, is calculable (see next section).

The correlators in this section have been rewritten into matrix elements 
of non-local products of fields. They involve both quark and gluon fields
as well as hadronic states or hadronic creation and annihilation
operators. For them there is no systematic perturbative expansion in terms
of the strong coupling constant. The kinematic separation of soft and
hard, however, allows the integration over $p^- = p{\cdot}P$, leaving
a product of operators at the light-front, that is at equal light-cone
time $\xi^+ = \xi{\cdot}n = 0$. For such a product the time-ordering is
automatic, which means that the $p^-$-integrated parton correlators, thus,
can be considered as a cut anti-parton-hadron scattering amplitude,
i.e.\ a Green function, untruncated in the parton 
legs~\cite{Jaffe:1983hp}. This is the
case for both collinear and TMD correlators~\cite{Diehl:1998sm}. This 
identification has been very important in deep inelastic 
processes~\cite{Landshoff:1971xb}, allowing the
use of analyticity and unitarity properties of field theories, at least 
under the assumption that these properties apply to QCD. We will use it
later in this paper for fragmentation correlators.

\section{Color gauge invariance\label{section3}}

It is worthwhile to repeat the steps that lead to color gauge invariance
by including collinear gluon fields, which will also be the first step 
to obtain the Wilson lines in TMD correlators. 
We include in the diagrammatic approach matrix
elements with gluon fields for which we make a Sudakov expansion,
\be
A^\mu(\eta) = A^n(\eta)\,P^\mu + A_\st^\mu(\eta)
+\left(A^P(\eta) - A^n(\eta)\,M^2\right)n^\mu .
\label{A-sudnaive}
\ee
A similar expansion can be written down for $A^\mu(p)$. In order to
use Ward identities it will be convenient to look at the (collinear) 
gluon field component along parton momentum $p^\mu$, hence we write
\bea
A^\mu(p)
& = & \int d^4\eta\ e^{i\,p\cdot \eta}\,A^\mu(\eta)
\nonumber
\\ &=& \int d^4\eta\ e^{ip\cdot \eta}\,\Biggl[
\frac{A^n(\eta)}{p{\cdot}n}\,p^\mu
%\nonumber\\ & & \mbox{} \hspace{2.1cm}
+ \frac{(p{\cdot}n)\,A_\st^\mu(\eta) - p_{\st}^\mu\,A^n(\eta)}{p{\cdot}n}
%\nonumber\\ & & \mbox{} \hspace{2.1cm}
+\frac{(p{\cdot}n)\,A^P(\eta) - (p{\cdot}P)\,A^n(\eta)}
{p{\cdot}n}\,n^\mu \Biggr].
\eea
In the correlator the momentum $p^\mu \longrightarrow i\partial^\mu(\eta)$,
so
\bea
A^\mu(p)
& = & \frac{1}{p{\cdot}n}\int d^4\eta\ e^{i\,p\cdot \eta}
\,\Biggl[ A^n(\eta)\,p^\mu
%\nonumber\\ & & \mbox{} \hspace{2.9cm}
+ i\partial^n(\eta)\,A_\st^\mu(\eta)
- i\partial_{\st}^\mu (\eta)\,A^n (\eta)
%\nonumber\\ & & \mbox{} \hspace{2.9cm}
+ \left(i\partial^n(\eta) A^P(\eta)
- i\partial^P(\eta)A^n (\eta)\right)n^\mu \Biggr]
\label{A-sud0}
\\
& = & \frac{1}{p{\cdot}n}\,\Biggl[
A^n(p)\,p^\mu
+ i\,G_\st^{n\mu}(p)
+i\,G^{nP}(p)\,n^\mu\Biggr].
\label{A-sud}
\eea
Although the latter appears to be only true for the Abelian case,
we will find the same result in the non-abelian case, but to complete
that proof, we first need to incorporate the collinear gluons into
the matrix elements. Using the expansion in Eq.~\ref{A-sud} rather than
the one in Eq.~\ref{A-sudnaive} streamlines the inclusion of 
collinear gluons circumventing the explicit treatment of transverse momentum
dependent parts (as done in Ref.~\cite{Boer:2003cm}). The results are
of course identical.

\begin{figure}
\epsfig{file=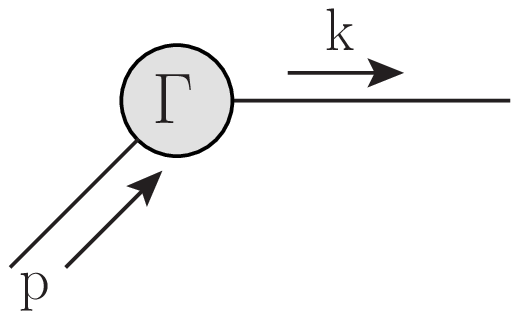,width=0.15\textwidth}
\hspace{1cm}
\epsfig{file=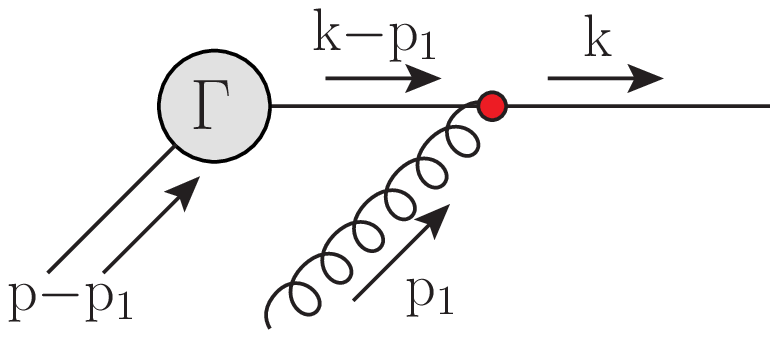,width=0.23\textwidth}
\hspace{1cm}
\epsfig{file=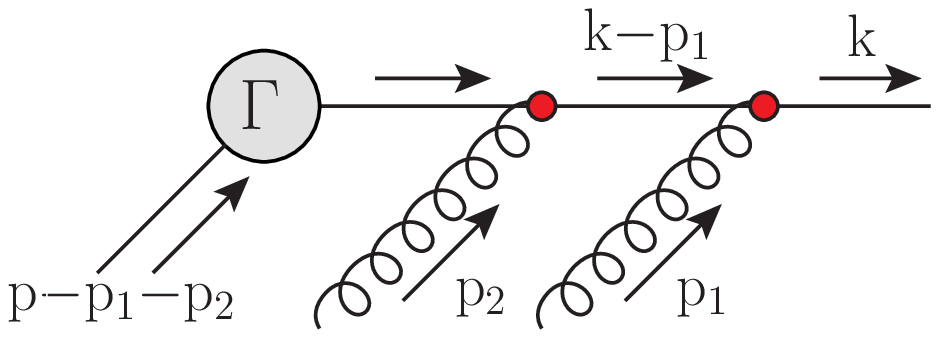,width=0.27\textwidth}
\\[0.2cm]
$A_0$\hspace{5.5cm} $A_1$\hspace{5cm} $A_2$
\caption{\label{fig3}
Inclusion of collinear gluons from 
$\Phi_{A\ldots A}(p-p_1\ldots - p_N,p_1,\ldots ,p_N)$ 
coupling to an outgoing (colored) quark line with momentum $k$.}
\end{figure}

\subsection{Collinear gluons}

Since matrix elements involving 
$\overline\psi(0)\,A^n(\eta_1)\ldots A^n(\eta_N)\,\psi(\xi)$ and
$G^{n\alpha}(0)\,A^n(\eta_1)\ldots A^n(\eta_N)\,G^{n\beta}(\xi)$ 
for quarks and gluons are as leading as the matrix elements without
collinear gluons, the contributions of correlators with gluon fields
of which the polarization is along its momentum $p^\mu$, i.e.\ the first
term in Eq.~\ref{A-sud} need to
be included in the leading expression for the cross section. 
Using Ward identities, only the contributions of gluons coupling to
external parton lines survive as discussed in detail in 
Ref.~\cite{Bomhof:2006dp}.
The resummation is best illustrated by looking at the example of
gluons attached to a $\Phi(p)$ correlator attached to an outgoing
quark line with momentum $k$ (see Fig.~\ref{fig3}). The results of these
parts within the amplitude are
\[
A_0 = \overline\psi(k)\Gamma\psi(p),
\]
where the $\psi(p)$ and $\overline \psi(k)$ are fields belonging to the
correlators of initial (momentum $p$) and final state quark (momentum $k$)
respectively. The one-gluon contribution to the link is
\[
A_1 = \int \frac{d^4p_1}{(2\pi)^4}\ \overline\psi(k)
\,\frac{-i\,\slash p_1\,A^k(p_1)}{p_1\cdot k}
\,\frac{i(\slash k - \slash p_1)}{(-2k\cdot p_1+i\epsilon)}
\,\Gamma\,\psi(p-p_1).
\]
The numerator becomes $\slash p_1(\slash k -\slash p_1)$ =
$\slash p_1\slash k$ = $\{\slash k,\slash p_1\}$ = $2\,k\cdot p_1$. The added
term is zero since $\overline \psi(k)\slash k \approx 0$. Thus one has
(note that the sign of $k\cdot p_1$ is positive),
\bea
A_1 &=& \overline\psi(k) \int \frac{d^4p_1}{(2\pi)^4}
\ \frac{A^k(p_1)}{(-k\cdot p_1+i\epsilon)}
\,\Gamma\,\psi(p-p_1)
= {\bcontraction{\overline\psi(k)\,}{U}{{}_+^{[k](1)}\,\Gamma}{\psi}
\overline\psi(k)\,U_+^{[k](1)}\,\Gamma\psi(p)},
%= \overline\psi(k) \,T^a\,\Gamma\,U_+^{a[k](1)}\psi(p),
\eea
which is the one-gluon contribution
to a path-ordered exponential (see Appendix~\ref{GL}).
The two-gluon term becomes
\bea
A_2 &=&
\int \frac{d^4p_1}{(2\pi)^4}\,\frac{d^4p_2}{(2\pi)^4}\,
\ \overline\psi(k)
\,\frac{A^k(p_1)}{p_1\cdot k}\,\frac{A^k(p_2)}{p_2\cdot k}
\slash p_1\,\frac{1}{\slash k - \slash p_1}
\slash p_2\,\frac{1}{\slash k - \slash p_1-\slash p_2}
\,\Gamma\,\psi(p-p_1-p_2)
\nonumber \\ &=&
\int \frac{d^4p_1}{(2\pi)^4}\,\frac{d^4p_2}{(2\pi)^4}\,
\ \overline\psi(k)
\,\frac{A^k(p_1)}{p_1\cdot k}\,\frac{A^k(p_2)}{p_2\cdot k}
\,\frac{\slash p_1\,(\slash k - \slash p_1)}{(-2k\cdot p_1+i\epsilon)}
%\nonumber \\ &&\mbox{}\hspace{3cm}\times
\,\frac{\slash p_2\,(\slash k - \slash p_1-\slash p_2)}
{(-2k\cdot p_1-2k\cdot p_2+i\epsilon)}
\,\Gamma\,\psi(p-p_1-p_2)
\nonumber \\
&=&
\overline \psi(k) \int \frac{d^4p_1}{(2\pi)^4}\,\frac{d^4p_2}{(2\pi)^4}
\ \frac{A^k(p_1)}{(-k\cdot p_1+i\epsilon)}
\,\frac{A^k(p_2)}{(-k\cdot p_1-k\cdot p_2+i\epsilon)}
\,\Gamma\,\psi(p-p_1-p_2)
\nonumber
\\&=&
{\bcontraction{\overline\psi(k)\,}{U}{{}_+^{[k](2)}\,\Gamma}{\psi}
\overline\psi(k)\,U_+^{[k](2)}\,\Gamma\psi(p)} .
\eea
This is the two-gluon contribution to the path-ordered exponential. 
This term illustrates the recursive procedure that gives
the all-gluon results to the full path-ordered exponential or Wilson line
(see Appendix~\ref{GL}),
\be
\sum_{N=0}^\infty A_N
= {\bcontraction{\overline\psi(k)\,}{U}{{}_+^{[k]}\,\Gamma}{\psi}
\overline\psi(k)\,U_+^{[k]}\,\Gamma\psi(p)},
\ee
where
\be
\bcontraction{}{U}{{}_+^{[n]}}{\psi}
U_+^{[n]}\psi(p) = 
\int d^4\xi\ \exp\left(i\,p\cdot \xi\right)\,
\mathscr{P}\exp\left(-ig\int_{\infty}^{\xi\cdot P}
d(\eta\cdot P)\ A^n(\eta)\right)\,\psi(\xi)
\ee
is the Fourier transform of the field including the Wilson line
of which the indices of $U_+^{[n]}$ indicate its direction, namely
a Wilson line running 
from light-cone $+\infty$ to $\xi$ along the light-like direction $n$.
We will instead of the notation with the contraction, mostly use the notation
\be
{\bcontraction{\overline\psi(k)\,}{U}{{}_+^{[k]}\,\Gamma}{\psi}
\overline\psi(k)\,U_+^{[k]}\,\Gamma\psi(p)}
= \overline\psi(k)\,U_+^{[k]}[p]\,\Gamma\psi(p).
\ee
The second expression gives the Wilson line a label $[p]$, indicating
that all fields in it belong to the correlator $\Phi(p)$, 
to which also $\psi(p)$ belongs. The problem with this path-ordered
exponential is that it is a unitary matrix in color space that is stuck 
in the respective traces, although 
it is in fact only the (symmetric) color charge operators 
$T^{a_1}\ldots T^{a_s}$ 
of the terms $U_+^{[n](s)}$ in the expansion of the $U_+^{[n]}$ that are
stuck there. 

For the gluon insertions coming from a particular correlator and
coupling to an incoming 
fermion line one finds a Wilson line connecting to light-cone $-\infty$,
which is a consequence of the sign $k\cdot p_i$ being negative in
that situation.
Including all multi-gluon interactions originating from $\Phi(p_1)$ in
Fig.~\ref{fig2}(b), we get the diagrammatic result
\bea
d\sigma & \sim & {\rm Tr}_c\bigl[\Phi(p_1)\,\Gamma_1^\ast
\,U_+^{[k_1]\dagger}[p_1]\Delta(k_1)U_+^{[k_1]}[p_1]\,\Gamma_1\bigr]
\nonumber \\&& \mbox{}\times
{\rm Tr}_c\bigl[U_-^{[p_2]\dagger}[p_1]\,\Phi(p_2)\,U_-^{[p_2]}[p_1]
\,\Gamma_2^\ast
\,U_+^{[k_2]\dagger}[p_1]\,\Delta(k_2)\,U_+^{[k_2]}[p_1]\,\Gamma_2\bigr],
\label{facto2}
\eea
with the (color charge of the) Wilson line stuck in the color traces
at the `positions' corresponding to the external parton lines.
Note that in Eq.~\ref{facto2} the Wilson lines in sub-expressions like 
$U_+^{[k_1]\dagger}[p_1]\Delta(k_1)U_+^{[k_1]}[p_1]$ 
are part of the Fourier transform in the correlator $\Phi(p_1)$. 
In coordinate space, taking $\xi_1$ to be the coordinate conjugate 
to $p_1$, the link and conjugate link
actually run between different points, in this case $U_+$ in
$U_+^{[k_1]}[p_1]\ldots \psi(p_1)$ corresponds actually to 
$U^{[k_1]}_{[+\infty,\xi_1]}$, while the conjugate link $U_+^\dagger$ in
$\overline \psi(p_1)\ldots U_+^{[k_1]\dagger}[p_1]$ corresponds to
$U^{[k_1]}_{[0_1,+\infty]}$. 
Wilson lines between $0_{1\st}$ and $\xi_{1\st}$ are still lacking,
so Eq.~\ref{facto2} is certainly not color gauge-invariant.

The next step in our treatment is the inclusion of gluon 
interactions coming from different correlators, say $\Phi(p_1)$
and $\Phi(p_2)$, coupling to the same quark line with
momentum $k_2$, such as shown e.g.\ in Fig.~\ref{knot}(a). 
They give rise to intertwined Wilson lines.
Two examples of this have been given in Appendix~\ref{GLB},
for the case of two gluons coming from different correlators and
for the case of three gluons, two coming from the same correlator 
and one from a different correlator. These examples
illustrate the recursive procedure.
The result for all insertions to a particular leg is a color symmetric
combination of the insertions from all correlators, which since
all gluon polarizations are identical is just
\be
U_+^{[k_2]}[p_1,p_2,k_1]=
{\mathcal S}\{U_+^{[k_2]}[p_1]U_+^{[k_2]}[p_2]U_+^{[k_2]}[k_1]\},
\label{linkbreak1}
\ee
in which the ordering of the three connections on the right-hand side is 
irrelevant (fully symmetrized). This is illustrated in Fig.~\ref{knot}.
\begin{figure}
\epsfig{file=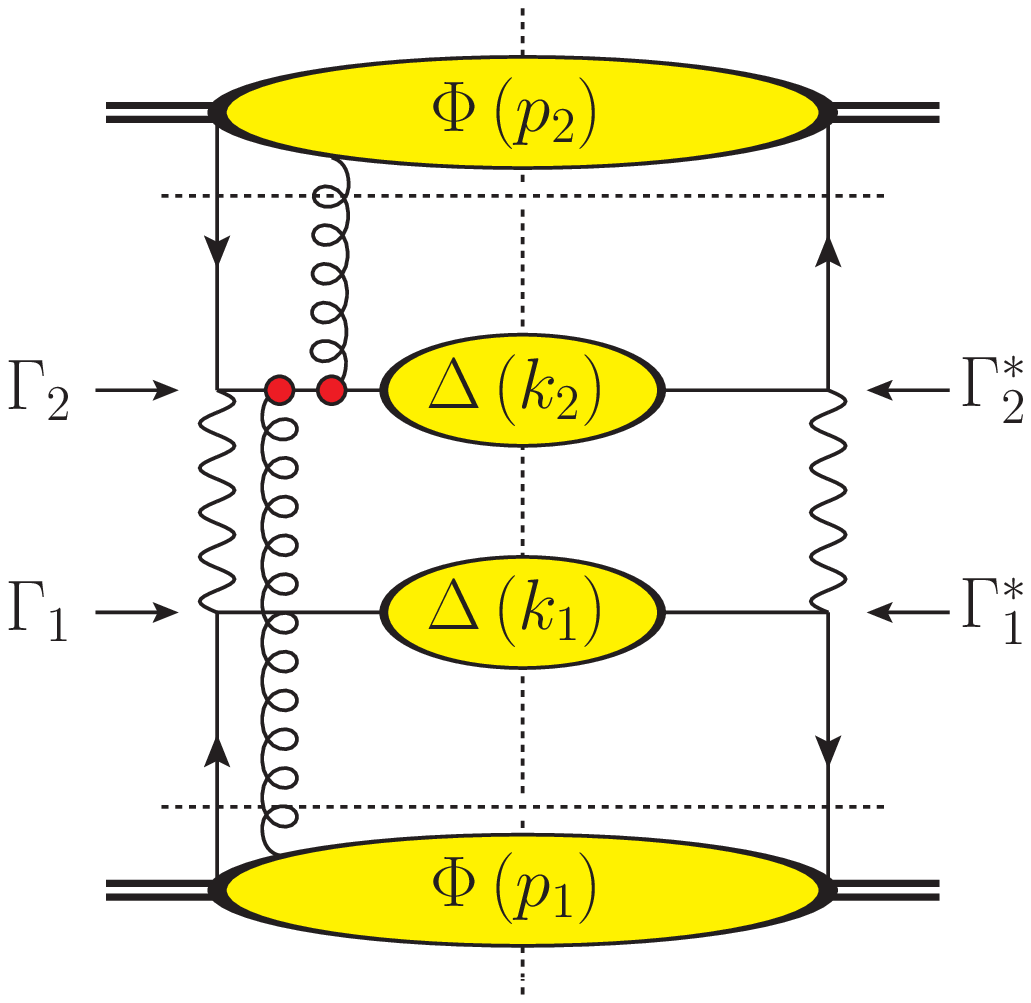,width=0.3\textwidth}
\hspace{1cm}
\epsfig{file=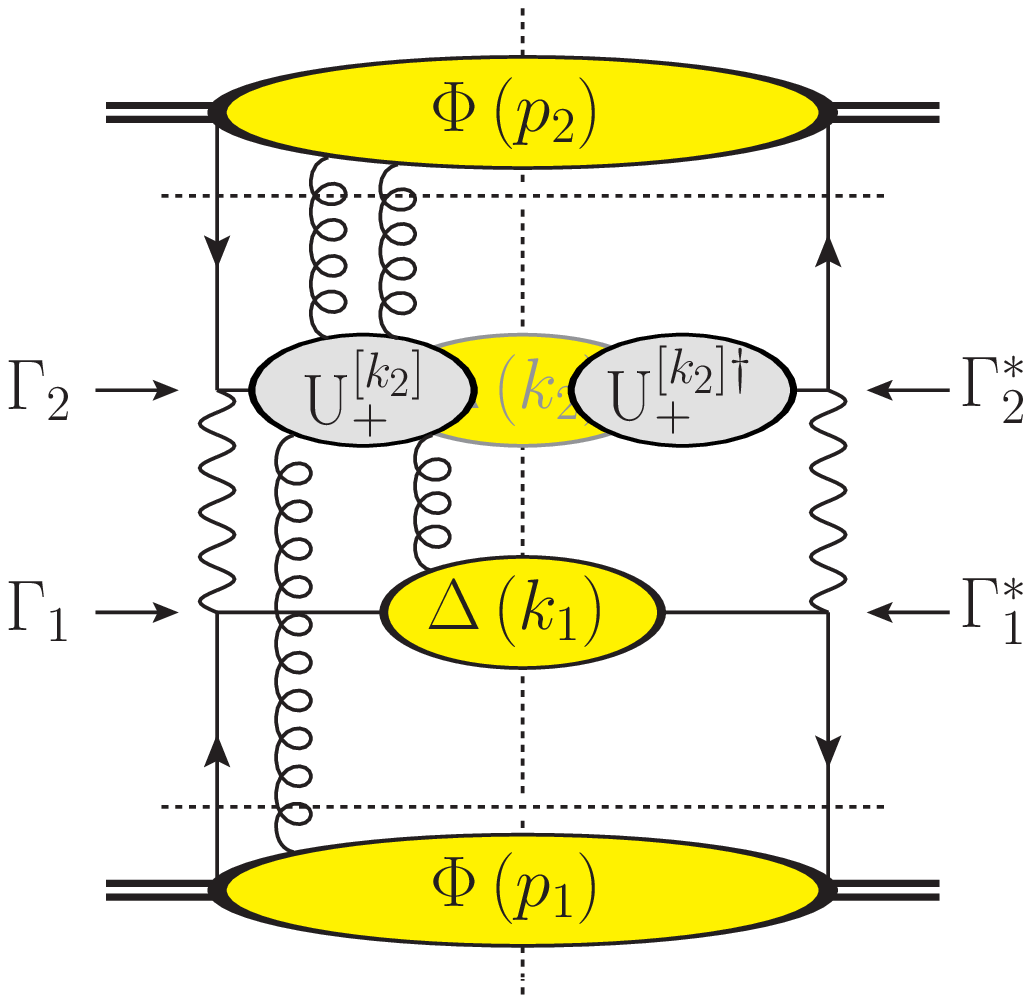,width=0.3\textwidth}
\\[0.2cm]
(a)\hspace{7cm} (b)
\caption{\label{knot}
(a) An example of a diagram with two gluons attaching to the same outgoing
line with momentum $k_2$. (b) In more detail these couplings are 
discussed in the appendices 
and they result into one gauge connection $U^{[k_2]}_+[p_1,p_2,k_1]$,
which combines all collinear gluons coming from $\Phi(p_1)$, $\Phi(p_2)$
and $\Delta(k_1)$. 
Together with transverse pieces, and combining it with gauge connections
at other legs one will get a
full color gauge invariant result, discussed in Section~\ref{section3}.}
\end{figure}
Including all multi-gluon interactions from $\Phi(p_1)$, $\Delta(k_1)$, 
$\Phi(p_2)$ and $\Delta(k_2)$ onto all legs, Eq.~\ref{facto2} generalizes to
\bea
d\sigma & \sim & {\rm Tr}_c\bigl[U_-^{[p_1]\dagger}[p_2,k_1,k_2]\,\Phi(p_1)
\,U_-^{[p_1]}[p_2,k_1,k_2]\,\Gamma_1^\ast
\,U_+^{[k_1]\dagger}[p_1,p_{2},k_{2}]\Delta(k_1)
\,U_+^{[k_1]}[p_{1},p_{2},k_{2}]\,\Gamma_1\bigr]
\nonumber \\&& \mbox{}\times
{\rm Tr}_c\bigl[U_-^{[p_2]\dagger}[p_1,k_1,k_2]
\,\Phi(p_2)\,U_-^{[p_2]}[p_1,k_1,k_2]
\,\Gamma_2^\ast\,U_+^{[k_2]\dagger}[p_1,p_2,k_1]\,\Delta(k_2)
\,U_+^{[k_2]}[p_1,p_2,k_1]\,\Gamma_2\bigr],
\label{facto3}
\eea
illustrated in Fig.~\ref{cgi-result}(a).
We note that the Wilson lines in Eq.~\ref{facto3} have different 
light-like directions, which originate from the fact that we in
the decomposition of gluon fields in Eq.~\ref{A-sud} simply made the
most convenient choice depending on the particular correlator.
For two different Sudakov decompositions of the gluon field, one finds that 
\be
\frac{A^{n}(p)}{p{\cdot}n}-\frac{A^{n^\prime}(p)}{p{\cdot}n^\prime}
= \frac{i\,G^{\,nn^\prime}(p)}{(p{\cdot}n)(p{\cdot}n^\prime)}. 
\ee
The field $G^{\,nn^\prime}(p)$, however, appearing in a correlator 
$\Phi^{n\,n^\prime}_G$ will not contribute at leading order, but
at subleading order ($1/Q$). This allows one to replace all the
light-like direction dependence in Eq.~\ref{facto3} by a generic
null-vector $n$. Even after choosing one light-like direction, the 
result in Eq.~\ref{facto3} is not yet gauge-invariant, since transverse 
gauge connections are still missing. 

\subsection{Collinear correlators}

The missing transverse pieces don't matter when one takes a collinear
approach, implying integration over transverse
momenta $p_\st$ besides the integration over $p\cdot P$. In that case
there is no transverse separation of the fields. 
After integration over transverse momenta,
one has a color gauge-invariant result,
\bea
\sigma & \sim & {\rm Tr}_c\bigl[U_-^{[n]\dagger}[p_2,k_1,k_2]\,\Phi(x_1)
\,U_-^{[n]}[p_2,k_1,k_2]\,\Gamma_1^\ast
\,U_+^{[n]\dagger}[p_1,p_{2},k_{2}]\Delta(z_1)
\,U_+^{[n]}[p_{1},p_{2},k_{2}]\,\Gamma_1\bigr]
\nonumber \\&& \mbox{}\times
{\rm Tr}_c\bigl[U_-^{[n]\dagger}[p_1,k_1,k_2]\,\Phi(x_2)\,U_-^{[n]}[p_1,k_1,k_2]
\,\Gamma_2^\ast 
\,U_+^{[n]\dagger}[p_1,p_2,k_1]\,\Delta(z_2)
\,U_+^{[n]}[p_1,p_2,k_1]\,\Gamma_2\bigr].
\label{facto3a}
\eea
One can combine the Wilson lines to and from light-cone $\pm \infty$, 
all made up of $A^n$ fields, into finite Wilson lines, 
e.g.\ $W^{[n]}[p_1] = U_+^{[n]\dagger}[p_1]\,U_+^{[n]}[p_1]$ since
after the integration over $p_{1\st}$ they not only both run along $n$, 
but they coincide since one also has $0_\st = \xi_\st$. 
Furthermore, it is irrelevant if one composes $W^{[n]}$ from Wilson lines 
running via plus or via minus infinity, and also the direction $n$ 
is in fact irrelevant. It is 
just the direction of the straight line connecting $0$ and $\xi$. 
We recall that the argument $p_1$ or $x_1$, given to the Wilson lines, 
is simply needed to indicate that the fields in that Wilson line belong 
to the correlator 
$\Phi(x_1)$, which is the Fourier transform of the matrix element
$\langle \overline \psi(0)\psi(\xi)\rangle$. 
Thus in coordinate space one just has the Wilson line in Eq.~\ref{glcol}, 
which connects the points $0$ and $\xi$ in $\Phi(x_1)$ composed of
$W^{[n]}_{[0,\xi]} = U_{[0,\infty]}^{[n]}\,U_{[\infty,\xi]}^{[n]}$.
As far as relevant for $\Phi(x_1)$, the Wilson lines
in the first trace form a gauge link, those in
the second trace form a closed loop,
which in the collinear situation (when $0_\st = \xi_\st$) becomes a
unit operator in color space.
One is left with
\bea
\sigma & \sim & {\rm Tr}_c\bigl[U_-^{[n]\dagger}[k_1]\,\Phi(x_1)\,U_-^{[n]}[k_1]
\,\Gamma_1^\ast\,U_-^{[n]\dagger}[p_1]\Delta(z_1)\,U_-^{[n]}[p_1]\bigr]
\nonumber \\&& \mbox{}\times
\,{\rm Tr}_c\bigl[U_-^{[n]\dagger}[k_2]\,\Phi(x_2)\,U_-^{[n]}[k_2]
\,\Gamma_2^\ast\,U_-^{[n]\dagger}[p_2]\,\Delta(z_2)\,U_-^{[n]}[p_2]\,\Gamma_2\bigr].
\eea
The way of turning the gauge connections into gauge links
at the collinear stage is actually just
applying gauge transformations $U^{[n]}_{[a,\xi]}$ (with a
fixed point $a$) to all fields. This can actually directly be applied
to Eq.~\ref{facto3a}.
One obtains
\be
\sigma  \sim  
\Phi^{[W]}(x_1) \Phi^{[W]}(x_2)
\,\underbrace{\Gamma_1 \,\Gamma_1^\ast \,\Gamma_2 \,\Gamma_2^\ast}_{\hat\Sigma}
\,\Delta^{[W]}(z_1) \,\Delta^{[W]}(z_2),
\label{facto3b}
\ee
where
\bea
\Phi^{[W]}(x) & = & {\rm Tr}_c\bigl[U_\pm^{[n]}[p]\Phi(x)
\,U_\pm^{[n]\dagger}[p]\bigr]
= \left.\int \frac{d\,\xi{\cdot}P}{2\pi}\ e^{i\,p\cdot \xi}
\ \langle P\vert \overline\psi(0)\,U_{[0,\pm\infty]}^{[n]}
\,U_{[\pm\infty,\xi]}^{[n]} \,\psi(\xi)\vert P\rangle
\right|_{LC} 
\nonumber \\ &=&
{\rm Tr}_c\bigl[W^{[n]}[p]\,\Phi(x)\bigr]
=  \left.\int \frac{d\,\xi{\cdot}P}{2\pi}\ e^{i\,p\cdot \xi}
\ \langle P\vert \overline\psi(0)\,W^{[n]}_{[0,\xi]}
\,\psi(\xi)\vert P\rangle
\right|_{LC} 
\eea
and
\bea
\Delta^{[W]}(z) & = & \frac{1}{N_c}\,{\rm Tr}_c\bigl[U_\pm^{[n]}[k]\Delta(z)
\,U_\pm^{[n]\dagger}[k]\bigr]
= \frac{1}{N_c}\,{\rm Tr}_c\bigl[\Delta(z)\,W^{[n]\dagger}[k]\bigr]
\nonumber \\ & = &
\left. \int \frac{d\,\xi{\cdot}K_h}{2\pi}\ e^{-ik\cdot \xi}\,
\frac{1}{N_c}\,{\rm Tr}_c \langle 0\vert U_{[\pm\infty,0]}^{[n]}\psi (0) 
a_h^\dagger a_h \overline \psi(\xi)\,U_{[\xi,\pm\infty]}^{[n]} \vert 0 \rangle
\right|_{LC} 
\eea
are the color gauge-invariant collinear correlators,
including unique gauge links $W$ along the light-like separation.
The gauge link being unique, it is usually omitted.
These (color gauge-invariant) correlators can be
expanded in terms of the standard parton distribution functions and
fragmentation functions, respectively. 

For the correlator $\Phi(x)$ it is also possible to circumvent 
manipulating with Wilson lines by taking moments in $x_1 = p_1{\cdot}n$, 
$x_2 = p_2{\cdot}n$, $1/z_1 = k_1{\cdot}n$ and
$1/z_2 = k_2{\cdot}n$. Using
\be
i\,\partial^n_\xi\,W^{[n]}_{[\eta,\xi]} = W^{[n]}_{[\eta,\xi]}\,iD^n(\xi),
\label{glderivative}
\ee
one easily verifies the factorized expression in Eq.~\ref{facto3b}.
In the process of taking moments one then encounters $x^N\,\Phi^{[W]}(x)$ 
involving matrix elements with covariant derivatives $D^n$,
\be
x^N\Phi^{[W]}(x) 
=  \left.\int \frac{d\,\xi{\cdot}P}{2\pi}\ e^{i\,p\cdot \xi}
\ \langle P\vert \overline\psi(0)\,W^{[n]}_{[0,\xi]}
(iD^n)^N\psi(\xi)\vert P\rangle
\right|_{LC} .
\label{momentscollinear}
\ee
We have shown everything for one correlator, but 
one has similar expressions for the other correlators.
Since the gauge link is unique for the collinear correlators, we will
just write $\Phi(x_1)$, $\Phi(x_2)$, $\Delta(z_1)$ and $\Delta(z_2)$.

\subsection{Including transverse gauge connections}

\begin{figure}[tb]
\epsfig{file=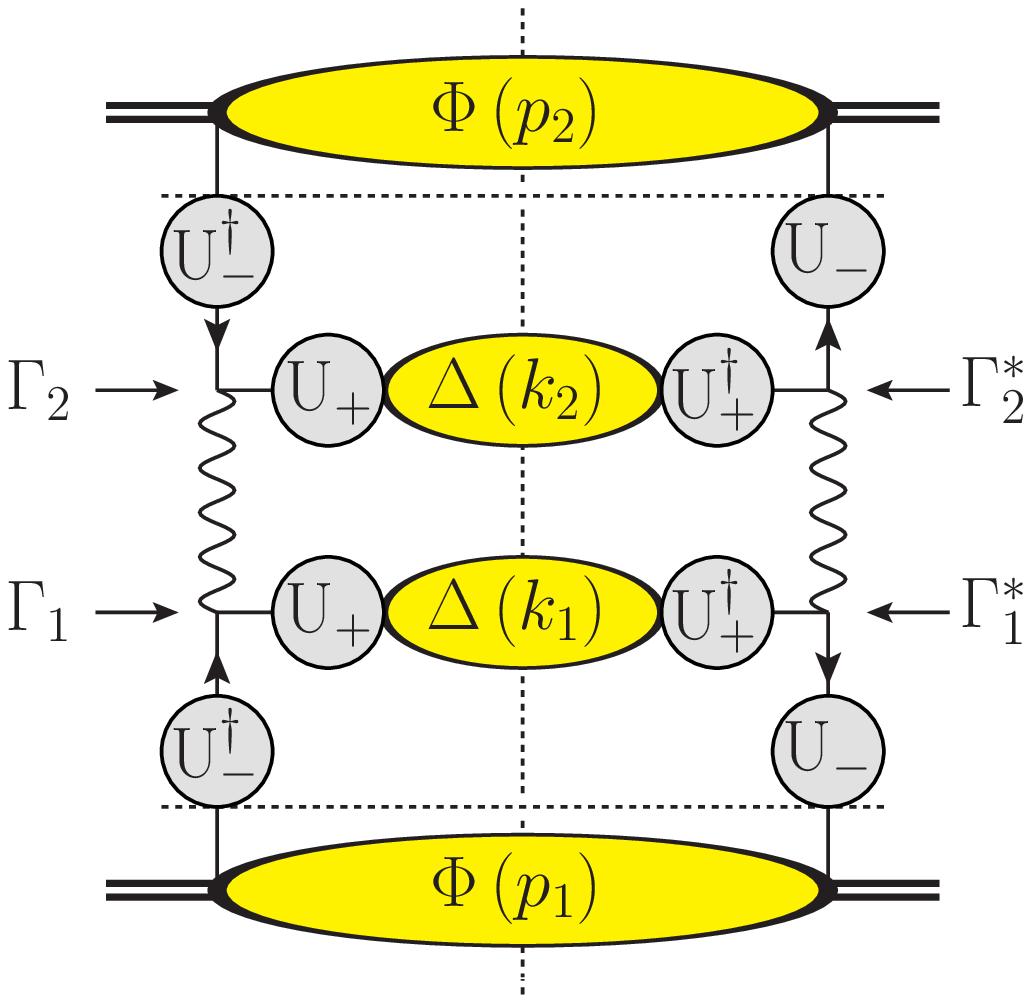,width=0.3\textwidth}
\hspace{1cm}
\epsfig{file=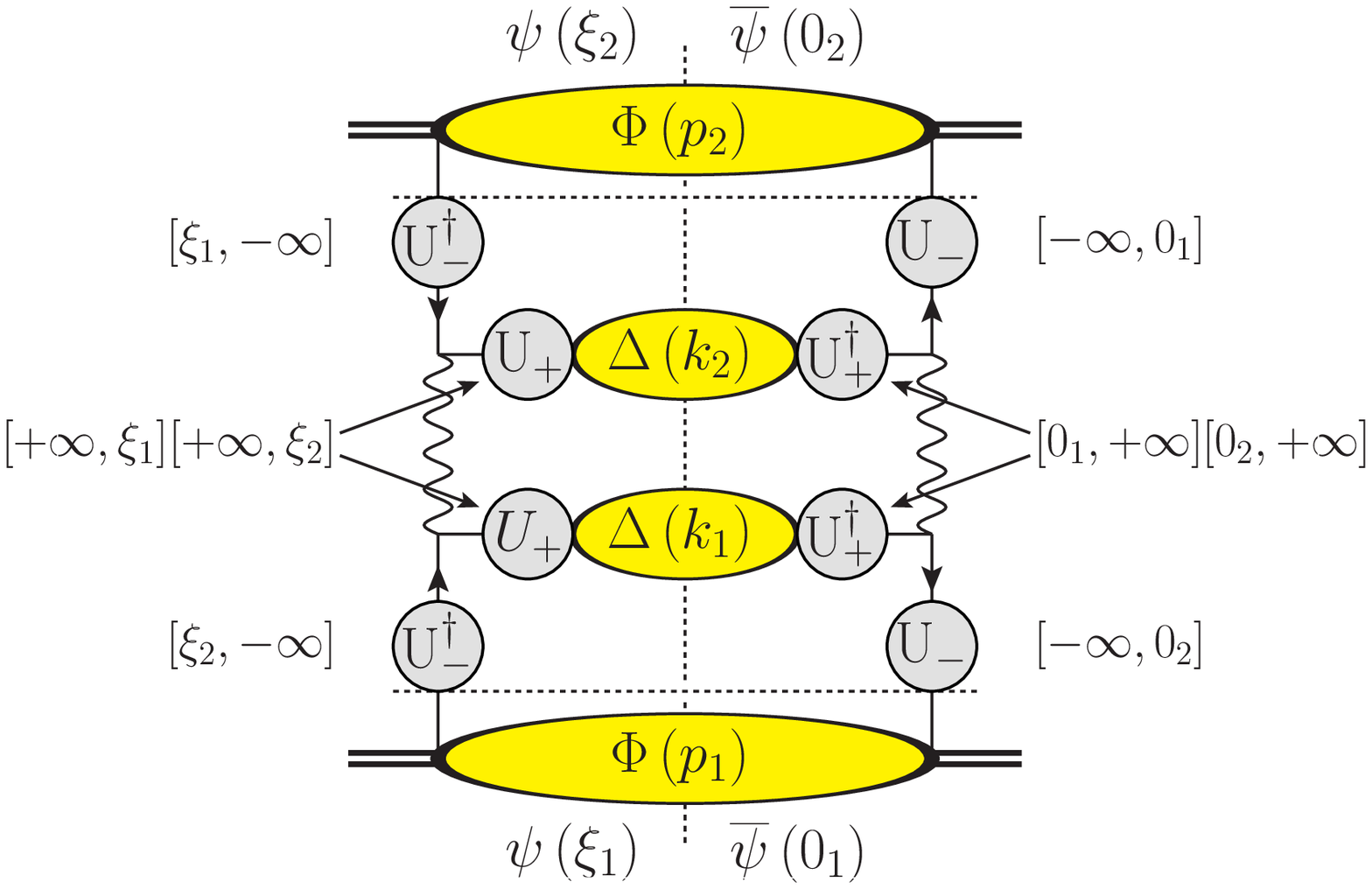,width=0.5\textwidth}
\\[0.2cm]
(a)\hspace{5.5cm} (b)
\caption{\label{cgi-result}
(a) The gauge connections from all collinear gluons from the various
soft correlators produce (entangled) gauge connections with color
charges located on the external legs of the hard part.
(b) We have indicated for the correlators and gauge connections also the 
actual space-time points they are bridging limiting ourselves for simplicity
to the coordinates conjugate to $p_1$ (points $0_1$ and $\xi_1$; see also
the discussion following Eq.~\ref{facto2}) 
and $p_2$ (points $0_2$ and $\xi_2$), leaving out
the space-time structure for the fragmentation correlators
for which we would have to include also the coordinates conjugate 
to $k_1$ and $k_2$.}
\end{figure}
To see how transverse gauge connections arise, we jump back to Eq.~\ref{facto3}
and note that not only the
$\langle \ldots \overline \psi(0)\,\psi(\xi)\ldots\rangle$ correlator
has acquired gauge connections along the $n$-direction, but also other
matrix elements involving the other gluon components in Eq.~\ref{A-sud0},
e.g.\ the operator combination 
\[
\langle \ldots
i\,\partial^n(\eta) U^{[n]}_{[\infty,\eta]}\,A_\st^\alpha(\eta)
\,U^{[n]}_{[\eta,\infty]}
-i\,\partial_\st^\alpha(\eta) U^{[n]}_{[\infty,\eta]}\,A^n(\eta)
\,U^{[n]}_{[\eta,\infty]}\ldots \rangle,
\]
for which we can use the non-abelian relation (directly based on
Eq.~\ref{glderivative}),
\be
U^{[n]}_{[\infty,\eta]}\,A_\st^\alpha(\eta)\,U^{[n]}_{[\eta,\infty]}
- A_\st^\alpha(\eta)\big|_{\eta^P{=}\infty}
= \left. \int_\infty^{\,\eta^P} d\zeta^P
\ U^{[n]}_{[\infty,\zeta]}\biggl(G^{n\alpha}(\zeta)
+ [\partial_\st^\alpha,A^n(\zeta)]\biggr)
U^{[n]}_{[\zeta,\infty]}\right|_{\zeta^n = \eta^n,\zeta_\st = \eta_\st}.
\label{AT1}
\ee
Differentiation with respect to $\partial^n_\eta = \partial/\partial \eta^P$
gives
\[
\langle \ldots
i\,\partial^n(\eta) U^{[n]}_{[\infty,\eta]}\,A_\st^\alpha(\eta)
\,U^{[n]}_{[\eta,\infty]}
-i\,\partial_\st^\alpha(\eta) U^{[n]}_{[\infty,\eta]}\,A^n(\eta)
\,U^{[n]}_{[\eta,\infty]}\ldots \rangle
= \langle \ldots
i\,U^{[n]}_{[\infty,\eta]}\,G_\st^{n\alpha}(\eta)
\,U^{[n]}_{[\eta,\infty]} \ldots \rangle,
\]
showing that the transition of Eq.~\ref{A-sud0} to Eq.~\ref{A-sud} 
also works in the non-abelian case, if the appropriate Wilson lines
are included. Similarly, one has
\be
U^{[n]}_{[\infty,\eta]}\,A^P(\eta)\,U^{[n]}_{[\eta,\infty]}
- A^P(\eta)\big|_{\eta^P{=}\infty}
= \left. \int_\infty^{\,\eta^P} d\zeta^P
\ U^{[n]}_{[\infty,\zeta]}\biggl(G^{nP}(\zeta)
+ [\partial^P,A^n(\zeta)]\biggr)
U^{[n]}_{[\zeta,\infty]}\right|_{\zeta^n = \eta^n,\zeta_\st = \eta_\st}.
\label{AP1}
\ee
The subtractions at light-cone infinity are important to produce the
missing transverse Wilson lines.
It is shown in detail in Ref.~\cite{Boer:2003cm} 
how these subtractions $A_\st^\alpha(\eta)\vert_{\eta^P{=}\infty}$ 
from the `higher twist' matrix elements with $A_\st$ (and $A^P$) fields 
provide the missing transverse gauge connections 
$U^{\st}_{[0_\st,\xi\st]} =
U^{\st}_{[0_\st,\infty_\st]}\,U^{\st}_{[\infty_\st,\xi_\st]}$, 
which we will indicate 
as $U^\st[p]$ with the argument $p$ again just indicating that
the link connection involves the (transverse) endpoints $0_\st$ and 
$\xi_\st$ conjugate to parton momentum $p$ and fields belonging to $\Phi(p)$. 
Incorporating all transverse gauge connections, we obtain (symmetrized) pieces
\be
U^\st[p_1,p_2,k_1] = {\mathcal S}\{U^\st[p_1]\,U^\st[p_2]\,U^\st[k_1]\},
\ee
with irrelevant ordering among themselves. 
In the ordering of these pieces and the combination with collinear pieces
of the gauge link, we will neglect contributions from commutators
involving $A^{\mu}(p_1)$ and $A^{\nu}(p_2)$ since $(p_1-p_2)^2 \sim Q^2$.
It is clear that in kinematic regions where all parton momenta are small
this cannot be used, complicating the full QCD analysis.
Combined with the connections along the light-like direction, we now have
for our tree-level analysis a set of entangled Wilson lines that
bridge the non-locality of parton fields in the various correlators
(see Fig.~\ref{cgi-result}(b)).
The resulting expression for the cross section is again of the
form in Equation Eq.~\ref{facto3}, but including 
transverse pieces, denoted $U_+[p,\ldots] = U^\st[p,\ldots]\,U^{[n]}[p,\ldots]$,
etc. One gets for the cross section an expression in terms of {\em
unintegrated correlators}, which are only integrated over $p{\cdot}P$,
\bea
d\sigma & \sim & {\rm Tr}_c\bigl[
U_-^{\dagger}[p_2,k_1,k_2]\,\Phi(x_1,p_{1\st})
\,U_-[p_2,k_1,k_2]\,\Gamma_1^\ast
\,U_+^{\dagger}[p_1,p_{2},k_{2}]\Delta(z_1,k_{1\st})
\,U_+[p_{1},p_{2},k_{2}]\,\Gamma_1\bigr]
\nonumber \\&& \mbox{}\times
{\rm Tr}_c\bigl[U_-^{\dagger}[p_1,k_1,k_2]\,\Phi(x_2,p_{2\st})
\,U_-[p_1,k_1,k_2]\,\Gamma_2^\ast 
\,U_+^{\dagger}[p_1,p_2,k_1]\,\Delta(z_2,k_{2\st})
\,U_+[p_1,p_2,k_1]\,\Gamma_2\bigr].
\label{facto4}
\eea
This resulting expression is now color gauge-invariant. 
The Wilson lines can be taken along a generic $n$-direction, 
which even could be chosen different for each of the gauge 
connections in Eq.~\ref{facto4}, 
but its color structure is fully entangled and it does not allow for
a factorized expression with universal correlators that have their 
own gauge links. Viewing it as a factorized expression it contains
hard amplitudes, soft correlators and gauge connections, where the
gauge connections take care of a `color resetting' which feels all
hadrons that are involved.

\section{Disentangling the color flow dependence\label{section4}}

\begin{figure}
\epsfig{file=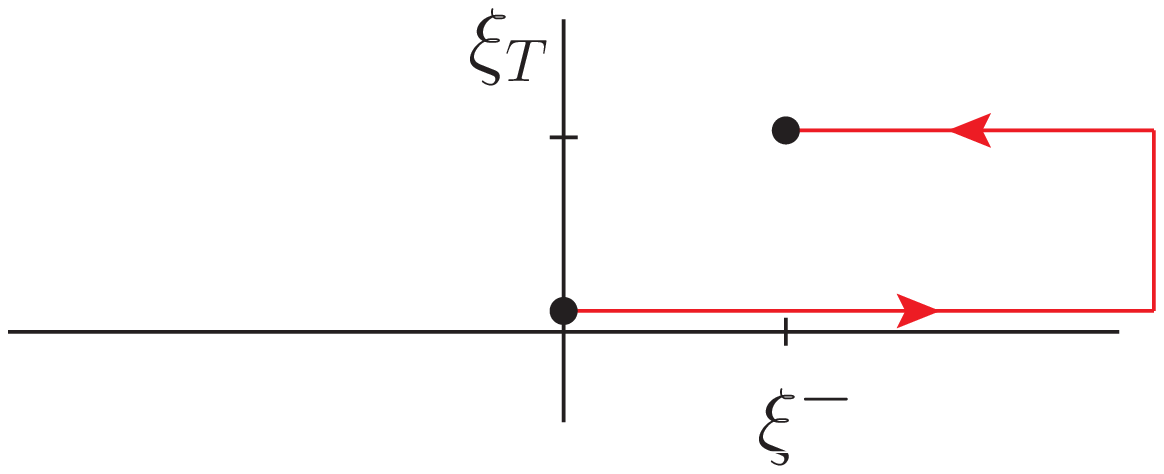,width=0.22\textwidth}
\hspace{1cm}
\epsfig{file=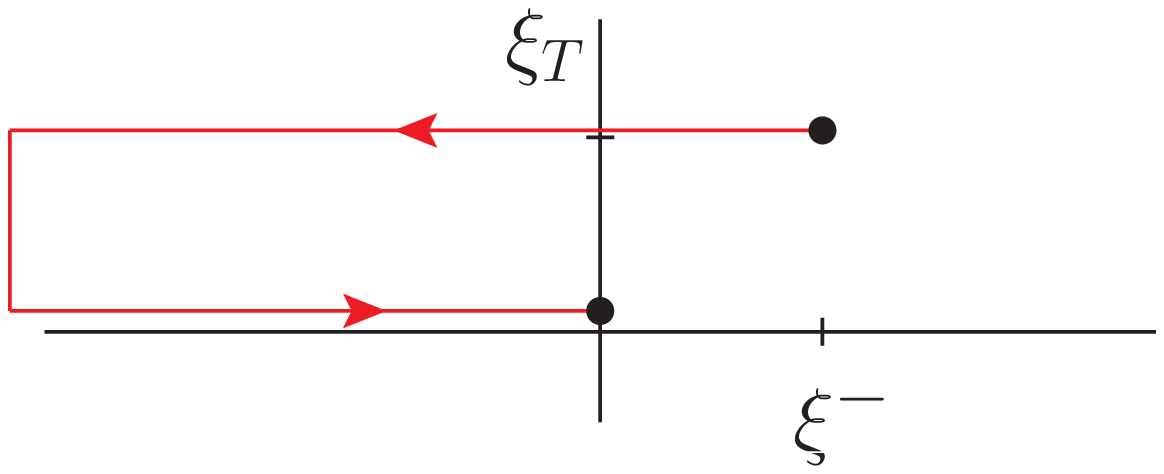,width=0.22\textwidth}
\hspace{1cm}
\epsfig{file=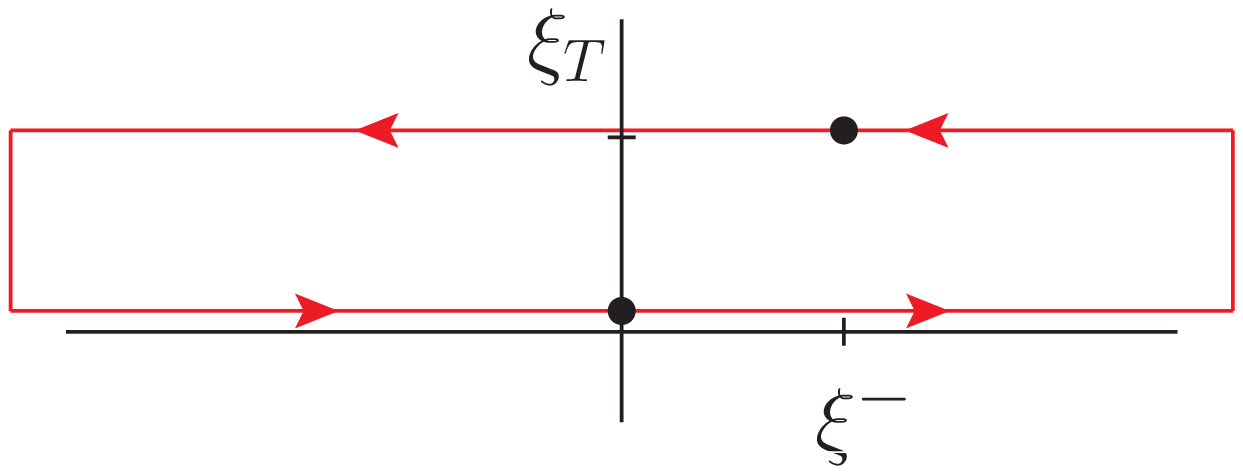,width=0.22\textwidth}
\\[0.2cm]
(a)\hspace{5.5cm} (b)\hspace{5.5cm} (c)
\caption{\label{figlink} The gauge connections present in the expression for 
the unintegrated cross section in Eq.~\ref{facto4}, 
(a) $U_+^\dagger\,U_+ \Rightarrow W_{+}^{[n]}
= U^{[n]}_{[0,\infty]}\,U^\st_{[0_\st,\xi_\st]}\,U^{[n]}_{[\infty,\xi]}$, 
(b) $U_-^\dagger\,U_- \Rightarrow W_{-}^{[n]\dagger}
= U^{[n]}_{[\xi,-\infty]}\,U^\st_{[\xi_\st,0_\st]}\,U^{[n]}_{[-\infty,0]}$, 
and (c) $W_{\Box}^{[n]} = W^{[n]}_+\,W^{[n]\dagger}_-$.}
\end{figure}

Starting with the expression of Eq.~\ref{facto4} for the cross section,
we first look at the case that the hard process doesn't affect the color flow 
as in Fig.~\ref{fig2}(a). This means that the vertices
$\Gamma_1$ and $\Gamma_2$ are color independent. 
In our expression we have floating around in the full expression pieces,
like $ U^\st[p]\,U_+^{[n]}[p]\ldots \Phi(x,p_\st)
\ldots U_+^{[n]\dagger}[p]\,U^{\st\dagger}[p]\ldots $, which only
when they are combined in a single trace would yield
gauge links appropriate for TMDs,
\bea
\Phi^{[\pm]}(x,p_\st) 
& = & {\rm Tr}_c\bigl[U^\st[p]\,U_+^{[n]}[p]\,\Phi(x,p_\st)
\,U_+^{[n]\dagger}[p]\,U^{\st\dagger}[p]\bigr]
\nonumber \\ &=&
\left.\int \frac{d\,\xi{\cdot}P\,d^2\xi_\st}{(2\pi)^3}\ e^{i\,p\cdot \xi}
\ \langle P\vert \overline\psi(0)\,U_{[0,\infty]}^{[n]}
\,U^\st_{[0_\st,\xi_\st]}
\,U_{[\infty,\xi]}^{[n]} \,\psi(\xi)\vert P\rangle
\right|_{LF}
\label{TMD-1}
\\ &=&
\left.\int \frac{d\,\xi{\cdot}P\,d^2\xi_\st}{(2\pi)^3}\ e^{i\,p\cdot \xi}
\ \langle P\vert \overline\psi(0)\,W_{\pm[0,\xi]}^{[n]}
\,\psi(\xi)\vert P\rangle
\right|_{LF}
= {\rm Tr}_c\bigl[W_\pm^{[n]}[p]\,\Phi(x,p_\st)\bigr],
\eea
or
\bea
\Delta^{[\pm]}(z,k_\st) 
& = & \frac{1}{N_c}\,{\rm Tr}_c\bigl[U^\st[k]\,U_\pm^{[n]}[k]\Delta(z,k_\st)
\,U_\pm^{[n]\dagger}[k]\,U^{\st\dagger}[k]\bigr]
= \frac{1}{N_c}\,{\rm Tr}_c\bigl[\Delta(z,k_\st)\,W^{[n]\dagger}_\pm[k]\bigr]
\nonumber \\ & = &
\left. \int \frac{d\,\xi{\cdot}K_h\,d^2\xi_\st}{(2\pi)^3}
\ e^{-ik\cdot \xi}\,
\frac{1}{N_c}\,{\rm Tr}_c \langle 0\vert U^\st_{[\infty_\st,0_\st]}
\,U_{[\pm\infty,0]}^{[n]}\psi (\xi) 
a_h^\dagger a_h \overline 
\psi(0)\,U_{[\xi,\pm\infty]}^{[n]}\,U^{\st}_{[\xi_\st,\infty_\st]} 
\vert 0 \rangle
\right|_{LF} ,
\eea
with the coordinate space structure of $W_{\pm[0,\xi]}^{[n]}$ 
shown in Figs~\ref{figlink}(a) and (b).
Even if one is in the lucky situation that one
can combine the relevant Wilson lines in a single trace, there still would
be combinations left in other color traces which form Wilson loops
$W^{[n]}_{\Box} = W_+^{[n]}\,W_-^{[n]\dagger}$
as illustrated in Fig.~\ref{figlink}(c).

\subsection{TMDs in `elementary' processes}

\begin{figure}[t]
\epsfig{file=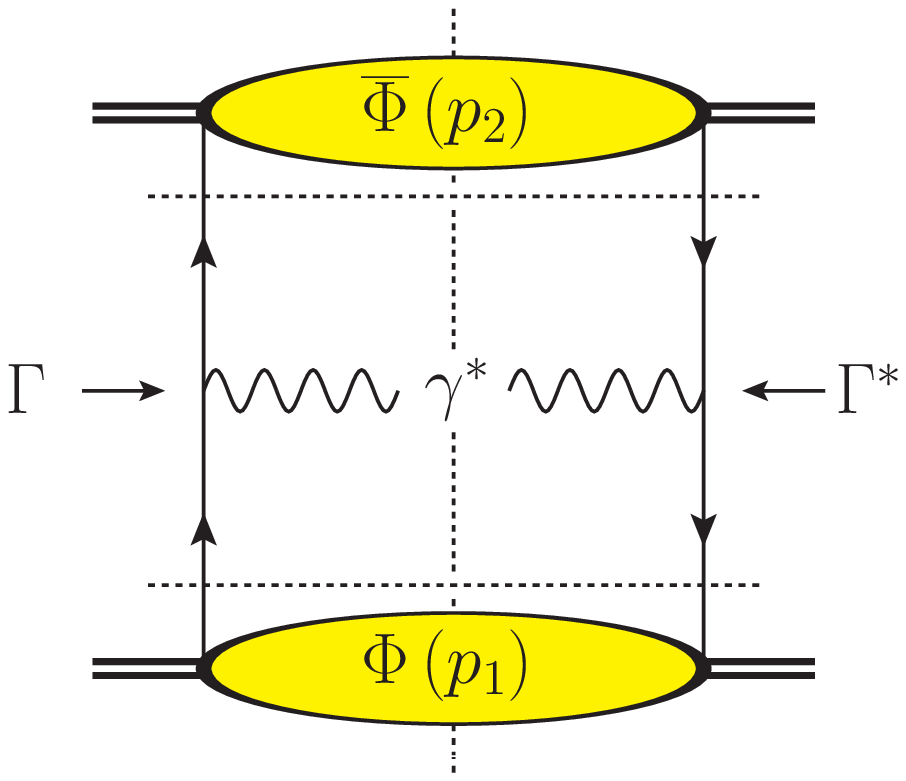,width=0.28\textwidth}
\hspace{1.5cm}
\epsfig{file=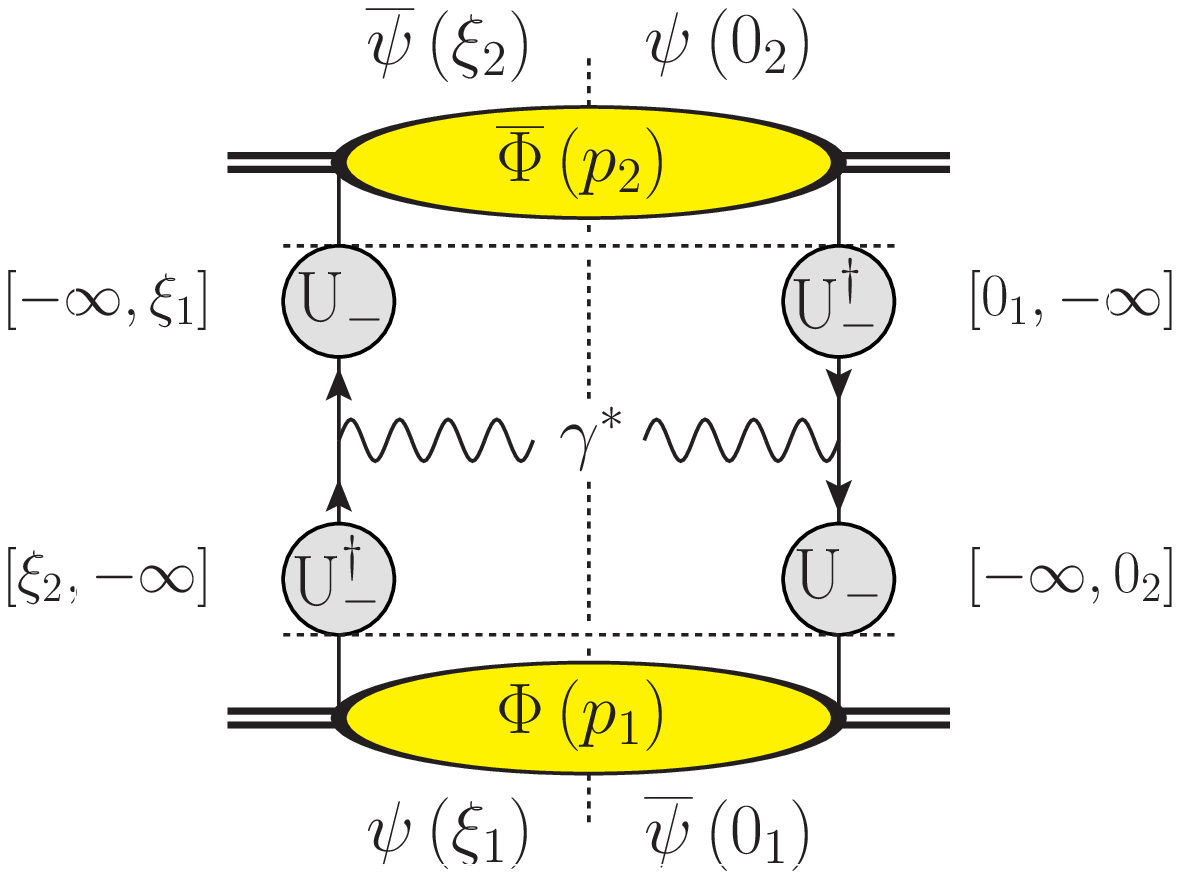,width=0.34\textwidth}
\\[0.2cm]
(a) \hspace{6cm} (b)
\caption{\label{figDY} (a) The color-flow in the Drell-Yan process and (b)
the gauge connections including their coordinate space structure.}
\end{figure}

Considering, as a reference, first the `elementary' Drell-Yan (DY) process, 
which like other electroweak processes such as leptoproduction
or electron-positron annihilation, is relatively simple, because
the color flow through initial and final states consists
of a single color loop.  In that case, taking the DY process as an
example (Fig.~\ref{figDY}), the cross section is given by
\bea
d\sigma_{DY} &\sim & {\rm Tr}_c\bigl[
U_-^{[n]\dagger}[p_2]\,\Phi(x_1,p_{1\st})\,U_-^{[n]}[p_2]
\,\Gamma^\ast
\,U_-^{[n]\dagger}[p_1]\,\overline\Phi(x_2,p_{2\st})\,U_-^{[n]}[p_1]\,\Gamma
\bigr]
\nonumber \\ & = &
{\rm Tr}_c\bigl[
W_-^{[n]\dagger}[p_2]\,\Phi(x_1,p_{1\st})\,\Gamma^\ast
\overline\Phi(x_2,p_{2\st})\,W_-^{[n]}[p_1]\,\Gamma
\bigr].
\eea
The gauge connections contain collinear and transverse fields
and one has to be careful in disentangling. 
We note in particular that in general
\bea
d\sigma_{DY} & \ne & 
{\rm Tr}_c\bigl[W_-^{[n]}[p_1]\,\Phi(x_1,p_{1\st})\bigr]
\,{\rm Tr}_c\bigl[\overline\Phi(x_2,p_{2\st})\,W_-^{[n]\dagger}[p_2]\bigr]
\,\tfrac{1}{N_c}\,\Gamma\,\Gamma^\ast .
\eea
However, if one of the hadrons can be treated collinear, i.e.\ integrating 
over one of the transverse momenta, say $p_{2\st}$, we can disentangle the 
traces and obtain
\bea
\frac{d\sigma_{DY}}{d^2p_{1\st}} & \sim & 
{\rm Tr}_c\bigl[W_-^{[n]}[p_1]\,\Phi(x_1,p_{1\st})\bigr]
\,\overline\Phi{}^{[W]}(x_2)
\underbrace{\tfrac{1}{N_c}\,\Gamma\,\Gamma^\ast}_{\hat\Sigma_{DY}}
\nonumber \\ & = &
\Phi^{[-]}(x_1,p_{1\st})
\,\overline\Phi(x_2)
\,\hat\Sigma_{DY} ,
\eea
where
$\Phi^{[-]}(p_1) = {\rm Tr}_c\bigl[W_-^{[n]}[p_1]\,\Phi(p_1)\bigr]$.
This illustrates the TMD factorization at tree-level and the
relevant gauge links to be used, which was considered in detail in
Ref.~\cite{Boer:1999si}. Beyond tree-level one has the factorization
formalism of Collins, Soper and Sterman~\cite{Collins:1981tt}, which has
proven to be very successful for many applications, but does not
catch all subtleties at small transverse momenta~\cite{Aybat:2011zv}.
We will return to the problems with finding a full tree-level factorization 
in two TMDs in Section~\ref{section5}.

\subsection{TMDs in 1-parton unintegrated processes}

In this subsection, we will turn to the situation of a
more complicated color flow and we will show that
TMDs with non-trivial gauge links appear in 1-parton unintegrated
(1PU) contributions to the cross section, by which we mean 
contributions in which one has 
integrated over the parton transverse momenta in all but one 
of the hadrons. At this point we do not worry about experimental 
feasibility of measuring such a cross section.
First assume as an intermediate step
that only the transverse momentum of correlators $\Phi(p_1)$ and
$\Phi(p_2)$ are left unintegrated, i.e.\ we look at jet production 
or non-hadronic final states. In that case one integrates 
over $k_{1\st}$ and $k_{2\st}$
and the collinear correlators $\Delta(z_1)$ and $\Delta(z_2)$ just involve 
unique collinear gauge links, while $W_{\Box}[k_1]$ and $W_{\Box}[k_2]$
become unity (in color space).
The result is an expression of the form
\bea
d\sigma_1 & \sim & {\rm Tr}_c\bigl[
U_-^{\dagger}[p_2]\,\Phi(x_1,p_{1\st})
\,U_-[p_2]\,\Gamma_1^\ast
\,U_+^{\dagger}[p_1,p_{2}]\Delta^{[W]}(z_1)
\,U_+[p_{1},p_{2}]\,\Gamma_1\bigr]
\nonumber \\&& \mbox{}\times
{\rm Tr}_c\bigl[U_-^{\dagger}[p_1]\,\Phi(x_2,p_{2\st})
\,U_-[p_1]\,\Gamma_2^\ast 
\,U_+^{\dagger}[p_1,p_2]\,\Delta^{[W]}(z_2)
\,U_+[p_1,p_2]\,\Gamma_2\bigr],
\label{facto5b}
\eea
which is still entangled. This was the example discussed in 
Ref.~\cite{Rogers:2010dm}.

Integrating over $p_{2\st}$, the result reduces to
\bea
d\sigma_1 & \sim & {\rm Tr}_c\bigl[
\Phi(x_1,p_{1\st})
\,\Gamma_1^\ast
\,U_+^{\dagger}[p_1]\Delta^{[W]}(z_1)
\,U_+[p_{1}]\,\Gamma_1\bigr]
\,{\rm Tr}_c\bigl[U_-^{\dagger}[p_1]\,\Phi^{[W]}(x_2)
\,U_-[p_1]\,\Gamma_2^\ast 
\,U_+^{\dagger}[p_1]\,\Delta^{[W]}(z_2)
\,U_+[p_1]\,\Gamma_2\bigr]
\nonumber \\ & \sim & {\rm Tr}_c\bigl[
\Phi(x_1,p_{1\st})
\,\Gamma_1^\ast\,\Delta^{[W]}(z_1)
\,W_+[p_{1}]\,\Gamma_1\bigr]
\,{\rm Tr}_c\bigl[W_-^{\dagger}[p_1]\,\Phi^{[W]}(x_2)
\,\Gamma_2^\ast\,\Delta^{[W]}(z_2)
\,W_+[p_1]\,\Gamma_2\bigr]
\nonumber \\ & = &
{\rm Tr}_c\bigl[W_+^{[n]}[p_1]\,\Phi(x_1,p_{1\st})
\,\Gamma_1^\ast\,\Delta(z_1)\,\Gamma_1\bigr]
\,{\rm Tr}_c\bigl[W_{\Box}^{[n]}[p_1]\,\Phi(x_2)
\,\Gamma_2^\ast\,\Delta(z_2)\,\Gamma_2\bigr],
\label{facto6a}
\eea
and gives
\bea
d\sigma_1  & \sim  &
{\rm Tr}_c\bigl[
\Phi(x_1,p_{1\st})\,W_+^{[n]}[p_1]\,\Gamma_1^\ast\,\Delta(z_1)\,\Gamma_1
\bigr] 
\,\tfrac{1}{N_c}\,{\rm Tr}_c\bigl[ W_{\Box}^{[n]}[p_1] \bigr]
\,{\rm Tr}_c\bigl[
\Phi(x_2)\,\Gamma_2^\ast \,\Delta(z_2)\,\Gamma_2
\bigr] 
\label{facto6b}
\\ &=& 
\Phi^{[+(\Box)]}(x_1,p_{1\st}) \,\Phi(x_2)
\,\underbrace{\Gamma_1\,\Gamma_1^\ast\,\Gamma_2\,\Gamma_2^\ast}_{\hat\Sigma_1} 
\,\Delta(z_1) \,\Delta(z_2),
\label{facto6c}
\eea
where Wilson lines combined with the correlators can be combined 
into TMD correlators with gauge links, such as
$\Phi^{[+]}(p_1) = {\rm Tr}_c\bigl[\Phi(p_1)\,W_+^{[n]}[p_1]\bigr]$.
In the step from \ref{facto6a} to \ref{facto6b} we have split the
second trace into two parts containing
the trace of $W_\Box^{[n]}$ (appropriately averaged) and the trace
of the ($n$-independent) collinear part. In
the last step the trace of the Wilson loop (non-local in the 
coordinate $\xi$ conjugate to the parton momentum $p_1$) has been 
absorbed into a more complicated gauge link for 
$\Phi^{[+(\Box)]}(x_1,p_{1\st})$
= $\frac{1}{N_c}\,{\rm Tr}_c\bigl[W^{[n]}_\Box[p_1]\bigr]
\,{\rm Tr}_c\bigl[\Phi(p_1)\,W_+^{[n]}[p_1]\bigr]$. Note that, in 
spite of what the notation might suggest, $\Phi^{[+(\Box)]}$ is not a
multiplicative factor times $\Phi^{[+]}$ because the fields
in both factors belong to the same matrix element (indicated through
argument $[p_1]$).
These complex gauge links are the process-dependent gauge links discussed in
Refs~\cite{Bomhof:2004aw,Bacchetta:2005rm,Bomhof:2006dp,Bomhof:2006ra,Bomhof:2007xt}.

\begin{figure}
\begin{minipage}{0.45\textwidth}
\epsfig{file=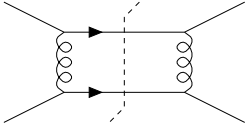,width=0.40\textwidth}
\end{minipage}
\begin{minipage}{0.5\textwidth}
\caption{\label{fig5}
The squared amplitude for the hard interaction between two quarks
coming from a one-gluon-exchange diagram.  We consider 
for simplicity two quarks with different flavors.}
\end{minipage}
\end{figure}
As a further illustration, we include the color flow possibilities
in the case of a hard process where a gluon 
is exchanged in the hard part (see Fig.~\ref{fig5}) instead of 
a colorless boson as in Fig.~\ref{fig2}. 
One gets different color flow patterns that contribute
%(see Fig.~\ref{fig6})
giving rise to two different color contractions 
\bea
d\sigma & \sim &
{\rm Tr}_c\bigl[
U_-^{[n]\dagger}[p_2,k_1,k_2]\,\Phi(p_1)\,U_-^{[n]}[p_2,k_1,k_2]
\,\Gamma_1^\ast\,T^a
U_+^{[n]\dagger}[p_1,p_2,k_2]\,\Delta(k_1)\,U_+^{[n]}[p_1,p_2,k_2]
\,\Gamma_2\,T^b\bigr]
\nonumber \\ && \mbox{}\times
{\rm Tr}_c\bigl[
U_-^{[n]\dagger}[p_1,k_1,k_2]\,\Phi(p_2)\,U_-^{[n]}[p_1,k_1,k_2]
\,\Gamma_2^\ast\,T^a
\,U_+^{[n]\dagger}[p_1,p_2,k_1]\,\Delta(k_2)\,U_+^{[n]}[p_1,p_2,k_1]
\,\Gamma_1\,T^b
\bigr]
\\ & = &
\tfrac{N_c^2+1}{N_c^2-1}
\,{\rm Tr}_c\bigl[
U_-^{[n]\dagger}[p_2,k_1,k_2]\,\Phi(p_1)\,U_-^{[n]}[p_2,k_1,k_2]
\,\Gamma_1^\ast
\,U_+^{[n]\dagger}[p_1,p_2,k_2]\,\Delta(k_1)\,U_+^{[n]}[p_1,p_2,k_2]
\,\Gamma_2\bigr]
\nonumber \\ && \mbox{}\hspace{1.5cm} \times
{\rm Tr}_c\bigl[
U_-^{[n]\dagger}[p_1,k_1,k_2]\,\Phi(p_2)\,U_-^{[n]}[p_1,k_1,k_2]
\,\Gamma_2^\ast
\,U_+^{[n]\dagger}[p_1,p_2,k_1]\,\Delta(k_2)\,U_+^{[n]}[p_1,p_2,k_1]
\,\Gamma_1
\bigr]
\nonumber \\&& \mbox{}
-\tfrac{2}{N_c^2-1}\,{\rm Tr}_c\bigl[
U_-^{[n]\dagger}[p_2,k_1,k_2]\,\Phi(p_1)\,U_-^{[n]}[p_2,k_1,k_2]
\,\Gamma_1^\ast
\,U_+^{[n]\dagger}[p_1,p_2,k_2]\,\Delta(k_1)\,U_+^{[n]}[p_1,p_2,k_2]
\nonumber \\ && \mbox{}\hspace{1.8cm}\times
U_-^{[n]\dagger}[p_1,k_1,k_2]\,\Phi(p_2)\,U_-^{[n]}[p_1,k_1,k_2]
\,\Gamma_2^\ast
\,U_+^{[n]\dagger}[p_1,p_2,k_1]\,\Delta(k_2)\,U_+^{[n]}[p_1,p_2,k_1]
\,\Gamma_1
\bigr]
\\ & = &
\tfrac{N_c^2+1}{N_c^2-1}\,d\sigma_1  
-\tfrac{2}{N_c^2-1}\,d\sigma_2,
\eea
normalized by the usual color factor, such that the coefficients in
front of these terms add to one.
The first part is the one considered above.
For the second part, one gets upon integration
over $k_{1\st}$ and $k_{2\st}$ 
\bea
d\sigma_2 & \sim &
{\rm Tr}_c\bigl[
U_-^{[n]\dagger}[p_2]\,\Phi(x_1,p_{1\st})\,U_-^{[n]}[p_2]
\,\Gamma_1^\ast
\,U_+^{[n]\dagger}[p_1,p_2]\,\Delta(z_1)\,U_+^{[n]}[p_1,p_2]
\,\Gamma_2
\nonumber \\ && \mbox{}\hspace{0.7cm}\times
U_-^{[n]\dagger}[p_1]\,\Phi(x_2,p_{2\st})\,U_-^{[n]}[p_1]
\,\Gamma_2^\ast
\,U_+^{[n]\dagger}[p_1,p_2]\,\Delta(z_2)\,U_+^{[n]}[p_1,p_2]
\,\Gamma_1
\bigr] ,
\label{facto15b}
\eea
leaving an entangled situation. 
If only one transverse momentum remains unintegrated, one again can
combine the result into a single complicated gauge link,
\bea
d\sigma_2 & \sim & 
{\rm Tr}_c\bigl[
\Phi(x_1,p_{1\st})\,\Gamma_1^\ast
\,U_+^{[n]\dagger}[p_1]\,\Delta(z_1)\,U_+^{[n]}[p_1]
\,\Gamma_2
\,U_-^{[n]\dagger}[p_1]\,\Phi(x_2)\,U_-^{[n]}[p_1]
\,\Gamma_2^\ast
\,U_+^{[n]\dagger}[p_1]\,\Delta(z_2)\,U_+^{[n]}[p_1]
\,\Gamma_1
\bigr] 
\nonumber
\\ & = &
{\rm Tr}_c\bigl[
\Phi(x_1,p_{1\st})\,W^{[n]}_+[p_1]\,\Gamma_1^\ast
\,\Delta(z_1)\,\Gamma_2
\,\Phi(x_2)\,W_{\Box}^{[n]}[p_1]\,\Gamma_2^\ast
\,\Delta(z_2)\,\Gamma_1 
\bigr] 
\label{facto16a}
\\ & = &
\Phi^{[+\Box]}(x_1,p_{1\st})
\,\Phi(x_2)
\,\underbrace{\Gamma_1\,\Gamma_1^\ast\,\Gamma_2\,\Gamma_2^\ast}_{\hat\Sigma_2} 
\,\Delta(z_1)
\,\Delta(z_2),
\eea
where
$\Phi^{[+\Box]}(p_1) = {\rm Tr}_c\bigl[\Phi(p_1)\,W_+^{[n]}[p_1]
\,W^{[n]}_\Box[p_1]\bigr]$.
Combining both contributions one gets the result~\cite{Bomhof:2006dp},
\be
d\sigma \sim 
\Bigl(
\tfrac{N_c^2+1}{N_c^2-1}\Phi^{[+(\Box)]}(x_1,p_{1\st})\,\Phi(x_2)\,\hat\Sigma_1
-\tfrac{2}{N_c^2-1}\Phi^{[+\Box]}(x_1,p_{1\st})\,\Phi(x_2)\,\hat\Sigma_2
\Bigr)
\,\Delta(z_1) \,\Delta(z_2) .
\label{jungle}
\ee
In this particular case $\hat\Sigma_1$ and $\hat\Sigma_2$ are the
same, but in general one has more diagrams contributing to the hard process, 
each contribution gets split up into its 
own color patterns. The results for such 1PU
processes have been tabulated in Ref.~\cite{Bomhof:2006dp}.

\section{Analysis in terms of transverse moments\label{section5}}

In this section, we will start with the unintegrated result in 
Eq.~\ref{facto4} and take transverse moments, which will give
collinear results, which include correlators with covariant
derivatives $iD_\st^\alpha$ and $A_\st^\alpha$ or in gauge-invariant
form integrals over $G^{n\alpha}$. The transverse moments are
obtained by looking at weighted cross sections of the form
\be
\langle p_\st^{\alpha_1}\ldots p_\st^{\alpha_N}\,\sigma\rangle
= \int d^2p_\st\ p_\st^{\alpha_1}\ldots p_\st^{\alpha_N}
\,\frac{d\sigma}{d^2p_\st}.
\ee
We will apply this to the cross section $\sigma$, built from the
two pieces $\sigma_1$ and $\sigma_2$, discussed in the previous
paragraph. The simplest, lowest, transverse moment is the 
integrated cross section,
\bea
\langle \sigma \rangle & = & \int d^2p_\st\ \frac{d\sigma}{d^2p_\st}
%\nonumber \\ & = &
= \Phi(x_1)\,\Phi(x_2)
\,\underbrace{\bigl(
\tfrac{N_c^2+1}{N_c^2-1}\,\hat\Sigma_1
-\tfrac{2}{N_c^2-1}\,\hat\Sigma_2\bigr)}_{\hat\Sigma}
\,\Delta(z_1) \,\Delta(z_2) ,
\eea
which has between brackets the {\em standard} partonic cross section
$\hat\Sigma$.

\subsection{Single weighted asymmetries}

To see what happens with the first transverse moment, we need
to investigate at the level of correlators what happens when
one applies $i\partial_\xi^\alpha$ to the matrix elements.
To clarify the role of the gauge link in a correlator, in particular
the contribution at infinity, we consider
\be
i\partial_\st^\alpha \bigl[\ldots U^{[n]}_{[0,\infty]}
\,U^{\st}_{[0_\st,\xi_\st]}\,U^{[n]}_{[\infty,\xi]}\ldots \psi(\xi)\bigr]
= \bigl[\ldots U^{[n]}_{[0,\infty]}\,U^{\st}_{[0_\st,\xi_\st]}
\,iD_\st^\alpha\,U^{[n]}_{[\infty,\xi]}\ldots \psi(\xi)\bigr] .
\label{TM1}
\ee
We can evaluate explicitly
\be
iD_{\st}^\alpha\,U^{[n]}_{[\infty,\xi]}\ldots \psi(\xi)
= U^{[n]}_{[\infty,\xi]}\,iD_\st^\alpha(\xi)\ldots \psi(\xi)
- \int_{\infty}^{\xi} d\,\eta{\cdot}P
\ U^{[n]}_{[\infty,\eta]}\,G_\st^{n\alpha}(\eta)
U^{[n]}_{[\eta,\xi]}\ldots \psi(\xi),
\label{58}
\ee
noting that the second term depends through the integration limits
on the structure of the Wilson line, in particular on the transverse 
piece being at plus or minus infinity. 
We could simply use the second term as a definition of the 
$A_\st^\alpha$ field but such a definition would for instance 
not have the correct time-reversal property. 
We will use a time-reversal-definite expression
for $A^\alpha(\xi)$,
\be
A^\alpha(\xi) \equiv \int_{-\infty}^\infty d\,\eta{\cdot}P
\ \epsilon(\xi^P - \eta^P)
\,U^{[n]}_{[\xi,\eta]}\,G^{n\alpha}(\eta)
U^{[n]}_{[\eta,\xi]},
\label{ATP2}
\ee
which is gauge-equivalent to the expressions in Eqs~\ref{AT1} and \ref{AP1}.
The form in Eq.~\ref{ATP2} implicitly implies $A^n = 0$ and 
antisymmetric boundary conditions at $\xi^P = \pm \infty$ for
$A_\st^\alpha(\xi)$ and $A^P(\xi)$, 
\be
U^{[n]}_{[\xi,\infty]}\,A^\alpha(\infty,\xi_\st,\xi^n)
\,U^{[n]}_{[\infty,\xi]}
+ U^{[n]}_{[\xi,-\infty]}\,A^\alpha(-\infty,\xi_\st,\xi^n)
U^{[n]}_{[-\infty,\xi]} = 0,
\ee
while
\bea
U^{[n]}_{[\xi,\infty]}\,A^\alpha(\infty,\xi_\st,\xi^n)
\,U^{[n]}_{[\infty,\xi]}
- U^{[n]}_{[\xi,-\infty]}\,A^\alpha(-\infty,\xi_\st,\xi^n)
U^{[n]}_{[-\infty,\xi]} = 
\int_{-\infty}^\infty d\,\xi{\cdot}P
\ G^{n\alpha}(\xi)
= 2\pi\widetilde G^{n\alpha}(\xi_\st,\xi^n) ,
\label{gluonpole}
\eea
or combining these two equations,
\bea
U^{[n]}_{[\xi,\infty]}\,A^\alpha(\infty,\xi_\st,\xi^n)
\,U^{[n]}_{[\infty,\xi]}
= - U^{[n]}_{[\xi,-\infty]}\,A^\alpha(-\infty,\xi_\st,\xi^n)
U^{[n]}_{[-\infty,\xi]} 
= \pi\widetilde G^{n\alpha}(\xi_\st,\xi^n) .
\eea
The integral expression for $\pi\widetilde G^{n\alpha}$ 
represents actually the Fourier transform
of the field strength tensor at zero momentum in the $n$-component
(which is indicated by including a tilde over the $G$ and
omitting $\xi^P$ from the argument list).
Eq.~\ref{58} becomes
\be
iD_{\st}^\alpha\,U^{[n]}_{[\infty,\xi]}\ldots \psi(\xi)
= U^{[n]}_{[\infty,\xi]}\,\left(iD_\st^\alpha(\xi) - A_\st^\alpha(\xi)
+ \pi\widetilde G^{n\alpha}(\xi)\right)\ldots \psi(\xi) .
\label{64}
\ee
To implement this in correlators,
we consider besides the TMD correlator $\Phi(x,p_\st)$
in Eq.~\ref{TMD-1} the collinear matrix elements $\Phi_A^\alpha(x,x_1)$,
which are the integrated quark-quark-gluon correlators starting with
Eq.~\ref{quarkgluonquark}. Including Wilson lines, this collinear 
correlator is of the form
\be
\Phi_A^\alpha(x,x_1) =
\left.\int \frac{d\,\xi{\cdot}P}{2\pi}\,\frac{d\,\eta{\cdot}P}{2\pi}
\ e^{i\,(p-p_1)\cdot \xi}\ e^{i\,p_1\cdot \eta}
\ \langle P\vert \overline\psi(0)\,U^{[n]}_{[0,\eta]}\,A^\alpha(\eta)
\,U^{[n]}_{[\eta,\xi]}\,\psi(\xi)
\vert P\rangle\right|_{LC},
\ee
or integrated over $x_1$,
\be
\Phi_A^\alpha(x) = \int dx_1\,\Phi_A^\alpha(x,x_1) =
\left.\int \frac{d\,\xi{\cdot}P}{2\pi}
\ e^{i\,p\cdot \xi}
\ \langle P\vert \overline\psi(0)\,U^{[n]}_{[0,\xi]}\,A^\alpha(\xi)
\,\psi(\xi) \vert P\rangle\right|_{LC}.
\ee
Similarly we define
$\Phi_D^\alpha(x,x_1)$ and $\Phi_G^\alpha(x,x_1)$ 
using $iD^\alpha(\eta)$ and $G^{n\alpha}(\eta)$, respectively. 
The latter is given by
\be
\Phi_G^\alpha(x,x_1) =
\left.\int \frac{d\,\xi{\cdot}P}{2\pi}\,\frac{d\,\eta{\cdot}P}{2\pi}
\ e^{i\,(p-p_1)\cdot \xi}\ e^{i\,p_1\cdot \eta}
\ \langle P\vert \overline\psi(0)\,U^{[n]}_{[0,\eta]}\,G^{n\alpha}(\eta)
\,U^{[n]}_{[\eta,\xi]}\,\psi(\xi)
\vert P\rangle\right|_{LC}.
\label{67}
\ee
The relation in Eq.~\ref{ATP2} implies for these correlators
\be
\Phi_A^\alpha(x) = {\rm PV}\int dx_1\ \frac{i}{x-x_1}\,\Phi_G^\alpha(x,x_1).
\ee
The operator $\widetilde G^{n\alpha}(\xi_\st,\xi^n)$ in Eq.~\ref{gluonpole}
appears in the correlator with a gluon field strength, but at zero 
$n$-component of the momentum, i.e. at $x_1 = 0$. The other (among them
transverse) components of the gluon momentum are also integrated over
in Eq.~\ref{67}.
This particular correlator, referred to as a gluonic pole matrix element,
\be
\Phi_G^\alpha(x,x) =
\left.\int \frac{d\,\xi{\cdot}P}{2\pi}
\ e^{i\,p\cdot \xi}
\ \langle P\vert \overline\psi(0)\,U^{[n]}_{[0,\xi]}
\,\widetilde G^{n\alpha}\,\psi(\xi)
\vert P\rangle\right|_{LC},
\label{Gtilde}
\ee
will play an important role for transverse moments.
Note that we have not included any argument for the field strenght
in $\widetilde G^{n\alpha}\,\psi(x)\Bigr|_{LC}$, because all arguments
become zero after integration over $p_{1\st}$, $p_1{\cdot}P$
(implying $\xi_\st{=}\eta_\st$ and $\xi^n{=}\eta^n$) followed by the
integration over $p_\st$ and $p{\cdot}P$ (putting $\xi$ on the light-cone,
i.e.\ $\xi_\st{=}0_\st$ and $\xi^n{=}0$). 

Using the expressions in Eqs~\ref{TM1} and \ref{58}, we consider the following
unintegrated TMD functions starting with $\Phi^{[\pm]}(x,p_\st)$,
\bea
\Phi_{\partial}^{\alpha\,[\pm]}(x,p_\st)
&\equiv& p_\st^{\alpha}\,\Phi^{[\pm]}(x,p_\st)
\nonumber \\
&=&
\left.\int \frac{d\,\xi{\cdot}P\,d^2\xi_\st}{(2\pi)^3}\ e^{i\,p\cdot \xi}
\ \langle P\vert \overline\psi(0)\,U_{[0,\pm\infty]}^{[n]}
\,U^\st_{[0_\st,\xi_\st]}
\,iD_\st^{\alpha}
\,U_{[\pm\infty,\xi]}^{[n]} \,\psi(\xi)\vert P\rangle\right|_{LF} .
\eea
Using the result in Eq.~\ref{64} one gets
\be
\Phi^{\alpha\,[\pm]}_{\partial}(x,p_\st)
= \widetilde\Phi^{\alpha\,[\pm]}_{\partial}(x,p_\st)
\pm \pi\Phi_{G}^{\alpha\,[\pm]}(x,p_\st) ,
\ee
where
\bea
\pi\Phi_{G}^{\alpha\,[\pm]}(x,p_\st)
&=&\left. \pm\int \frac{d\,\xi{\cdot}P\,d^2\xi_\st}{(2\pi)^3}
\ e^{i\,p\cdot \xi}
\ \langle P\vert \overline\psi(0)\,U_{[0,\pm\infty]}^{[n]}
\,U^\st_{[0_\st,\xi_\st]}
A_\st^{\alpha}(\pm\infty,\xi_\st)
\,U_{[\pm\infty,\xi]}^{[n]} \,\psi(\xi)\vert P\rangle\right|_{LF}
\nonumber \\ 
&=&
\left.\int \frac{d\,\xi{\cdot}P\,d^2\xi_\st}{(2\pi)^3}\ e^{i\,p\cdot \xi}
\ \langle P\vert \overline\psi(0)\,W_{\pm[0,\xi]}^{[n]}
\,\pi\widetilde G_\st^{n\alpha}(\xi_\st)
\,\psi(\xi)\vert P\rangle\right|_{LF} ,
\\
\widetilde\Phi_{\partial}^{\alpha\,[\pm]}(x,p_\st)
&=&\left.\int \frac{d\,\xi{\cdot}P\,d^2\xi_\st}{(2\pi)^3}\ e^{i\,p\cdot \xi}
\ \langle P\vert \overline\psi(0)\,W_{\pm[0,\xi]}^{[n]}
\,i\partial_\st^{\alpha}
\,\psi(\xi)\vert P\rangle\right|_{LF} 
\nonumber \\
&=&\left.\int \frac{d\,\xi{\cdot}P\,d^2\xi_\st}{(2\pi)^3}\ e^{i\,p\cdot \xi}
\ \langle P\vert \overline\psi(0)\,W_{\pm[0,\xi]}^{[n]}
\left(iD_\st^{\alpha}(\xi) - A_\st^\alpha(\xi)\right)
\,\psi(\xi)\vert P\rangle\right|_{LF} ,
\eea
or after integration over $p_\st$,
\bea
\Phi_\partial^{\alpha\,[\pm]}(x) & \equiv & 
\int d^2p_\st\ p_\st^{\alpha}\,\Phi^{[\pm]}(x,p_\st)
= \Phi_D^\alpha(x) 
- \int dx_1\ \frac{i}{x-x_1\mp i\epsilon}\,\Phi_G^\alpha(x,x_1)
\\ & = &
\underbrace{\Phi_D^\alpha(x) 
- \Phi_A^\alpha(x)}_{\widetilde\Phi_\partial^\alpha(x)} 
\pm \pi\,\Phi_G^\alpha(x,x).
\eea
We note that the sole dependence on the direction of the gauge link
is in the sign in front of the gluonic pole correlator.

In the analysis presented here, we have tacitly assumed that the Wilson
line in combination with the quark fields could be combined into a
correlator with a gauge link and we only considered the cases of the
simplest TMDs $\Phi^{[\pm]}$ (without Wilson loops).
To see what is happening in general we must realize that the 
complications arise from differentiating the Wilson line and
obtaining covariant derivatives (in the above expressions the
$iD_\st^\alpha$) in $\Phi_\partial^\alpha$. This result is then
separated into two parts, the parts $\widetilde\Phi_\partial^\alpha$,
which has the same color structure as $\Phi$ itself and a 
gluonic pole part of which the color structure is 
$\Phi_G^\alpha(x,x) = {\rm Tr}_c[\Phi_8(x)\widetilde G^{n\alpha}[p]]$,
where the argument $[p]$ of $\widetilde G^{n\alpha}$ 
(see Eq.~\ref{Gtilde}) just indicates its (spatial) connection to
the coordinates in the quark-quark correlator $\Phi(p)$, similarly as
has been used for the Wilson lines $U[p]$. In the entangled expressions
for the cross sections we have to use
\bea
&& \left.
i\partial_{1\st}^\alpha\left(\Phi(p_1)\ldots U_-^{[n]\dagger}[p_1]
\,\Phi(p_2)\,U_-^{[n]}[p_1]\ldots\right)\right|_{LC} 
\nonumber \\ && \mbox{}\hspace{3cm} = 
\left(\widetilde\Phi_\partial^\alpha(p_1)
\ldots \Phi(p_2)\ldots\right) 
+ \left(\Phi_8(x_1)
\ldots \pi\widetilde G^{n\alpha}[p_1]\,\Phi(p_2)\ldots\right) .
%\\
%&& \left.
%i\partial_{1\st}^\alpha\left(\Phi(p_1)\ldots U_-^{[n]\dagger}[p_1]
%\,\Phi_8(p_2)\,U_-^{[n]}[p_1]\ldots\right)\right|_{LC} 
%\nonumber \\ && \mbox{}\hspace{3cm} = 
%\left(\widetilde\Phi_\partial^\alpha(p_1)
%\ldots \Phi_8(p_2)\ldots\right) 
%+ \left(\Phi_8(x_1)
%\ldots [\pi\widetilde G^{n\alpha}[p_1],\Phi_8(p_2)]\ldots\right) .
\eea
After further color disentangling of the right-hand-side, where one must
be careful because $\Phi(p_2)$ and possible other entries can have an octet 
structure, one then can recombine parts into
$\pi\Phi_G^{\alpha}(x,x) 
= {\rm Tr}_c[\Phi_8(x)\,\pi\widetilde G^{n\alpha}[p]]$, as explained
for a full quark-quark-gluon correlator following Eq.~\ref{quarkgluonquark}.
For an antiquark correlator $\overline\Phi(p_2)$ one obtains
\bea
&& \left.
i\partial_{1\st}^\alpha\left(\Phi(p_1)\ldots U_-^{[n]\dagger}[p_1]
\,\overline\Phi(p_2)\,U_-^{[n]}[p_1]\ldots\right)\right|_{LC} 
\nonumber \\ && \mbox{}\hspace{3cm} = 
\left(\widetilde\Phi_\partial^\alpha(p_1)
\ldots \overline\Phi(p_2)\ldots\right) 
- \left(\Phi_8(x_1)\ldots \overline\Phi(p_2)
\,\pi\widetilde G^{n\alpha}[p_1] \ldots\right) .
%\\
%&& \left.
%i\partial_{1\st}^\alpha\left(\Phi(p_1)\ldots U_-^{[n]\dagger}[p_1]
%\,\overline\Phi_8(p_2)\,U_-^{[n]}[p_1]\ldots\right)\right|_{LC} 
%\nonumber \\ && \mbox{}\hspace{3cm} = 
%\left(\widetilde\Phi_\partial^\alpha(p_1)
%\ldots \overline\Phi_8(p_2)\ldots\right) 
%- \left(\Phi_8(x_1)
%\ldots [\overline\Phi_8(p_2),\pi\widetilde G^{n\alpha}[p_1]]\ldots\right) .
\eea
Starting with the cross section $\sigma_1$ in Eq.~\ref{facto6a} we find
that weighting with transverse momentum gives
\bea
\langle p_{1\st}^\alpha\,\sigma_1 \rangle  & \sim  &
{\rm Tr}_c\bigl[
\widetilde\Phi_\partial^\alpha(x_1)
\,\Gamma_1^\ast\,\Delta(z_1)\,\Gamma_1
\bigr]
\,{\rm Tr}_c\bigl[\Phi(x_2)
\,\Gamma_2^\ast\,\Delta(z_2)\,\Gamma_2
\bigr] 
\nonumber \\ && \mbox{} +
{\rm Tr}_c\bigl[
\Phi_8(x_1)\,\pi \widetilde G^{n\alpha}[p_1]
\,\Gamma_1^\ast\,\Delta(z_1)\,\Gamma_1
\bigr]
\,{\rm Tr}_c\bigl[
\Phi(x_2)\,\Gamma_2^\ast
\,\Delta(z_2)\,\Gamma_2
\bigr]
\nonumber \\ && \mbox{} +
{\rm Tr}_c\bigl[
\Phi_8(x_1)\,\Gamma_1^\ast\,\Delta(z_1)\,\Gamma_1
\bigr]
\,{\rm Tr}_c\bigl[
\Phi(x_2)\,2\pi \widetilde G^{n\alpha}[p_1]\,\Gamma_2^\ast
\,\Delta(z_2)\,\Gamma_2
\bigr].
\eea
The traces in the last term are zero and we are left with
\bea
\langle p_{1\st}^\alpha\,\sigma_1\rangle  & \sim  &
\bigl(\widetilde\Phi_\partial^\alpha(x_1)
+\pi\Phi_G^\alpha(x_1,x_1)\bigr)\,\Phi(x_2)\,\hat\Sigma_1
\,\Delta(z_1)\,\Delta(z_2).
\eea
The weighted cross section of Eq.~\ref{facto16a} gives
\bea
\langle p_{1\st}^\alpha\,\sigma_2\rangle & \sim &
{\rm Tr}_c\bigl[
\widetilde\Phi_\partial^\alpha(x_1)\,\Gamma_1^\ast
\,\Delta(z_1)\,\Gamma_2
\,\Phi(x_2)\,\Gamma_2^\ast
\,\Delta(z_2)\,\Gamma_1 
\bigr] 
\nonumber \\ && \mbox{} +
{\rm Tr}_c\bigl[
\Phi_8(x_1)\,\pi \widetilde G^{n\alpha}[p_1]\,\Gamma_1^\ast
\,\Delta(z_1)\,\Gamma_2
\,\Phi(x_2)\,\Gamma_2^\ast
\,\Delta(z_2)\,\Gamma_1 
\bigr] 
\nonumber \\ && \mbox{} +
{\rm Tr}_c\bigl[
\Phi_8(x_1)\,\Gamma_1^\ast
\,\Delta(z_1)\,\Gamma_2
\,\Phi(x_2)\,2\pi \widetilde G^{n\alpha}[p_1]
\,\Gamma_2^\ast
\,\Delta(z_2)\,\Gamma_1 
\bigr] .
\eea
All terms survive and we are left with
\bea
\langle p_{1\st}^\alpha\,\sigma_2\rangle  & \sim  &
\bigl(\widetilde\Phi_\partial^\alpha(x_1)
+ 3\pi\Phi_G^\alpha(x_1,x_1)\bigr)\,\Phi(x_2)\,\hat\Sigma_2
\,\Delta(z_1)\,\Delta(z_2) .
\eea
Combining both contributions in the same way as in Eq.~\ref{jungle}, 
one obtains the result~\cite{Bomhof:2006dp},
\bea
\langle p_{1\st}^\alpha\,\sigma\rangle  &\sim & 
\Bigl(\tfrac{N_c^2+1}{N_c^2-1}\Phi_\partial^{\alpha[+(\Box)]}(x_1)
\,\Phi(x_2)\,\hat\Sigma_1
-\tfrac{2}{N_c^2-1}\Phi_\partial^{\alpha[+\Box]}(x_1)
\,\Phi(x_2)\,\hat\Sigma_2
\Bigr)
\,\Delta(z_1) \,\Delta(z_2)
\nonumber \\ &=&
\widetilde\Phi_\partial^\alpha(x_1)\,\Phi(x_2)
\underbrace{\Bigl(\tfrac{N_c^2+1}{N_c^2-1}\,\hat\Sigma_1
-\tfrac{2}{N_c^2-1}\,\hat\Sigma_2
\Bigr)}_{\hat\Sigma}
\,\Delta(z_1) \,\Delta(z_2)
\nonumber\\ && \mbox{} +
\pi\Phi_G^{\alpha}(x_1,x_1)\,\Phi(x_2)
\underbrace{\Bigl(\tfrac{N_c^2+1}{N_c^2-1}\,\hat\Sigma_1
-\tfrac{6}{N_c^2-1}\,\hat\Sigma_2
\Bigr)}_{\hat\Sigma_{GP}}
\,\Delta(z_1) \,\Delta(z_2),
\label{gpxsection}
\eea
where the combination of hard squared amplitudes in the second term
of Eq.~\ref{gpxsection} is referred to as the {\em gluonic pole
cross section}.
In the case of quark-quark scattering with distinguishable quarks
($\hat\Sigma_1 = \hat\Sigma_2$) one thus has $\hat\Sigma_{GP}$
= $\tfrac{N_c^2-5}{N_c^2-1}\hat\Sigma$, and for $N_c = 3$ we get
\bea
\langle p_{1\st}^\alpha\,\sigma\rangle  &\sim & 
\Bigl(\widetilde\Phi_\partial^\alpha(x_1)
\,\Phi(x_2) \,\hat\Sigma
+\pi\Phi_G^{\alpha}(x_1,x_1)\,\Phi(x_2) 
\,\tfrac{1}{2}\,\hat\Sigma\Bigr)
\,\Delta(z_1) \,\Delta(z_2).
\eea
In this result the two pieces in the cross section experimentally can be
distinguished because of their time-reversal behavior. For instance
single spin asymmetries have opposite time reversal behavior as 
compared to spin-averaged or double spin asymmetries. This time reversal
behavior affects the parametrization of the gluonic pole matrix element 
(containing T-odd distribution functions such as Sivers and Boer-Mulders 
functions). These functions appear 
convoluted with the gluonic pole cross section rather than the 
standard partonic cross section. For hadron-hadron scattering
the partonic cross sections and gluonic pole cross sections have
been tabulated in Ref.~\cite{Bomhof:2006ra}.

At this point we want to comment on the usefulness of our diagrammatic
approach starting with an assumed convolution of soft correlators 
$\Phi(x,p_\st)$ (integrated over $p\cdot P$) and hard amplitudes. For
this we have to realize that the operators involved have particular
canonical dimensions $d$ and twist $t$. Assignment of definite twist is only
possible for local matrix elements or collinear correlators,
e.g.\ the quark-quark (or gluon-gluon) correlators $\Phi^{[t=2]}(x)$,
$\Phi^{[t=3]}(x)$, \ldots or quark-quark-gluon correlators
$\Phi_D^{[t=3]}(x)$ and $\Phi_A^{[t=3]}(x)$. We have schematically
\[
\Phi^{[d=2]}(x,p_\st) = \Phi^{[t\ge 2]}(x,p_\st),
\]
indicating on the right-hand-side that operators are involved 
of arbitrary twist with the minimal twist being the canonical dimension.

Now, let us look at an observable at a high-energy scale $Q$ 
including its dependence on collinear 
fractions and transverse momenta assuming appropriate identification
of parton variables with kinematical variables, which for transverse
momenta in hadron-hadron scattering would be $q_\st = p_{1\st} + p_{2\st}$.
We have
\bea
d\sigma (x_1,x_2,q_\st;Q) & = &
\Phi^{[d=2]}(x_1,p_{1\st}) \otimes \Phi^{[d=2]}(x_2,p_{2\st}) 
\otimes \hat \Sigma 
\nonumber \\ & + &
\frac{M}{Q}\bigl[\Phi^{[d=2]}(x_1,p_{1\st}) \otimes \Phi^{[d=2]}(x_2,p_{2\st}) 
\otimes \hat \Sigma \bigr]
\nonumber \\ & + &
\frac{1}{Q}\bigl[\Phi^{[d=3]}(x_1,p_{1\st}) \otimes \Phi^{[d=2]}(x_2,p_{2\st}) 
\otimes \hat \Sigma 
+ \Phi^{[d=2]}(x_1,p_{1\st}) \otimes \Phi^{[d=3]}(x_2,p_{2\st}) 
\otimes \hat \Sigma \bigr] 
\nonumber \\ & + & \ldots ,
\label{diagrammatic}
\eea
of which only the first term survives at $Q\rightarrow \infty$.
Looking at the $q_\st$-averaged (i.e.\ integrated) cross section, we have
as leading contribution
\be
\langle \sigma \rangle (x_1,x_2;Q)
\quad \stackrel{Q\rightarrow \infty}{\longrightarrow} \quad
\Phi^{[t=2]}(x_1) \otimes \Phi^{[t=2]}(x_2) 
\otimes \hat \Sigma ,
\ee
which summarizes the collinear approach. The leading contribution in the
$q_\st$-weighted cross section is given by 
\be
\langle q_\st\,\sigma \rangle (x_1,x_2;Q)
\quad \stackrel{Q\rightarrow \infty}{\longrightarrow} \quad
\langle p_{1\st}\,\Phi^{[d=2]}\rangle (x_1) \otimes \Phi^{[t=2]}(x_2) 
\otimes \hat \Sigma +
\Phi^{[t=2]}(x_1) \otimes
\langle p_{2\st}\,\Phi^{[d=2]}\rangle (x_2) 
\otimes \hat \Sigma ,
\ee
where the weighted TMD correlators contain actually twist three and higher twist
operators, schematically 
\[
\langle p_{\st}\,\Phi^{[d=2]}\rangle (x) = 
\Phi^{[t\ge 3]}(x) . 
\]
Among the twist three correlators we have T-even correlators 
$\Phi_D^{[t=3]}$, $\Phi_A^{[t=3]}$ in the combination 
$\widetilde\Phi_\partial^{[t=3]}$ and the T-odd gluonic pole correlator
$\Phi_G^{[t=3]}(x,x)$. The usefulness of the diagrammatic approach
starting with Eq.~\ref{diagrammatic} is in its ability to provide us
in a straightforward way with the tree-level coefficients of the relevant
(combinations of) twist-3 collinear correlators through the intermediate
step of TMD correlators with a complex (process-dependent) gauge link. 
Similar results for the weighted asymmetries can also be obtained 
using collinear functions including the gluonic pole or
ETQS functions directly from the start~\cite{Beppu:2010qn,Beppu:2010sp}.
The tree-level results of the diagrammatic approach, of course, need 
to be improved upon by including NLO QCD contributions.
In line with our proof of absorbing all gauge connections into a 
gauge link in 1PU processes, the diagrammatic approach also provides
the tree-level results for higher transverse moments as long as only
one initial state hadron is involved. This is useful first of all for
leptoproduction processes involving a single incoming hadron and,
furthermore, for selected asymmetries in hadron-hadron scattering.

\subsection{Double weighted asymmetries}

We want to use the analysis in terms of transverse moments to 
indicate what is happening when the transverse momentum of two
hadrons is involved, e.g.\ a double Sivers or double Boer-Mulders effect.
The weightings with $p_{1\st}^\alpha\,p_{1\st}^\beta$
or with $p_{2\st}^\alpha\,p_{2\st}^\beta$ can straightforwardly
be obtained from Eq.~\ref{jungle}, although the splitting of TMDs 
with more complicated gauge links in T-even and T-odd parts is 
certainly not trivial.
We will investigate the weighting with 
$p_{1\st}^\alpha\,p_{2\st}^\beta$ following the same steps as done 
in the above for the single moments.

We will first turn to an `elementary' process, namely DY scattering.
Having one color loop, one finds that in single and double weighting
one needs
\bea
&&
{\rm Tr}_c[I] = \frac{1}{N_c}\,{\rm Tr}_c[I]\,{\rm Tr}_c[I] ,
\nonumber
\\ &&
{\rm Tr}_c[T^aT^a] = \frac{1}{N_c}\,{\rm Tr}_c[T^aT^a]\,{\rm Tr}_c[I] ,
\nonumber
\\ &&
%{\rm Tr}_c[[T^a,T^b][T^b,T^a]] = 
%\frac{N_c}{N_c^2-1}\,{\rm Tr}_c[T^aT^a]\,{\rm Tr}_c[T^bT^b] 
{\rm Tr}_c[T^aT^bT^aT^b] = 
-\frac{1}{N_c(N_c^2-1)}\,{\rm Tr}_c[T^aT^a]\,{\rm Tr}_c[T^bT^b] ,
\nonumber
\eea
yielding 
\bea
\langle p_{1\st}^\alpha\,p_{2\st}^\beta\sigma_{DY}\rangle 
& = &
\biggl(\widetilde\Phi_\partial^\alpha(x_1)
\,\widetilde{\overline\Phi}{}_\partial^\beta(x_2)
- \pi\Phi_G^\alpha(x_1,x_1)\,\widetilde{\overline\Phi}{}_\partial^\beta(x_2)
- \widetilde\Phi_\partial^\alpha(x_1)\,\pi\overline\Phi{}_G^\beta(x_2,x_2)
\nonumber \\ &&
\mbox{} \hspace{1cm} - \tfrac{1}{N_c^2-1}
\,\pi\Phi_G^\alpha(x_1,x_1)
\,\pi\overline\Phi{}_G^\beta(x_2,x_2)\biggr)
\,\hat\sigma_{DY} .
\eea
This indicates a breaking of universality for the double gluonic pole
contribution. This breaking is consistent with non-factorizability of
DY at twist four (actually double twist-3), which would be the level
needed to consider similar asymmetries in a collinear treatment that
employs gluonic pole matrix elements and where one ought to find the same breaking. 
It also shows that TMD factorization
of DY does not hold for the T-odd parts in the quark-quark correlators.
The actual outcome is a strong suppression of the double T-odd 
contributions and a sign change (-1/8 as compared to 1).

In the same way as for DY, we can analyze
higher moments in other processes for which we will consider 
the example of quark-quark scattering. 
From the unintegrated result in Eq.~\ref{facto5b} one gets the 
double-weighted result,
\bea
\langle p_{1\st}^\alpha\,p_{2\st}^\beta\,\sigma_1 \rangle  &\sim &
{\rm Tr}_c\bigl[
\widetilde\Phi_\partial^\alpha(x_1)
\,\Gamma_1^\ast\,\Delta(z_1)\,\Gamma_1 \bigr]
\,{\rm Tr}_c\bigl[
\widetilde\Phi_\partial^\beta(x_2)
\,\Gamma_2^\ast\,\Delta(z_2)\,\Gamma_2 \bigr]
\nonumber \\ && \mbox{} +
{\rm Tr}_c\bigl[
\widetilde\Phi_\partial^\alpha(x_1)
\,\Gamma_1^\ast\,\Delta(z_1)\,\Gamma_1 \bigr]
\,{\rm Tr}_c\bigl[
\Phi_8(x_2)\,\pi\widetilde G^{n\beta}[p_2]
\,\Gamma_2^\ast\,\Delta(z_2)\,\Gamma_2 \bigr]
\nonumber \\ && \mbox{} +
{\rm Tr}_c\bigl[
\Phi_8(x_1)\,\pi\widetilde G^{n\alpha}[p_1]
\,\Gamma_1^\ast\,\Delta(z_1)\,\Gamma_1 \bigr]
\,{\rm Tr}_c\bigl[
\widetilde\Phi_\partial^\beta(x_2)
\,\Gamma_2^\ast\,\Delta(z_2)\,\Gamma_2 \bigr]
\nonumber \\ && \mbox{} +
{\rm Tr}_c\bigl[
\widetilde\Phi_\partial^\alpha(x_1)\,2\pi\widetilde G^{n\beta}[p_2]
\,\Gamma_1^\ast\,\Delta(z_1)\,\Gamma_1 \bigr]
\,{\rm Tr}_c\bigl[
\Phi_8(x_2)\,\Gamma_2^\ast\,\Delta(z_2)\,\Gamma_2 \bigr]
\nonumber \\ && \mbox{} +
{\rm Tr}_c\bigl[
\Phi_8(x_1)\,\Gamma_1^\ast\,\Delta(z_1)\,\Gamma_1 \bigr]
\,{\rm Tr}_c\bigl[
\widetilde\Phi_\partial^\beta(x_2)\,2\pi\widetilde G^{n\alpha}[p_1]
\,\Gamma_2^\ast\,\Delta(z_2)\,\Gamma_2 \bigr]
\nonumber \\ && \mbox{} +
{\rm Tr}_c\bigl[
\Phi_8(x_1)\,\tfrac{1}{2}\{\pi\widetilde G^{n\alpha}[p_1],
\pi\widetilde G^{n\beta}[p_2]\}
\,\Gamma_1^\ast\,\Delta(z_1)\,\Gamma_1 \bigr]
\,{\rm Tr}_c\bigl[
\Phi_8(x_2)\,\Gamma_2^\ast\,\Delta(z_2)\,\Gamma_2 \bigr]
\nonumber \\ && \mbox{} +
{\rm Tr}_c\bigl[
\pi\widetilde G^{n\beta}[p_2]\,\Phi_8(x_1)\,\pi\widetilde G^{n\alpha}[p_1]
\,\Gamma_1^\ast\,\Delta(z_1)\,\Gamma_1 \bigr]
\,{\rm Tr}_c\bigl[
\Phi_8(x_2)\,\Gamma_2^\ast\,\Delta(z_2)\,\Gamma_2 \bigr]
\nonumber \\ && \mbox{} +
{\rm Tr}_c\bigl[
\Phi_8(x_1)\,\Gamma_1^\ast\,\Delta(z_1)\,\Gamma_1 \bigr]
\,{\rm Tr}_c\bigl[
\Phi_8(x_2)\,\tfrac{1}{2}\{\pi\widetilde G^{n\beta}[p_2],
\pi\widetilde G^{n\alpha}[p_1]\}
\,\Gamma_2^\ast\,\Delta(z_2)\,\Gamma_2 \bigr]
\nonumber \\ && \mbox{} +
{\rm Tr}_c\bigl[
\Phi_8(x_1)\,\Gamma_1^\ast\,\Delta(z_1)\,\Gamma_1 \bigr]
\,{\rm Tr}_c\bigl[
\pi\widetilde G^{n\alpha}[p_1]\,\Phi_8(x_2)\,\pi\widetilde G^{n\beta}[p_2]
\,\Gamma_2^\ast\,\Delta(z_2)\,\Gamma_2 \bigr]
\nonumber \\ && \mbox{} +
{\rm Tr}_c\bigl[
\Phi_8(x_1)\,\pi\widetilde G^{n\alpha}[p_1]
\,\Gamma_1^\ast\,\Delta(z_1)\,\Gamma_1 \bigr]
\,{\rm Tr}_c\bigl[
\Phi_8(x_2)\,\pi\widetilde G^{n\beta}[p_2]
\,\Gamma_2^\ast\,\Delta(z_2)\,\Gamma_2 \bigr]
\nonumber \\ && \mbox{} +
{\rm Tr}_c\bigl[
\Phi_8(x_1)\,\pi\widetilde G^{n\beta}[p_2]
\,\Gamma_1^\ast\,\Delta(z_1)\,\Gamma_1 \bigr]
\,{\rm Tr}_c\bigl[
\Phi_8(x_2)\,\pi\widetilde G^{n\alpha}[p_1]
\,\Gamma_2^\ast\,\Delta(z_2)\,\Gamma_2 \bigr]
\nonumber \\ && \mbox{} +
{\rm Tr}_c\bigl[
\Phi_8(x_1)\,\pi\widetilde G^{n\beta}[p_2]
\,\Gamma_1^\ast\,\Delta(z_1)\,\Gamma_1 \bigr]
\,{\rm Tr}_c\bigl[
\pi\widetilde G^{n\alpha}[p_1]\,\Phi_8(x_2)
\,\Gamma_2^\ast\,\Delta(z_2)\,\Gamma_2 \bigr]
\nonumber \\ && \mbox{} +
{\rm Tr}_c\bigl[
\pi\widetilde G^{n\beta}[p_2]\,\Phi_8(x_1)
\,\Gamma_1^\ast\,\Delta(z_1)\,\Gamma_1 \bigr]
\,{\rm Tr}_c\bigl[
\Phi_8(x_2)\,\pi\widetilde G^{n\alpha}[p_1]
\,\Gamma_2^\ast\,\Delta(z_2)\,\Gamma_2 \bigr]
\nonumber \\ && \mbox{} +
{\rm Tr}_c\bigl[
\pi\widetilde G^{n\beta}[p_2]\,\Phi_8(x_1)
\,\Gamma_1^\ast\,\Delta(z_1)\,\Gamma_1 \bigr]
\,{\rm Tr}_c\bigl[
\pi\widetilde G^{n\alpha}[p_1]\,\Phi_8(x_2)
\,\Gamma_2^\ast\,\Delta(z_2)\,\Gamma_2 \bigr].
\eea
Several terms are trivially zero after color tracing (four through nine).
The others give particular
transverse moments, where in particular for the last terms one must
evaluate the extra color factor because of the specific color flow.
To be specific for terms eleven through fourteen one needs
\[
\frac{{\rm Tr}_c[T^aT^b]\,{\rm Tr}_c[T^bT^a]}
{{\rm Tr}_c[T^aT^a]\,{\rm Tr}_c[T^bT^b]} = \frac{1}{N_c^2-1},
\]
as compared to term ten. The result then is
\bea
\langle p_{1\st}^\alpha\,p_{2\st}^\beta\,\sigma_1 \rangle  &\sim &
\Bigl(\widetilde\Phi_\partial^\alpha(x_1) \,\widetilde\Phi_\partial^\beta(x_2)
+ \widetilde\Phi_\partial^\alpha(x_1) \,\pi\Phi_G^\beta(x_2,x_2)
+ \pi\Phi_G^\alpha(x_1,x_1) \,\widetilde\Phi_\partial^\beta(x_2)
\nonumber \\ && \mbox{} +
\tfrac{N_c^2+3}{N_c^2-1}\,\pi\Phi_G^\alpha(x_1,x_1)
\,\pi\Phi_G^\beta(x_2,x_2)\Bigr)\,\hat\Sigma_1
\,\Delta(z_1) \Delta(z_2).
\eea
From the unintegrated result in Eq.~\ref{facto15b} one gets
\bea
\langle p_{1\st}^\alpha\,p_{2\st}^\beta\,\sigma_2 \rangle & \sim &
{\rm Tr}_c\bigl[
\widetilde\Phi_\partial^\alpha(x_1)
\,\Gamma_1^\ast\,\Delta(z_1)\,\Gamma_2
\,\widetilde\Phi_\partial^\beta(x_2)
\,\Gamma_2^\ast\,\Delta(z_2)\,\Gamma_1\bigr]
\nonumber \\ && \mbox{} +
{\rm Tr}_c\bigl[
\widetilde\Phi_\partial^\alpha(x_1)
\,\Gamma_1^\ast\,\Delta(z_1)\,\Gamma_2
\,\Phi_8(x_2)\,\pi\widetilde G^{n\beta}[p_2]
\,\Gamma_2^\ast\,\Delta(z_2)\,\Gamma_1\bigr]
\nonumber \\ && \mbox{} +
{\rm Tr}_c\bigl[
\Phi_8(x_1)\,\pi\widetilde G^{n\alpha}[p_1]
\,\Gamma_1^\ast\,\Delta(z_1)\,\Gamma_2
\,\widetilde\Phi_\partial^\beta(x_2)
\,\Gamma_2^\ast\,\Delta(z_2)\,\Gamma_1\bigr]
\nonumber \\ && \mbox{} +
{\rm Tr}_c\bigl[
\widetilde\Phi_\partial^\alpha(x_1)\,2\pi\widetilde G^{n\beta}[p_2]
\,\Gamma_1^\ast\,\Delta(z_1)\,\Gamma_2
\,\Phi_8(x_2)\,\Gamma_2^\ast\,\Delta(z_2)\,\Gamma_1\bigr]
\nonumber \\ && \mbox{} +
{\rm Tr}_c\bigl[
\Phi_8(x_1)\,\Gamma_1^\ast\,\Delta(z_1)\,\Gamma_2
\,\widetilde\Phi_\partial^\beta(x_2)\,2\pi\widetilde G^{n\alpha}[p_1]
\,\Gamma_2^\ast\,\Delta(z_2)\,\Gamma_1\bigr]
\nonumber \\ && \mbox{} +
{\rm Tr}_c\bigl[
\Phi_8(x_1)
\,\tfrac{1}{2}\{\pi\widetilde G^{n\alpha}[p_1],\pi\widetilde G^{n\beta}[p_2]\}
\,\Gamma_1^\ast\,\Delta(z_1)\,\Gamma_2
\,\Phi_8(x_2)\,\Gamma_2^\ast\,\Delta(z_2)\,\Gamma_1\bigr]
\nonumber \\ && \mbox{} +
{\rm Tr}_c\bigl[
\pi\widetilde G^{n\beta}[p_2]\,\Phi_8(x_1)\,\pi\widetilde G^{n\alpha}[p_1]
\,\Gamma_1^\ast\,\Delta(z_1)\,\Gamma_2
\,\Phi_8(x_2)\,\Gamma_2^\ast\,\Delta(z_2)\,\Gamma_1\bigr]
\nonumber \\ && \mbox{} +
{\rm Tr}_c\bigl[
\Phi_8(x_1)\,\Gamma_1^\ast\,\Delta(z_1)\,\Gamma_2
\,\Phi_8(x_2)
\,\tfrac{1}{2}\{\pi\widetilde G^{n\beta}[p_2],\pi\widetilde G^{n\alpha}[p_1]\}
\,\Gamma_2^\ast\,\Delta(z_2)\,\Gamma_1\bigr]
\nonumber \\ && \mbox{} +
{\rm Tr}_c\bigl[
\Phi_8(x_1)\,\Gamma_1^\ast\,\Delta(z_1)\,\Gamma_2
\,\pi\widetilde G^{n\alpha}[p_1]\,\Phi_8(x_2)\,\pi\widetilde G^{n\beta}[p_2]
\,\Gamma_2^\ast\,\Delta(z_2)\,\Gamma_1\bigr]
\nonumber \\ && \mbox{} +
{\rm Tr}_c\bigl[
\Phi_8(x_1)\,\pi\widetilde G^{n\alpha}[p_1]
\,\Gamma_1^\ast\,\Delta(z_1)\,\Gamma_2
\,\Phi_8(x_2)\,\pi\widetilde G^{n\beta}[p_2]
\,\Gamma_2^\ast\,\Delta(z_2)\,\Gamma_1\bigr]
\nonumber \\ && \mbox{} +
{\rm Tr}_c\bigl[
\Phi_8(x_1)\,\pi\widetilde G^{n\beta}[p_2]
\,\Gamma_1^\ast\,\Delta(z_1)\,\Gamma_2
\,\Phi_8(x_2)\,\pi\widetilde G^{n\alpha}[p_1]
\,\Gamma_2^\ast\,\Delta(z_2)\,\Gamma_1\bigr] 
\nonumber \\ && \mbox{} +
{\rm Tr}_c\bigl[
\Phi_8(x_1)\,\pi\widetilde G^{n\beta}[p_2]
\,\Gamma_1^\ast\,\Delta(z_1)\,\Gamma_2
\,\pi\widetilde G^{n\alpha}[p_1]\,\Phi_8(x_2)
\,\Gamma_2^\ast\,\Delta(z_2)\,\Gamma_1\bigr] 
\nonumber \\ && \mbox{} +
{\rm Tr}_c\bigl[
\pi\widetilde G^{n\beta}[p_2]\,\Phi_8(x_1)
\,\Gamma_1^\ast\,\Delta(z_1)\,\Gamma_2
\,\Phi_8(x_2)\,\pi\widetilde G^{n\alpha}[p_1]
\,\Gamma_2^\ast\,\Delta(z_2)\,\Gamma_1\bigr] 
\nonumber \\ && \mbox{} +
{\rm Tr}_c\bigl[
\pi\widetilde G^{n\beta}[p_2]\,\Phi_8(x_1)
\,\Gamma_1^\ast\,\Delta(z_1)\,\Gamma_2
\,\pi\widetilde G^{n\alpha}[p_1]\,\Phi_8(x_2)
\,\Gamma_2^\ast\,\Delta(z_2)\,\Gamma_1\bigr] .
\eea
To rewrite everything in terms of standard transverse moments
$\Phi_G^\alpha$ we have to compare the color structure of the terms
with two gluonic poles with the standard color structure for 
$\sigma_2$. We need the ratio
\[
\frac{{\rm Tr}_c[T^aT^bT^aT^b]}
{{\rm Tr}_c[T^aT^aT^bT^b]} = -\frac{1}{N_c^2-1}.
\]
We note that the factor 
${\rm Tr}_c[T^aT^aT^bT^b]/{\rm Tr}_c[T^aT^a]\,{\rm Tr}_c[T^bT^b] = 1/N_c$
is already incorporated in the factors multiplying $\hat\Sigma_1$ 
and $\hat\Sigma_2$.
The result is
\bea
\langle p_{1\st}^\alpha\,p_{2\st}^\beta\,\sigma_2 \rangle & \sim &
\Bigl(\widetilde\Phi_\partial^\alpha(x_1)\,\widetilde\Phi_\partial^\beta(x_2)
+ 3\,\widetilde\Phi_\partial^\alpha(x_1)\,\pi\Phi_G^\beta(x_2,x_2)
+ 3\,\pi\Phi_G^\alpha(x_1,x_1)\,\widetilde\Phi_\partial^\beta(x_2)
\nonumber \\ && \mbox{} +
\tfrac{6N_c^2 - 9}{N_c^2-1}\,\pi\Phi_G^\alpha(x_1,x_1)\,\pi\Phi_G^\beta(x_2,x_2)
\Bigr)\,\hat\Sigma_2
\,\Delta(z_1)\,\Delta(z_2) .
\eea
Combining the terms we get
\bea
\langle p_{1\st}^\alpha\,p_{2\st}^\beta\,\sigma \rangle & \sim &
\widetilde\Phi_\partial^\alpha(x_1)
\,\widetilde\Phi_\partial^\beta(x_2)
\underbrace{\Bigl(\tfrac{N_c^2+1}{N_c^2-1}\,\hat\Sigma_1
-\tfrac{2}{N_c^2-1}\,\hat\Sigma_2
\Bigr)}_{\hat\Sigma}
\,\Delta(z_1) \,\Delta(z_2)
\nonumber\\ && \mbox{} +
\Bigl(\widetilde\Phi_\partial^\alpha(x_1)\,\pi\Phi_G^\beta(x_2,x_2)
+ \pi\Phi_G^\alpha(x_1,x_1)\,\widetilde\Phi_\partial^\beta(x_2)\Bigr)
\underbrace{\Bigl(\tfrac{N_c^2+1}{N_c^2-1}\,\hat\Sigma_1
-\tfrac{6}{N_c^2-1}\,\hat\Sigma_2
\Bigr)}_{\hat\Sigma_{GP}}
\,\Delta(z_1) \,\Delta(z_2)
\nonumber\\ && \mbox{} +
\pi\Phi_G^\alpha(x_1,x_1)\,\pi\Phi_G^\beta(x_2,x_2)
\Bigl(
\tfrac{(N_c^2+1)(N_c^2+3)}{(N_c^2-1)^2}\,\hat\Sigma_1
- \tfrac{12N_c^2-18}{(N_c^2-1)^2}\,\hat\Sigma_2\Bigr)
\,\Delta(z_1) \,\Delta(z_2),
\eea
which in our example of distinguishable quarks ($\hat\Sigma_1 = \hat\Sigma_2$)
gives
\bea
\langle p_{1\st}^\alpha\,p_{2\st}^\beta\,\sigma \rangle & \sim &
\widetilde\Phi_\partial^\alpha(x_1)
\,\widetilde\Phi_\partial^\beta(x_2)
\,\hat\Sigma
\,\Delta(z_1) \,\Delta(z_2)
\nonumber \\ && \mbox{} +
\bigl(\widetilde\Phi_\partial^\alpha(x_1)\,\pi\Phi_G^\beta(x_2,x_2)
+\pi\Phi_G^\alpha(x_1,x_1)\,\widetilde\Phi_\partial^\beta(x_2)\bigr)
\,\tfrac{N_c^2-5}{N_c^2-1}
\,\hat\Sigma
\,\Delta(z_1) \,\Delta(z_2)
\nonumber \\ && \mbox{} +
\pi\Phi_G^\alpha(x_1,x_1)
\,\pi\Phi_G^\beta(x_2,x_2)
\,\tfrac{(N_c^4 - 8N_c^2 + 21)}{(N_c^2-1)^2}
\,\hat\Sigma
\,\Delta(z_1) \,\Delta(z_2)
\\ &=&
\Bigl(\widetilde\Phi_\partial^\alpha(x_1)
\,\widetilde\Phi_\partial^\beta(x_2)\,\hat\Sigma
+\bigl(\widetilde\Phi_\partial^\alpha(x_1)
\,\pi\Phi_G^\beta(x_2,x_2)
+\pi\Phi_G^\alpha(x_1,x_1)
\,\widetilde\Phi_\partial^\beta(x_2)\bigr)
\,\tfrac{1}{2}\,\hat\Sigma
\nonumber \\ && \mbox{} +
\,\pi\Phi_G^\alpha(x_1,x_1)\,\pi\Phi_G^\beta(x_2,x_2)
\,\tfrac{15}{32}\,\hat\Sigma \Bigr)
\,\Delta(z_1) \,\Delta(z_2).
\eea
The result shows that while the effect of a single gluonic pole
is suppressed (factor 1/2) as compared to the naive TMD factorization, 
the effect of the double gluonic pole (double Sivers or double Boer-Mulders) 
is suppressed less than the naive expectation 
expected on the basis of single gluonic pole effects
that are expected to show up in single spin asymmetries
(factor 15/32 versus 1/4).

Although these examples for DY and hadron-hadron scattering show that
TMD factorization fails, we think that the starting point Eq.~\ref{facto4} 
at the end of Section~\ref{section3} appears to be useful, incorporating 
all leading matrix elements based on a counting of canonical dimensions.
From that expression, one can proceed and calculate higher weighted 
asymmetries. 

\section{Conclusions}

In this paper we have used a diagrammatic approach to analyse the 
leading contributions at tree-level in hard processes in which several
hadrons are involved. The diagrammatic approach combines correlators 
involving hadron states and parton fields with hard partonic amplitudes. 
The correlators depend on (on-shell) hadron momenta (initial state
hadrons $P$, produced hadrons $K_h$) and the parton momenta ($p$ and $k$). 
At high energies, momentum fractions $x\,P$ or $(1/z)\,K_h$ and 
transverse momenta can be used, while the components $p{\cdot}P$ and 
$k{\cdot}K_h$ are integrated over. At that stage one has
correlators that also have a natural link to light-cone wave 
functions~\cite{Pasquini:2009eb,Braun:2011aw}. 
In the usual collinear treatment, 
the transverse momenta are also integrated over, or one studies the limit 
in which these momenta become large, which just is a collinear 
treatment involving a partonic process with one or more additional
partons. Using TMD correlators one aims at incorporating all 
features related to (soft) transverse momenta, including in the
parametrization of TMDs the possibility of T-odd correlators, not 
forbidden by any symmetry. These incorporate the effects of initial
and final state interactions~\cite{Brodsky:2002cx} and they can explain 
the existence of single spin asymmetries at high energies.

The basic tree-level result was presented in Section~\ref{section3}.
It includes all Wilson lines originating from collinear gluons in
all of the correlators. These contributions are sufficient to study
the collinear correlators (integrating over transverse components of
the parton momenta) and give rise to straight light-like gauge links.
For TMD correlators, one has to include transverse gauge connections,
but after accounting for the collinear gluons, this all works out
nice and produces the transverse pieces at the right 
place~\cite{Belitsky:2002sm,Boer:2003cm}. One obtains
a color gauge-invariant expression (Eq.~\ref{facto4}), which is fully
entangled.

In order to disentangle color, one needs to consider for each diagrammatic
contribution in the hard amplitude, also its color-flow possibilities. 
This does not resolve the entanglement in general. Also, integrating 
over transverse momenta of final state partons, e.g.\ when one considers
jet production, does not resolve the problem. Then one remains with 
an entangled
situation if the initial state involves more than one hadron, as in
the case of hadron-hadron scattering. This was considered as an
explicit example of factorization breaking in Ref.~\cite{Rogers:2010dm}.
The full entangled tree-level result for this situation is given in
Eq.~\ref{facto5b}. Even in cases in which one just has a single color flow
such as in the electroweak processes of leptoproduction, electron-positron
annihilation or the Drell-Yan process, one gets TMD factorization with
correlators having specific process-dependent gauge links running via
$\pm$ light-cone infinity. For these cases one then has a starting point
with universal correlators and one can study the full QCD factorization.
There are several examples of similarly {\em simple} processes, such as 
two-photon production in Drell-Yan, which even for gluons has a 
simple color flow~\cite{Qiu:2011ai}.

The main aim of this paper was to show what happens in the special case 
of a 1PU process, in which only the transverse momentum 
in one specific hadron is considered.  In that situation, one can resolve
the entanglement (at least at tree-level). 
What remains is a color gauge-invariant expression (Eq.~\ref{jungle})
with a process-dependent gauge link. All-but-one collinear processes
may seem hard to realize at first sight, since the parton transverse
momentum is an integration variable, rather than an observable. 
Nevertheless, symmetry considerations in combination with polarization, 
in particular transverse polarization of target hadrons, may help to
create the conditions for a 1PU treatment. Furthermore one has in
leptoproduction processes a whole class of such processes.
Consideration of TMD correlators is possible for quark as well as 
gluon correlators. A recent example
of the latter situation was pointed out in Ref.~\cite{Boer:2010zf}
which considered heavy quark production in the Drell-Yan process.

The simplifications and steps towards a TMD factorizable result become
more transparant if one constructs transverse moments. This analysis
has been given in Section~\ref{section5} for single and double weighted
asymmetries. Single weighted asymmetries can use the results of the 
1PU situation and lead to the results where
T-even transverse moments are convoluted with the standard partonic
cross section and T-odd transverse moments are convoluted with
a different (but color gauge-invariant) combination of squared amplitudes, 
the so-called gluonic pole cross sections. The T-odd transverse
moments are weighted TMD correlators, involving $p_\st^2$-weighted
T-odd distribution functions like the Sivers or Boer-Mulders 
distribution functions of which the operator structure is a
gluonic pole matrix element, i.e.\ a quark-quark-gluon matrix element
at vanishing gluon momentum.

While the single weighted asymmetries exhibit factorization at the
TMD-level (with process-dependence through the link in the TMD) and 
at the collinear
level involve gluonic pole cross sections, rather than the standard
partonic cross sections, the explicit evaluation of the double weighted
asymmetry shows that such factorizations are not generally applicable.
Additional color factors come in at this level, which provide besides
double weighted TMDs for hadron 1 and hadron 2, products of single
weighted functions. The factors multiplying the final result for the 
cross terms, however, is 
not related to the factors appearing in the single weighted results. 
It does involve
products of the T-even and T-odd transverse moments, but each of
these terms with their own factors. 

The analysis in terms of transverse moments, however, also provides us
with an important result for correlators belonging to final state hadrons.
In analogy to the analysis of distribution correlators, the transverse 
moments for fragmentation correlators involve a part 
$\widetilde\Delta_\partial^\alpha$
with a T-even combination of operators and a part
$\Delta_G^\alpha$ with a T-odd combination of operators, the
latter involving an integration over transverse gluon fields
corresponding to vanishing gluon momentum. 
Such gluonic pole matrix elements (and also multi-gluonic pole
matrix elements appearing in higher transverse moments)
vanish for fragmentation. This has been shown in models as well
using spectral analysis of final states~\cite{Gamberg:2008yt,Meissner:2008yf} 
or studies of analytic properties~\cite{Gamberg:2010uw} 
based on the field theoretical correspondence to amplitudes 
mentioned at the end of section~\ref{section2}. 
Since the color structure of the remaining
$\widetilde\Delta_\partial^\alpha$ part is simple, there are no 
complications to absorb the appropriate gluons into gauge links.
This implies that the result of Eq.~\ref{facto5b} also applies for
unintegrated situations, where the dependence on transverse
momenta in the final state is kept. Thus we can  use in Eq.~\ref{facto5b} 
TMD correlators $\Delta(z,k_\st)$ rather than the collinear 
correlators $\Delta(z)$. Instead of Eq.~\ref{facto4} one 
has a tree-level result of the form
\bea
d\sigma & \sim & {\rm Tr}_c\bigl[
U_-^{\dagger}[p_2]\,\Phi(x_1,p_{1\st})
\,U_-[p_2]\,W_+^{\dagger}[p_1,p_{2}]\bigr]
{\rm Tr}_c\bigl[U_-^{\dagger}[p_1]\,\Phi(x_2,p_{2\st})
\,U_-[p_1]\,W_+^{\dagger}[p_1,p_2]\bigr]
\nonumber \\&& \mbox{}\times
\Gamma_1^\ast\,\Delta(z_1,k_{1\st})\,\Gamma_1
\,\Gamma_2^\ast\,\Delta(z_2,k_{2\st})\,\Gamma_2 ,
\label{facto4final}
\eea
with entanglement only involving the initial state TMD correlators,
including dependence on color flow in initial and final state.
In spite of having a T-even operator structure, 
the universal TMD fragmentation correlators $\Delta(z,k_\st)$ or its
transverse moment $\widetilde\Delta_\partial^\alpha(z)$
allows for T-odd fragmentation functions in its parametrization 
such as the Collins function because the non-plane-wave nature of the 
states $\vert K_h, K_X\rangle$ in the matrix elements of Eq.~\ref{frag}
prevents using time-reversal symmetry constraints.
 
The formalism in this paper allows a rich phenomenology, including many
results on single transverse momentum weighted cross sections that 
have already been obtained. It can be used for quark as well as gluon
TMDs, although the latter involve a number of additional complications.
However, we do want to emphasize once more that our results are strictly 
tree-level, which does provide insight into the universality of TMDs 
and which is a necessary condition to study factorization issues,
but it does not provide proof of factorization in QCD. For a discussion
of those aspects we refer to Ref.~\cite{Collins:2011}, to 
Ref.~\cite{Cherednikov:2010uy} on subtleties with gauge link structures at
higher orders and to Ref.~\cite{Aybat:2011zv}
for attempts to bridge the gap between the phenomenology and the more
formal definitions.

\section*{Acknowledgements}
We would like to acknowledge early stage contributions of Christiaan Mantz
and discussions with Daniel Boer, Mert Aybat, Ted Rogers. 
Also acknowledged are discussions with Jan-Wei Qiu, John Collins, Yuji Koike 
and several others at the stimulating Workshop on Opportunities for Drell-Yan 
at RHIC at Brookhaven National Laboratory (11-13 May 2011). 
This work is part of research programs of the foundation for Fundamental 
Research of Matter (FOM) and the National Organization for Scientific 
Research (NWO) as well as the FP 7 EU-program Hadron Physics (No. 227431).
All figures were made using Jaxodraw~\cite{Binosi:2003yf,Binosi:2008ig}.

\appendix
\section{\label{GL}Collinear gauge links}

In this appendix we incorporate the collinear gluons, i.e.\ the first term in
the Sudakov expansion of $A^\mu(p)$ given in Eq.~\ref{A-sud}. This will 
produce the parts of the gauge link along light-like directions conjugate
to the momentum $P$ and in the collinear situation (when all transverse
momenta are integrated over) it produces the gauge link as given in
Eq.~\ref{glcol} in the light-cone correlators.

Looking at the Fourier transformed fields with collinear Wilson lines starting
at $\pm \infty$, we consider the Fourier transform of the field 
$U^{[n]}_{[\pm \infty,\xi]}\,\psi(\xi)$. For the field with Wilson line 
starting at minus infinity we get,
\bea
\bcontraction{}{U}{{}_-^{[n]}}{\psi}
U_-^{[n]}\psi(p) &=&
\int d^4\xi\ \exp\left(i\,p\cdot \xi\right)\,
\mathscr{P}\exp\left(-ig\int_{-\infty}^{\xi\cdot P}
d(\eta\cdot P)\ n\cdot A(\eta)\right)\,\psi(x) .
\eea
Looking only at the relevant component $p^+ = p\cdot n$ one gets,
\bea
\bcontraction{}{U}{{}_-^{[n]}}{\psi}
U_-^{[n]}\psi(p) &=&
\sum_{N=0}^\infty (-i)^N\int_{-\infty}^\infty d\xi^-
\int_{-\infty}^\xi d\eta^-_N \int_{\eta^-_N}^\xi d\eta^-_{N-1}
\ldots\int_{\eta^-_2}^\xi d\eta^-_1
\ A^+(\eta^-_N)\ldots A^+(\eta^-_1)\psi(\xi)\,e^{i\,p^+\xi^-},
\eea
where the arguments run between $-\infty<\eta^-_N < \eta^-_{N-1}< \ldots <
\eta^-_1 < \xi^-$, implemented through $\theta$ functions
$\theta(\eta^-_{N-1}-\eta^-_N)\ldots\theta(\eta^-_1-\eta^-_2)
\theta(\xi^--\eta^-_1)$,
which can be rewritten as momentum-space integrations,
\bea
\bcontraction{}{U}{{}_-^{[n]}}{\psi}
U_-^{[n]}\psi(p^+) &=&
\sum_{N=0}^\infty (-i)^N\int_{-\infty}^\infty d\xi^-
\int_{-\infty}^\infty d\eta^-_N \int_{-\infty}^\infty d\eta^-_{N-1}
\ldots\int_{-\infty}^\infty d\eta^-_1
\nonumber\\
&&\mbox{}\times
\int \frac{dp^+_N}{-2\pi\,i}
\,\ldots\int \frac{dp^+_1}{-2\pi\,i}
\ \frac{e^{-i\,p^+_N(\eta^-_{N-1}-\eta^-_N)}}{p_N^++i\epsilon}\, \ldots
\frac{e^{-i\,p^+_1(\xi^--\eta^-_1)}}{p^+_1+i\epsilon}\,
\ A^+(\eta^-_N)\ldots A^+(\eta^-_1)\psi(\xi^-)\,e^{i\,p^+\xi^-} .
\nonumber
\eea
Including again the dependence on other momenta, we find
\bea
\bcontraction{}{U}{{}_-^{[n]}}{\psi}
U_-^{[n]}\psi(p) &=&
\sum_{N=0}^\infty
\int \frac{d^4p_N}{(2\pi)^4}\ldots \int \frac{d^4p_1}{(2\pi)^4}
\frac{A^n(p_N)}{(x_N+i\epsilon)}
\,\frac{A^n(p_{N-1}-p_N)}{(x_{N-1}+i\epsilon)}
\ldots\frac{A^n(p_1-p_2)}{(x_1+i\epsilon)}\,\psi(p-p_1)
\nonumber
\\ &=&
\sum_{N=0}^\infty
\int \frac{d^4p_1}{(2\pi)^4}\ldots \int \frac{d^4p_N}{(2\pi)^4}
\frac{A^n(p_1)}{(x_1+i\epsilon)}
\,\frac{A^n(p_2)}{(x_1+x_2+i\epsilon)}
\ldots\frac{A^n(p_N)}{(x_1+\ldots +x_N+i\epsilon)}
\,\psi\left(p-\sum_{i=1}^N p_i\right),
\nonumber
\label{baselink-1}
\eea
where $x_i = p_i\cdot n$.
We thus find for the gauge link
\be
\bcontraction{}{U}{{}_\pm^{[n]}}{\psi}
\bcontraction{U_\pm^{[n]}\psi(p) = \sum_{M=0}^\infty }{U}{
{}_\pm^{[n](M)}}{\psi}
U_\pm^{[n]}\psi(p) = \sum_{M=0}^\infty U_\pm^{[n](M)}\psi(p),
\label{gauge-expansion}
\ee
where $U_\pm^{[n](0)} = 1$. The gauge link and the terms in
its expansion not only have a particular structure in coordinate or
momentum space, but they also have a charge structure. In particular for
applications in non-abelian gauge theories one has matrix-valued fields
$A^\mu = A^{\mu a}T^a$.

The gauge link can be written in a nicer symmetric form for the correlators.
Commutators  $[A^n(p_1),A^n(p_2)]$ don't matter here, neither in color
space, where they are `contained' in other matrix elements with less gluons, 
nor in Hilbert space, where one has commuting fields, which for
fields with the light-cone index $n$ imply vanishing commutators.
We can then use relations
\bea
\frac{1}{(x_1+x_2+i\epsilon)}
\left[\frac{1}{(x_1+i\epsilon)} + \frac{1}{(x_2+i\epsilon)}\right]
&=& \frac{1}{(x_1+i\epsilon)(x_2+i\epsilon)},
\\
\frac{1}{(x_1+x_2+x_3+i\epsilon)}
\underbrace{\left[\frac{1}{(x_1+x_2+i\epsilon)(x_1+i\epsilon)}
+ \ldots\right]}_{\mbox{6 permutations}}
&=&
\frac{1}{(x_1+x_2+x_3+i\epsilon)}
\underbrace{\left[\frac{1}{(x_1+i\epsilon)(x_2+i\epsilon)}
+ \ldots\right]}_{\mbox{3 permutations}}
\nonumber \\
&=& \frac{1}{(x_1+i\epsilon)(x_2+i\epsilon)(x_3+i\epsilon)},
\eea
and its generalization to more terms, to symmetrize the result.
This gives
\bea
\bcontraction{}{U}{{}_-^{[n]}}{\psi}
U_-^{[n]}\psi(p) &=&
\int d^4\xi\ \exp\left(i\,p\cdot \xi\right)\,
\mathscr{P}\exp\left(-ig\int_{-\infty}^{\xi\cdot P}
d(\eta\cdot P)\ n\cdot A(\eta)\right)\,\psi(x)
\nonumber
\\ &=&
\sum_{N=0}^\infty
\frac{1}{N!}\int \frac{d^4p_1}{(2\pi)^4}\ldots \int \frac{d^4p_N}{(2\pi)^4}
\,\frac{A^n(p_1)\,A^n(p_2) \ldots A^n(p_N)}
{(x_1+i\epsilon)(x_2+i\epsilon) \ldots (x_N+i\epsilon)}
\,\psi\left(p-\sum_{i=1}^N p_i\right).
\label{baselink-2}
\eea
From this expression one sees that the term $U_\pm^{[n](M)}$ is the
consecutive action of $M$ simple (commuting) $U_\pm^{[n](1)}$-connections

For a link along $n$ coming from $+\infty$ one has
\bea
\bcontraction{}{U}{{}_+^{[n]}}{\psi}
U_+^{[n]}\psi(p) &=&
\int d^4\xi\ \exp\left(i\,p\cdot \xi\right)\,
\mathscr{P}\exp\left(-ig\int_{\infty}^{\xi\cdot P}
d(\eta\cdot P)\ n\cdot A(\eta)\right)\,\psi(x)
\nonumber
\\ &=&
\sum_{N=0}^\infty
\int \frac{d^4p_N}{(2\pi)^4}\ldots \int \frac{d^4p_1}{(2\pi)^4}
\frac{A^n(p_N)}{(-x_N+i\epsilon)}
\,\frac{A^n(p_{N-1}-p_N)}{(-x_{N-1}+i\epsilon)}
\ldots\frac{A^n(p_1-p_2)}{(-x_1+i\epsilon)}\,\psi(p-p_1)
\nonumber
\\ &=&
\sum_{N=0}^\infty
\int \frac{d^4p_1}{(2\pi)^4}\ldots \int \frac{d^4p_N}{(2\pi)^4}
\frac{A^n(p_1)}{(-x_1+i\epsilon)}
%\,\frac{A^n(p_2)}{(-x_1-x_2+i\epsilon)}
\ldots\frac{A^n(p_N)}{(-x_1-\ldots -x_N+i\epsilon)}
\,\psi\left(p-\sum_{i=1}^N p_i\right)
\nonumber
\\ &=&
\sum_{N=0}^\infty
\frac{1}{N!}\int \frac{d^4p_1}{(2\pi)^4}\ldots \int \frac{d^4p_N}{(2\pi)^4}
\,\frac{A^n(p_1)\,A^n(p_2) \ldots A^n(p_N)}
{(-x_1+i\epsilon)(-x_2+i\epsilon) \ldots (-x_N+i\epsilon)}
\,\psi\left(p-\sum_{i=1}^N p_i\right).
\eea
The notation used in equations with many fields and links will be
$\bcontraction{}{U}{{}_+^{[n]}}{\psi}
U_+^{[n]}\psi(p) = U_+^{[n]}[p]\psi(p)$.

\section{\label{GLB}Intertwined gauge connections}

\begin{figure}[tb]
\epsfig{file=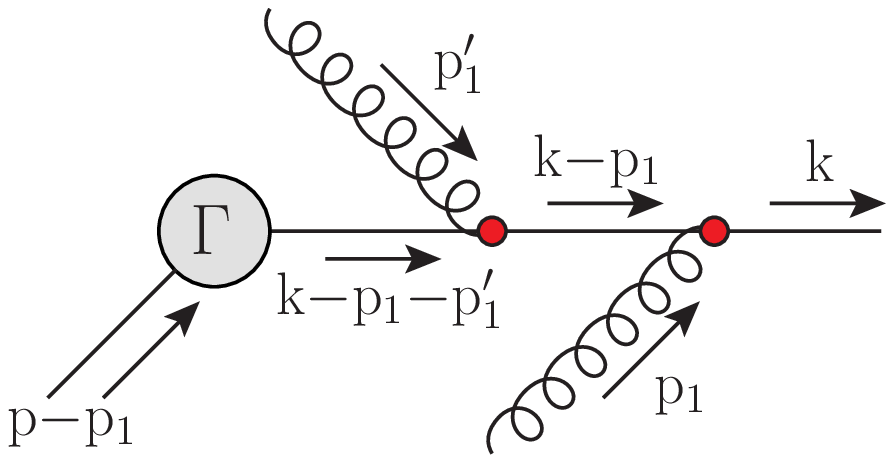,width=0.22\textwidth}
\hspace{1cm}
\epsfig{file=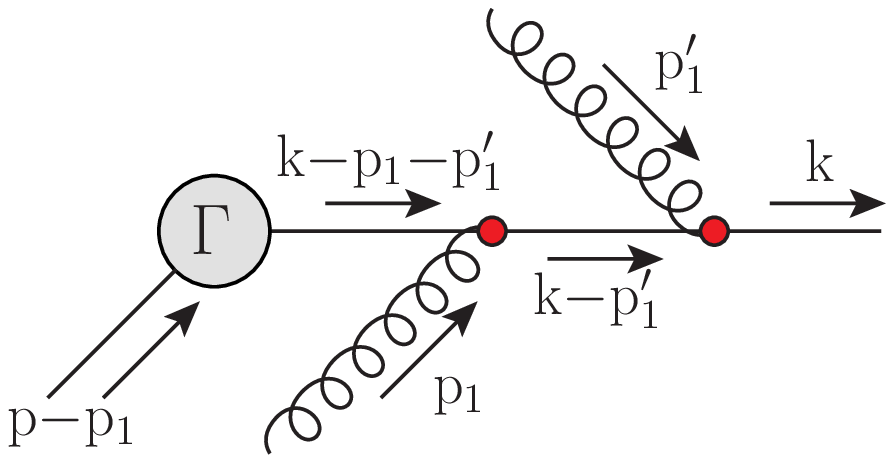,width=0.22\textwidth}
\hspace{1cm}
\epsfig{file=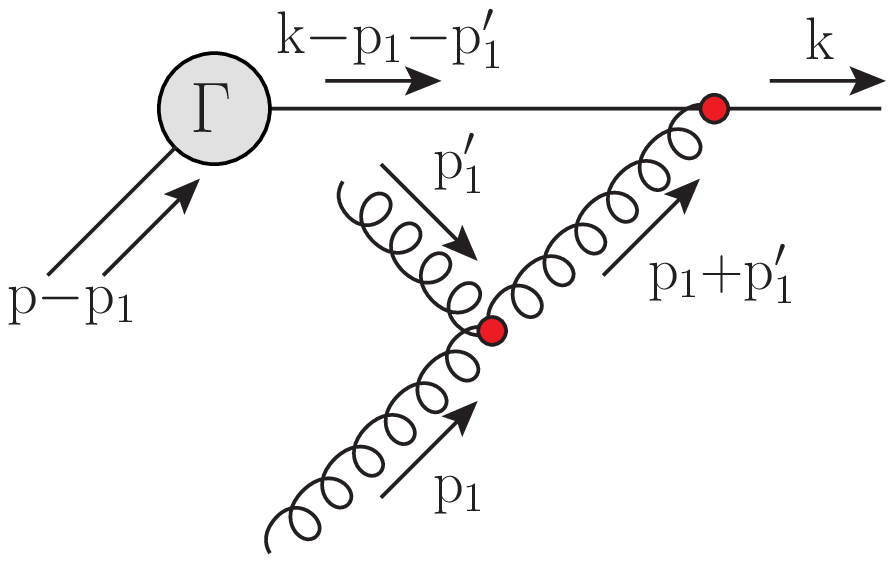,width=0.22\textwidth}
\\[0.2cm]
(a)\hspace{5.0cm} (b)\hspace{5.0cm} (c)
\caption{\label{figB1}
The gluon insertions on an outgoing quark line coming from two different
soft pieces, one from $\Phi(p)$ and one from $\Phi(p^\prime)$, respectively.}
\end{figure}
We consider the situation of gluon insertions on an outgoing quark line 
coming from two different soft pieces, one from $\Phi(p)$ and one from
$\Phi(p^\prime)$, respectively. There are three leading contributions
of $A^k(p_1)\,p_1^\mu$ and $A^k(p_1^\prime)\,p_1^{\prime\mu}$ gluon 
components, corresponding to the diagrams in Fig.~\ref{figB1}, one
of them involving a three-gluon vertex. The combined result of the
diagrams actually nicely adds up to 
\be
A_{11} =
\bigl[\overline\psi(k)\,U_+^{[k](11)}[p,p^\prime]
\,\Gamma\,\psi(p)\bigr]\ldots \psi(p^\prime) =
\tfrac{1}{2}\bigl[\overline\psi(k)
\,\bigl\{U_+^{[k](1)}[p^\prime],U_+^{[k](1)}[p]\bigr\}
\,\Gamma\,\psi(p)\bigr]\ldots \psi(p^\prime),
\ee
which is a gauge connection which is the (color) symmetrized product
of simple connections.

\begin{figure}
\epsfig{file=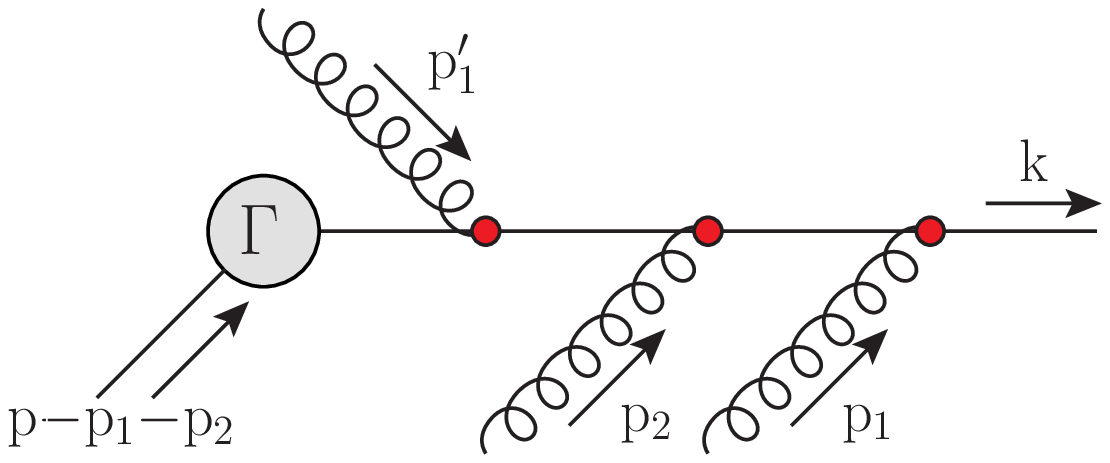,width=0.25\textwidth}
\hspace{1cm}
\epsfig{file=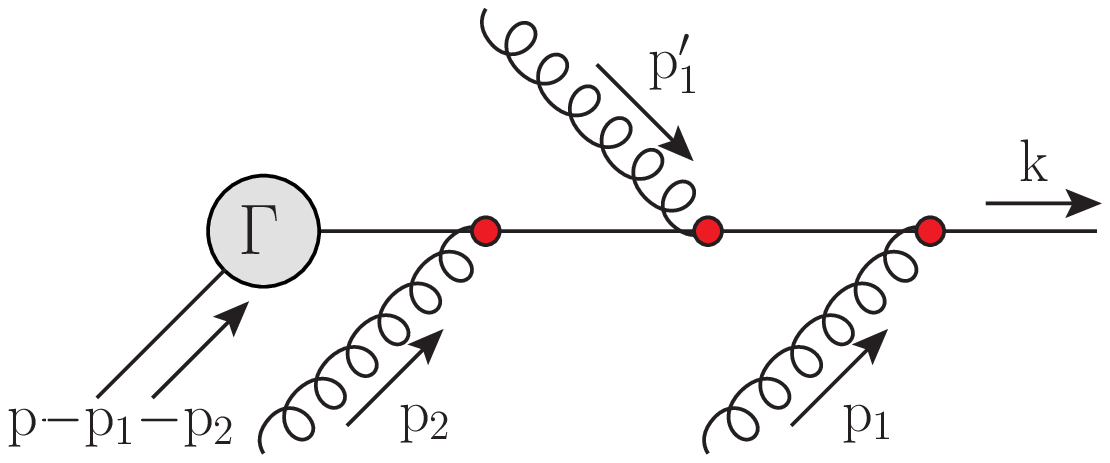,width=0.25\textwidth}
\hspace{1cm}
\epsfig{file=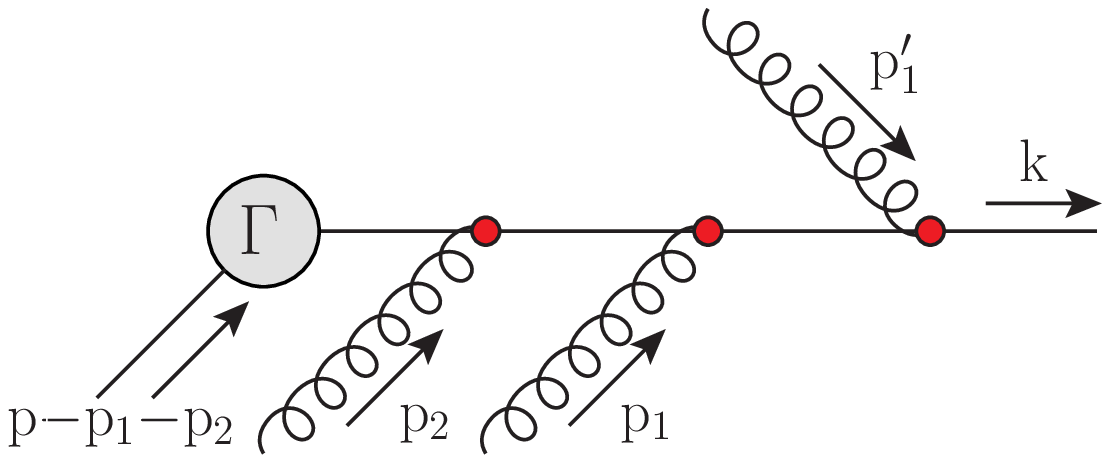,width=0.25\textwidth}
\\[0.2cm]
(a)\hspace{4.0cm} (b)\hspace{4.0cm} (c)
\\
\epsfig{file=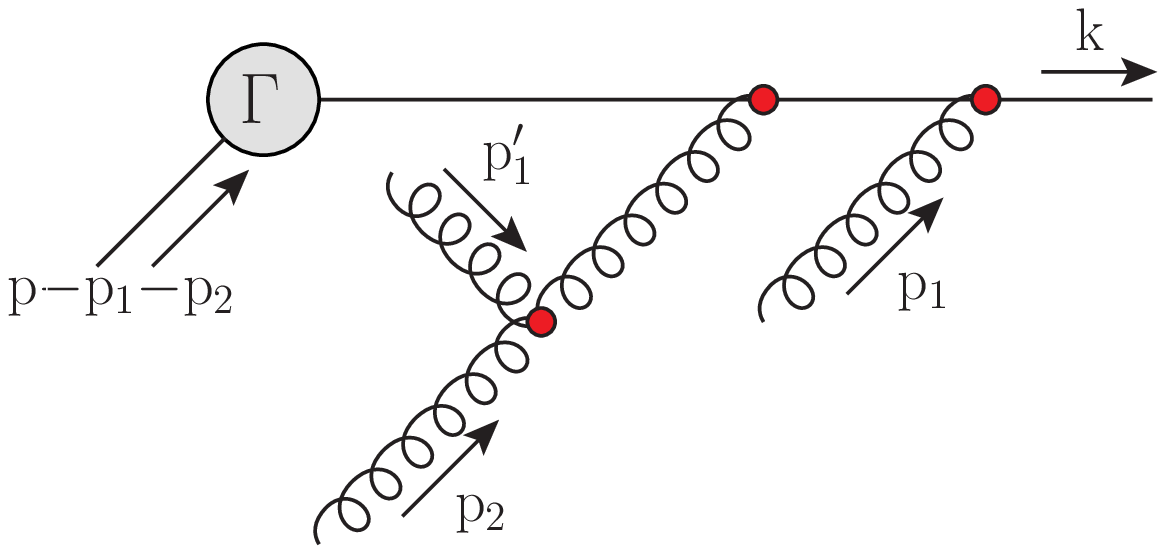,width=0.25\textwidth}
\hspace{1cm}
\epsfig{file=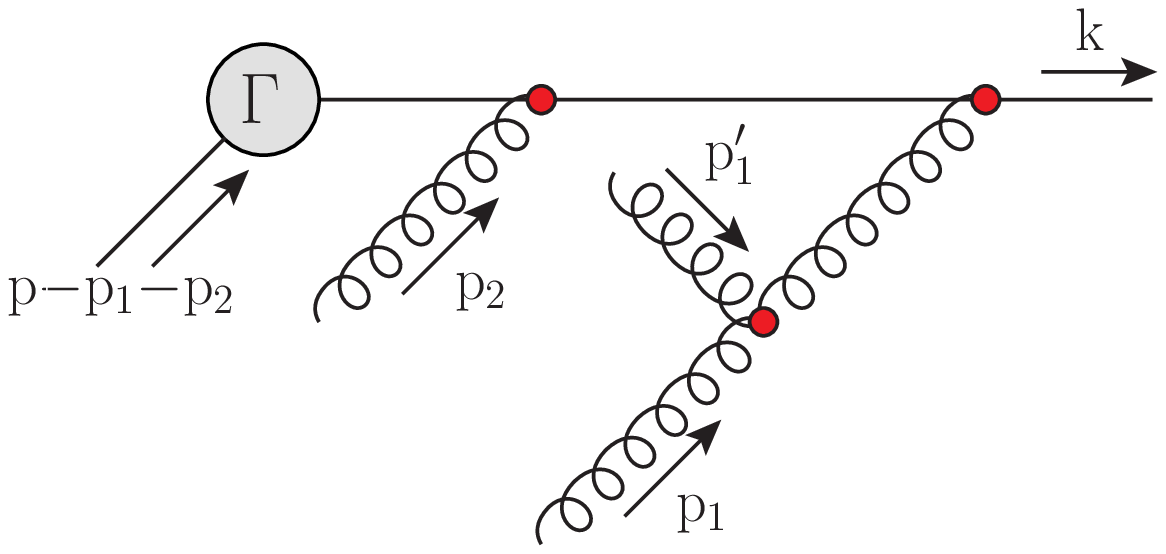,width=0.25\textwidth}
\hspace{1cm}
\epsfig{file=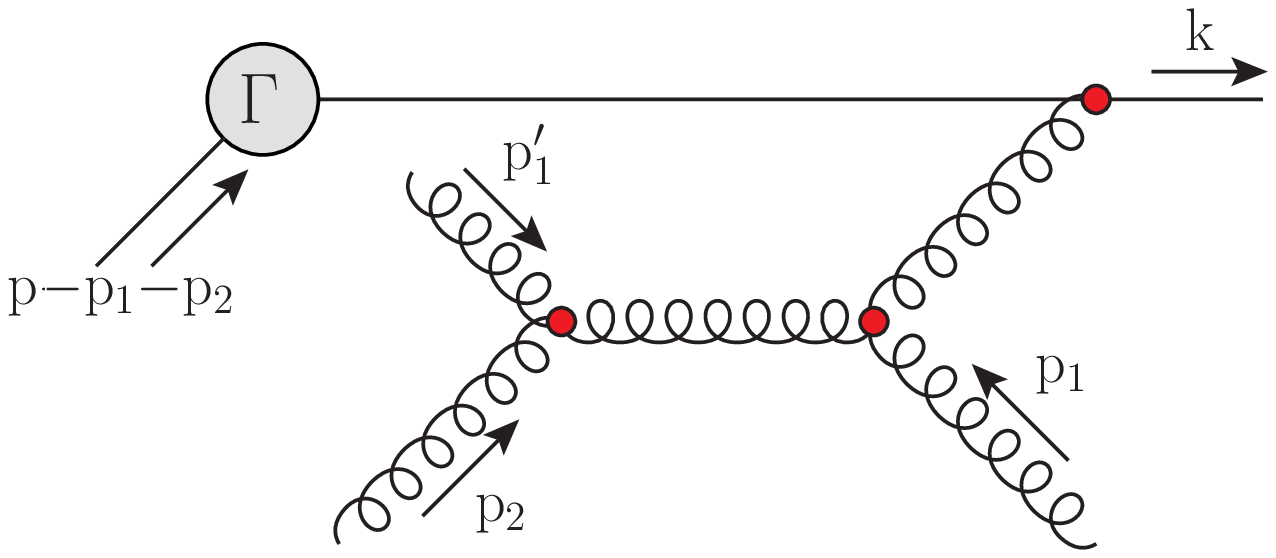,width=0.25\textwidth}
\\[0.2cm]
(d)\hspace{4.0cm} (e)\hspace{4.0cm} (f)
\\
\begin{minipage}{5cm}
\epsfig{file=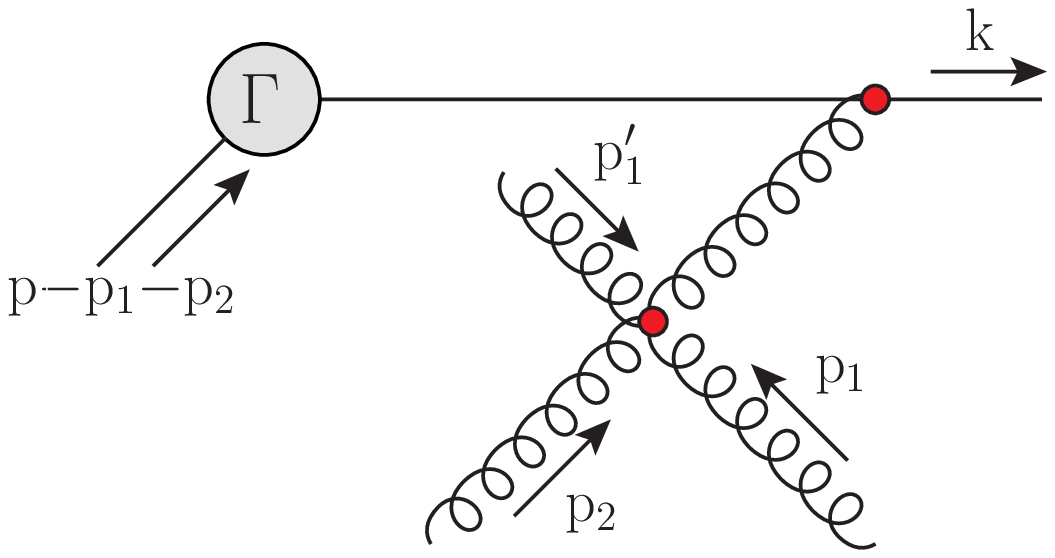,width=0.85\textwidth}
\\
(g)
\end{minipage}
\begin{minipage}{10cm}
\caption{\label{figB2}
The gluon insertions on an outgoing quark line for three gluons
coming from two different soft pieces, two gluons coming from $\Phi(p)$ 
and one coming from $\Phi(p^\prime)$, respectively. Note the absence
of a three-gluon vertex with gluon legs with momenta $p_1$ and $p_2$
coupling to the same soft part $\Phi(p)$.  Note the absence
of a three-gluon vertex with gluon legs with momenta $p_1$ and $p_2$
coupling to the same soft part $\Phi(p)$. It is actually already
included in the soft part with one gluon leg less.}
\end{minipage}
\end{figure}

In order to illustrate the recursive procedure, it is instructive to
give the result for the situation of insertions on an outgoing quark line 
for three gluons coming from two different soft pieces, two gluons coming 
from $\Phi(p)$ and one coming from $\Phi(p^\prime)$, respectively. This
is shown in Fig.~\ref{figB2} and involves three- and four-gluon couplings.
The result becomes
\bea
A_{21} &=&
\bigl[\overline\psi(k)
\,U_+^{[k](21)}[p,p^\prime]
\,\Gamma\ldots \psi(p)\bigr]\ldots \psi(p^\prime)
\\ &=&
\bigl[\overline\psi(k)\bigl(\tfrac{1}{4}\,U_+^{[k](2)}[p]
\,U_+^{[k](1)}[p^\prime]
+\tfrac{1}{4}\,U_+^{[k](1)}[p]\,U_+^{[k](1)}[p^\prime]\,U_+^{[k](1)}[p]
\nonumber \\ &&\mbox{}\hspace{2cm}
+\tfrac{1}{4}\,U_+^{[k](1)}[p^\prime]\,U_+^{[k](2)}[p]\bigr)
\Gamma\ldots \psi(p)\bigr]\ldots \psi(p^\prime)
\\ &=&
\bigl[\overline\psi(k)\bigl(
\tfrac{1}{8}\,U_+^{[k](1)}[p]\,U_+^{[k](1)}[p]\,U_+^{[k](1)}[p^\prime]
+\tfrac{1}{4}\,U_+^{[k](1)}[p]\,U_+^{[k](1)}[p^\prime]\,U_+^{[k](1)}[p]
\nonumber \\ &&\mbox{}\hspace{2cm}
+\tfrac{1}{8}\,U_+^{[k](1)}[p^\prime]\,U_+^{[k](1)}[p]\,U_+^{[k](1)}[p]
\bigr)
\Gamma\ldots \psi(p)\bigr]\ldots \psi(p^\prime) .
\eea
For a general $A_{ijk}$-gluon term, with $i$-, $j$- and $k$ gluons from 
three (or more) different correlators, one finds an 
expression where the $A$-fields from these different correlators are 
a color symmetrized product of $U^{[k](1)}$ factors containing commuting
gauge fields. 
Thus if one, as is the case in Eq.~\ref{linkbreak1}, looks at the result
from insertions on leg $k_2$ coming from correlators $\Phi(p_1)$, 
$\Phi(p_2)$ and $\Delta(k_1)$ it can be broken apart in a symmetrized
product of simple gauge links, 
\be
U_+^{[k_2]}[p_1,p_2,k_1]=
{\mathcal S}\{U_+^{[k_2]}[p_1]U_+^{[k_2]}[p_2]U_+^{[k_2]}[k_1]\},
\label{linkbreakB2}
\ee
of which the ordering is irrelevant.

\bibliographystyle{apsrev}
%\bibliography{references}

\end{document}